\documentclass[11pt]{article}

\usepackage[colorlinks,bookmarksopen,bookmarksnumbered,citecolor=black,urlcolor=red]{hyperref}
\usepackage{setspace}
\usepackage[margin=1in,right=1in,top=1in,bottom=1in]{geometry} 
\usepackage{amsmath,amsthm,amssymb,scrextend, enumerate, mathrsfs}
\usepackage{fancyhdr}
\usepackage{tikz}
\usepackage{pgfplots}
\usepackage{natbib}
\usepackage{amsmath, threeparttable}
\usepackage[labelfont=bf]{caption}
\usepackage[mathscr]{euscript}
 \let\mathscr\relax
\usepackage[scr]{rsfso}
\usepackage{amsthm}
\usepackage{amssymb}
\usepackage{multicol}
\usepackage{multirow}
\usepackage{booktabs}
\usepackage{lscape}
\usepackage{rotating}
\usepackage{mathtools}

\newtheorem{theorem}{Theorem}

\newtheorem{lemma}{Lemma}

\newtheorem{assumption}{Assumption}
\newtheorem{example}{Example}
\newtheorem{remark}{Remark}

\newcommand{\mc}{\mathcal}
\DeclareMathOperator\supp{supp}
\renewcommand{\subsectionmark}[1]{}

\title{Proximal Causal Inference for Modified Treatment Policies}
\author{}
\date{}

\begin{document}

\author{
Antonio Olivas-Martinez \\ Department of Biostatistics, University of Washington
\and
Peter B. Gilbert  \\ Vaccine and Infectious Disease Division, Fred Hutchinson Cancer Center \\ Public Health Sciences Division, Fred Hutchinson Cancer Center \\ Department of Biostatistics, University of Washington
\and
Andrea Rotnitzky \\ Department of Biostatistics, University of Washington
}

\maketitle


\begin{abstract}

The proximal causal inference framework enables the identification and estimation of causal effects in the presence of unmeasured confounding by leveraging two disjoint sets of observed strong proxies: negative control treatments and negative control outcomes. In the point exposure setting, this framework has primarily been applied to estimands comparing counterfactual outcomes under a static fixed intervention or, possibly randomized, regime that depends on baseline covariates. For continuous exposures, alternative hypothetical scenarios can enrich our understanding of causal effects, such as those where each individual receives their observed treatment dose modified in a pre-specified manner—commonly referred to as modified treatment regimes. In this work, we extend the proximal causal inference framework to identify and estimate the mean outcome under a modified treatment regime, addressing this gap in the literature. We propose a flexible strategy that does not rely on the assumption that all confounders have been measured—unlike existing estimators—and leverages modern debiased machine learning techniques using non-parametric estimators of nuisance functions to avoid restrictive parametric assumptions. Our methodology was motivated by immunobridging studies of COVID-19 vaccines aimed at identifying correlates of protection, where the individual's underlying immune capacity is an important unmeasured confounder. We demonstrate its applicability using data from such a study and evaluate its finite-sample performance through simulation studies.
    
\end{abstract}


\section{Introduction}
\label{s:intro}
The methodological developments in this article are motivated by challenges in vaccine research. Vaccine development requires rigorous assessment, including phase 3 randomized efficacy trials often involving lengthy follow-up, which delays rapid deployment of new, effective vaccines. To address this, researchers aim to identify early vaccine-induced immune markers that predict later outcomes and can be used to predict the level of vaccine protection---known as correlates of protection \citep*{plotkin2010correlates, fleming2012biomarkers}---which can serve as surrogates in future trials and potentially expedite vaccine development and approval. Identifying these correlates involves characterizing the causal effect of a continuous exposure---the immune marker---on the outcome of interest. One approach to assess this effect is via the causal dose-response curve, which assigns to each immune marker level the mean outcome that would be observed in a hypothetical world where everyone's marker level is set to that value. A limitation of this approach is that it may be impossible to assign everyone a common immune marker value due to unmanipulable  factors such as immunocompromised conditions.

An alternative approach that has gained popularity in recent years involves the shift-intervention-response curve \citep{hejazi2021efficient, huang2023stochastic}. This curve assigns to each possible shift the mean outcome that would be observed if everyone's level of immune markers were set to their actual vaccine-induced value, adjusted by the specified shift. These interventions are of interest because, for reasonable shift values, the modified immune marker levels could realistically be achieved with alternative vaccine formulations. A shift intervention is an example of a modified treatment policy (MTP) \citep*{robins1986new,diaz2012population,haneuse2013estimation}. More generally, an MTP is an intervention that assigns each individual a treatment dose corresponding to their observed dose in the study but modified in a pre-specified manner. Importantly, MTPs allow treatment assignment to depend on baseline covariates, enabling more flexible and personalized interventions. For example, the amount of dose shift could vary based on individual characteristics such as age, immunocompetence status, or other relevant factors. 

\begin{figure}\centering
    \includegraphics[width=4.5in]{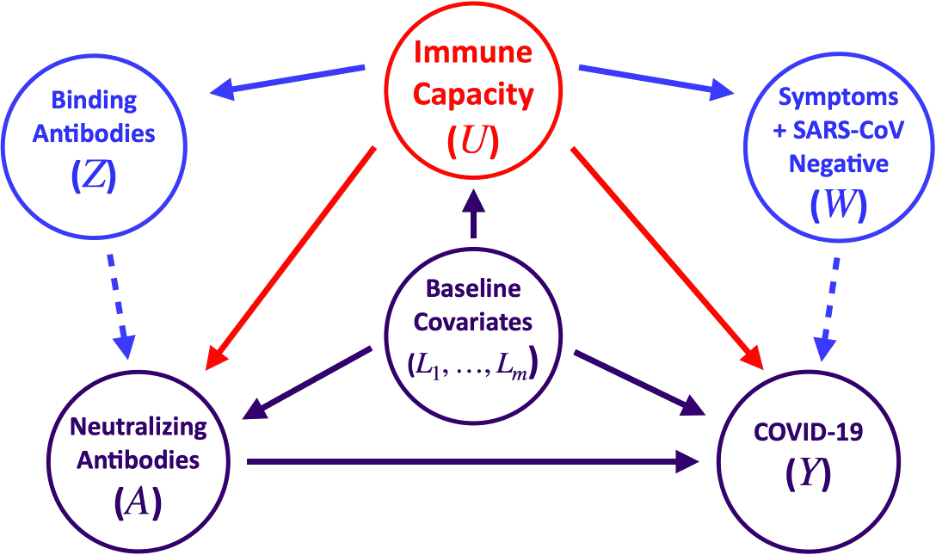}\caption{A causal directed acyclic graph of proximal causal inference. $A$ and $Y$ represent the exposure and outcome of interest; $L$ and $U$ represent the observed and unobserved confounders; and $Z$ and $W$ represent the negative control treatments and negative control outcomes, respectively. In the ENSEMBLE trial, $A$ is neutralizing antibody titers at the Day 29 post-vaccination visit, $Y$ is COVID-19 within 7 months post-Day 29, $L$ includes age, geographic region, and a COVID-19 risk score, $Z$ is anti-spike binding antibody concentration at Day 29, and $W$ is the burden of respiratory symptoms with negative SARS-CoV-2 test results.}\label{fig:dag}
\end{figure}

An additional complication when assessing immune markers as correlates of protection is that vaccine-induced immune marker levels cannot be randomized, making any analysis with immune marker levels as the exposure susceptible to confounding, possibly due to unmeasured confounders. In particular, an individual's underlying immunity capacity---involving unmeasured factors such as immunological genetics, microbiome status, and innate immunity factors that all impact an individual's ability to mount effect immune responses to vaccination or infection---is likely a strong unobserved confounder. A promising solution to this challenge is the proximal causal inference framework, recently introduced by \citet*{miao2018identifying, tchetgen2024introduction}; and \citet*{cui2024semiparametric}. This framework enables the identification and inference of causal effects in scenarios involving unobserved confounders, provided one has access to two disjoint sets of observed strong proxies for the unmeasured confounders. These proxies are commonly referred to as negative control treatments and negative control outcomes. Negative control treatments are proxies of the unmeasured confounders that lack a direct impact on the outcome of interest, while negative control outcomes are proxies that remain unaffected by both the negative control treatments and the primary treatment of interest. Figure~\ref{fig:dag} provides a visual representation of a causal diagram incorporating these negative control variables, illustrating their relationships within the framework. The proximal causal inference framework is well suited for identifying correlates of protection in vaccine efficacy studies, as these trials routinely measure variables that qualify as negative controls, given the antigenic specificity of the adaptive immune system (the lock-and-key mechanism), making the framework's application both feasible and relevant. While estimators of the counterfactual MTP mean under the assumption of no unmeasured confounders have been extensively studied \citep{diaz2012population,haneuse2013estimation,hejazi2021efficient,diaz2023nonparametric}, to our knowledge, estimators under the proximal causal inference assumptions have not yet been developed for this context. In this paper, we address this gap by proposing a novel estimator of the counterfactual MTP mean that leverages the proximal causal inference framework to assess immune markers as correlates of protection in vaccine efficacy studies.

In Section \ref{sec:mot}, we present our motivating example, the ENSEMBLE trial, highlighting the proxies used to address unmeasured confounding. Section \ref{sec:ident} introduces the causal estimand of interest and the assumptions that allow for its identification. In Section \ref{sec:est}, we describe our estimator and provide its asymptotic analysis. Section \ref{sec:exp} evaluates the finite-sample properties of our estimator through numerical experiments. In Section \ref{sec:appl}, we apply our estimator to identify correlates of protection in the ENSEMBLE trial. Finally, we offer general conclusions and future research directions in Section \ref{sec:discussion}.

\section{Motivating Example}\label{sec:mot}

We aim to investigate vaccine-induced neutralizing antibody titers at day 29 (D29) post-vaccination as correlates of protection against COVID-19 infection in the first four months after antibody measurement. This immune marker has been identified as a correlate of protection across various vaccine platforms through multiple research approaches \citep{feng2021correlates,gilbert2022covid, gilbert2022immune, fong2022immune, fong2023immune}. Our analysis utilizes data from the ENSEMBLE trial, a multinational, randomized, placebo-controlled phase 3 study assessing the efficacy of a single dose of Ad26.COV2.S in preventing symptomatic SARS-CoV-2 infection \citep{sadoff2022final}.

Potential observed confounders include age, participant region of residence, and a predefined risk score for COVID-19, while an individual's immunity capacity level is likely a strong unmeasured confounder. The ENSEMBLE trial measured strong proxies of the immunity capacity level that can serve as negative controls: the anti-spike binding antibody concentration at D29 as the negative control treatment and the burden of respiratory symptoms testing negative for SARS-CoV-2 during trial follow-up as the negative control outcome.

Measuring anti-spike binding antibodies is a more accessible alternative to assessing neutralizing antibodies, as it does not require biosafety level 2 or 3 facilities. However, the anti-spike binding antibody assay has a key limitation: it only detects antibodies that bind to the SARS-CoV-2 spike protein without distinguishing those with actual neutralizing activity. Thus, while it can identify the presence of antibodies, it cannot measure their capacity to neutralize the virus \citep{sholukh2020evaluation}. In our analysis, we treat anti-spike binding antibodies as suitable negative control treatments. This approach aligns with an immunological model in which binding to the spike protein is a necessary intermediate step in antibody-mediated protection, but only neutralizing antibodies directly block host cell infection. Under this model, anti-spike binding antibodies should not have an independent effect on the outcome once neutralizing antibodies are accounted for. 

Similarly, SARS-CoV-2-specific antibodies should not protect against other respiratory infections, making non-COVID-19 respiratory symptoms suitable negative control outcomes. The burden of these symptoms serves as a proxy for an individual's general susceptibility to respiratory infections, which may correlate with their underlying immunity capacity level, without being directly affected by SARS-CoV-2 antibodies.

\section{Target of Inference and Identification}
\label{sec:ident}

Consider a prospective cohort study of $n$ individuals randomly selected from a target population. For each subject $i=1,\dots,n$, we observe $m$ baseline covariates $L_i=(L_{i,1}, \dots, L_{i,m})$, a continuous non-randomized exposure $A_i$ assessed at a fixed follow-up time, and an outcome $Y_i$ defined over a prespecified follow-up period after the exposure is measured. Let $Y_i(a)$ be the counterfactual outcome when subject $i$ receives exposure level $A_i=a$. For a random variable or vector $V$, we write $\supp(V)$ to denote its support. We are interested in estimating
\[
\psi_0 = \mathbb{E}\left[Y_i\left\{q(A_i,L_i)\right\}\right],
\]
where $q:\supp{(A,L)}\rightarrow\mathbb{R}$ is a given function. For instance, if $q(a,l)=a+\delta$, then $\psi_0=\mathbb{E}[Y_i(A_i+\delta)]$ is the expected value of the outcome in a hypothetical scenario where every subject receives an exposure that is $\delta$ units higher than the exposure they actually received.

  In our motivating application, the sample consists of the participants randomized to the vaccine arm in the ENSEMBLE trial who actually received the vaccine and were seronegative at enrollment (i.e., had no detectable titers of antibodies against the SARS-CoV-2 virus). The variable $A$ is the vaccine-induced log10 neutralizing antibody titer measured at the Day 29 post-vaccination study visit (D29), $Y$ is the indicator of a new COVID-19 diagnosis between 7 and 115 days after D29, and $L$ is the vector of observed variables deemed as potential confounders. Thus, when $q(a,l)=a+\delta$, $\psi_0$ corresponds to the probability of COVID-19 in a hypothetical world where the neutralizing antibody titers of each participant are $10^\delta$ times higher than those observed in the trial. The shift-intervention-curve $\delta  \rightarrow \mathbb{E}\left[Y_{i} (A_{i} + \delta )\right]$ provides a useful perspective for assessing the causal effect of neutralizing antibody titers on the outcome. Unfortunately, when the support of $A$ is bounded, we cannot identify $\mathbb{E}\left[Y_{i} (A_{i} + \delta )\right]$ for any $\delta \neq 0$, even in the favorable scenario in which the measured covariates $L$ suffice to control for confounding. To see this, suppose \([c, d]\) is the support of \(A\), and use the g-formula to write $\mathbb{E}[Y(A + \delta)] = \int p(l) \, dl \int_c^d \mathbb{E}[Y(a + \delta) \mid A = a, L = l] p(a \mid l) \, da$. When \(L\) suffices to control for confounding, for any \(a\) such that $a$ and \(a + \delta\) are in \([c, d]\), we can infer the counterfactual mean $\mathbb{E}[Y(A + \delta) | A = a, L = l]$ from the observed data since 
\begin{align*}
  \mathbb{E}[Y(A + \delta) \mid A = a, L = l] 
  &= \mathbb{E}[Y(a + \delta) \mid A = a, L = l] \\ 
  &= \mathbb{E}[Y(a + \delta) \mid A = a + \delta, L = l] \\ 
  &= \mathbb{E}[Y \mid A = a + \delta, L = l],
\end{align*}  
where  the second equality follows from $Y(a^{*})$ being conditionally independent of $A$ given $L$, and the third follows if, as we assume throughout, $Y(A)=Y$. However, if \(a \in [c, d]\) but \(a + \delta \notin [c, d]\), the conditional mean $\mathbb{E}[Y \mid A = a + \delta, L = l]$ does not exist, and the conditional counterfactual mean cannot be inferred from the observed data. 

\begin{figure}\centering
    \includegraphics[width=6.5in]{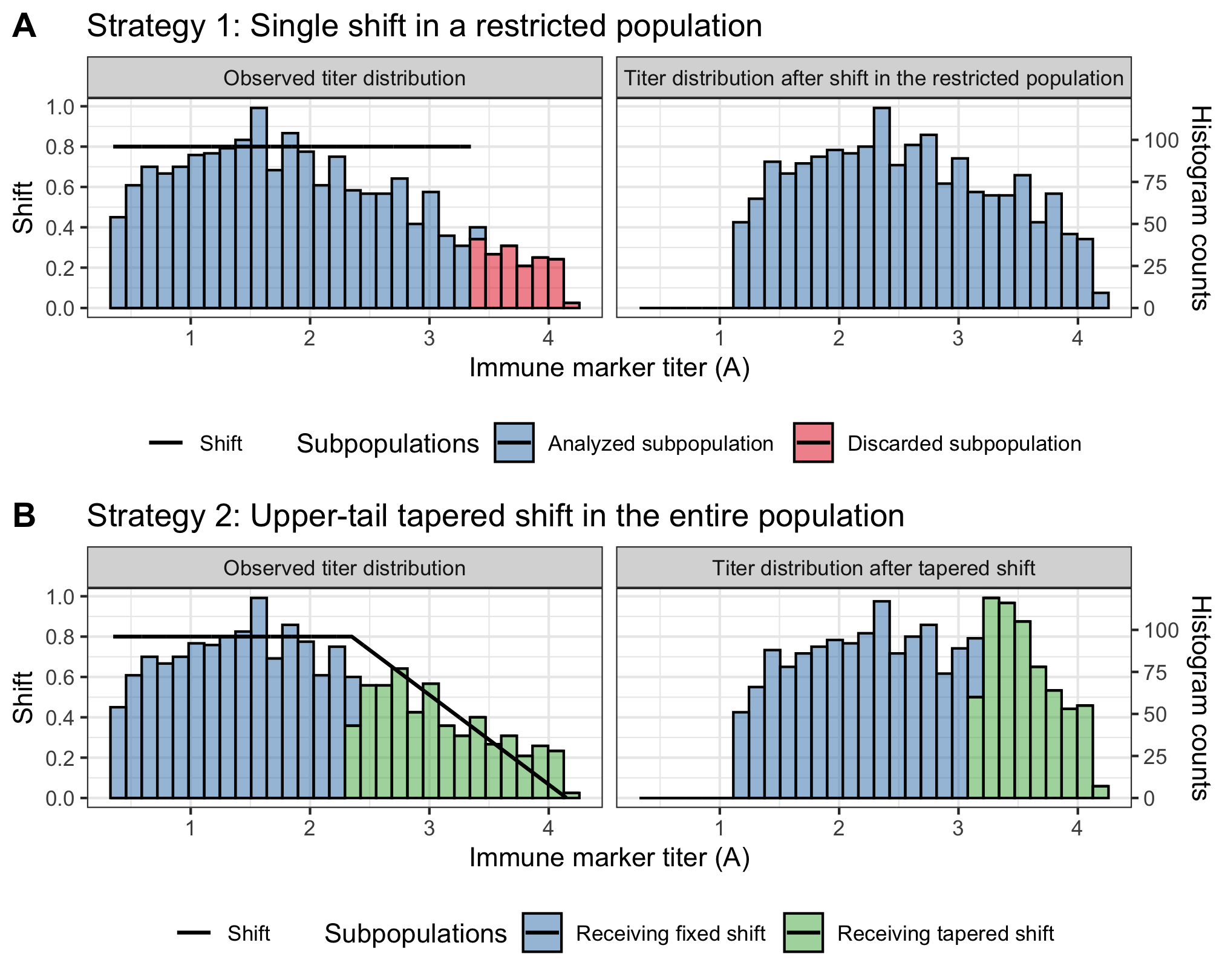}\caption{Two strategies to ensure positivity of the shift intervention policy for a hypothetical immune marker titer exposure. Panel A illustrates a population-restriction strategy, while Panel B illustrates a strategy that modifies the intervention via a tapered shift in the upper tail of the titer distribution.}\label{fig:policies}
\end{figure}

The preceding discussion illustrates the key point that, for an arbitrary $q$, $\mathbb{E}\left[Y\left\{q(A,L)\right\}\right]$ is not identified by the observed data if there exists $a\in\supp{(A)}$ such that $q(a,l)$ falls outside the support of $A$. Two strategies can be employed to address this: (i) restrict the population over which we compute the mean of $Y\{q(A,L)\}$, or (ii) change the policy in a meaningful way so that the image of the new policy always lies within the support of $A$. Figure~\ref{fig:policies} and the following examples illustrate these strategies for policies of the form $q(a,l)=a+\delta$.

\begin{example}[First strategy: Restrict the population]  
For a range of values \(\delta\) of interest, compute the mean of \(Y\{q(A,L)\}\) over a subpopulation defined by \((A,L) \in \mc{S}\), where \(\mc{S}\subset\supp(A,L)\) satisfies the condition that \((a+\delta, l) \) is in the support of $(A,L)$ whenever \((a, l) \in \mc{S}\). For example, if the support of the conditional distribution of \(A\) given \(L\) is constant and equal to \([c, d]\), and \(\delta \in [0, 10]\), we can choose \(\mc{S} = [c, d - 10] \times \supp(L)\), provided \(c < d - 10\).  
\label{ex1}  
\end{example}  

\begin{example}[Second strategy: Change the target policy] Suppose that, for each $l$ in the support of $L$, the support of $A$ is a known finite interval which we denote by $[c(l),d(l)]$. Then, we may consider target policies of the form 
\begin{equation*}
    q^\varepsilon(a,l)=\begin{cases}
         a+\delta &  \text{if } a\in[c(l),d(l)-\delta-\varepsilon(l)),  \\
         a+\frac{\delta}{\delta+\varepsilon(l)}\left\{d(l)-a\right\} & \text{if }a\in[d(l)-\delta-\varepsilon(l), d(l)].
    \end{cases}
\end{equation*}
for a specified positive function $l\mapsto \varepsilon (l)$ such that $c(l)<d(l)-\delta-\varepsilon(l)$ for all $l\in\supp(L)$. The policy $q^\varepsilon(a,l)$ shifts each individual's observed exposure value by $\delta$ if the observed exposure is lower than $d(l)-\delta-\varepsilon(l)$, and by $\frac{\delta}{\delta+\varepsilon(l)}\left\{d(l)-a\right\}$ otherwise. For individuals with observed exposures within $\delta+\varepsilon(l)$ units of $d(l)$, the shift decreases in proportion to the exposure value. We refer to $q^\varepsilon$ as the \textit{$\varepsilon$--upper-tail tapered shift intervention}. \label{ex2}
\end{example}

The trade-offs between these two strategies are evident. The first strategy focuses on simple and interpretable shift policies but restricts the analysis to a truncated subpopulation, limiting generalizability. In contrast, the second strategy evaluates effects across the entire population but introduces slightly more complex and less intuitive policies.

To minimize distracting notational complexity, in the main text we shall focus on target parameters defined using the second strategy; alternatively, parameters as in the first strategy in which \(\mc{S}\) is the full support of \((A, L)\). The more general case, corresponding to \(\mc{S}\) being a specified strict subset of the support of \((A, L)\), is deferred to Appendix~\ref{sec:app_mtp_identification}.  

\subsection{Proximal Identifying Assumptions}

In studies of correlates of protection, as in studies with a non-randomized exposure more broadly, the estimation of counterfactual MTP means is often hindered by unmeasured confounders, such as the unobserved immunity capacity level of individuals in vaccine trials.

Building on the foundational contributions of \citet{miao2018identifying, tchetgen2024introduction}; and \citet{cui2024semiparametric}, we demonstrate that, under certain regularity conditions, the counterfactual MTP mean \(\psi_0\) becomes identifiable if, in addition to baseline covariates \(L\), we also measure a negative control treatment \(Z\) and a negative control outcome \(W\). Heuristically, a negative control treatment \(Z\) is a variable influenced by the unmeasured confounder \(U\) that either has no direct effect on the outcome \(Y\) or has an effect on \(Y\) that is entirely mediated through the exposure \(A\). Similarly, a negative control outcome \(W\) is a variable influenced by the unmeasured confounder \(U\), and potentially by \(Y\), but which is unaffected by the treatment \(A\) and the negative control treatment \(Z\). These relationships are depicted in Figure~\ref{fig:dag}. 

We assume that $(U_i,L_i, Z_i, W_i, A_i,Y^{c}_i)$, where $Y^{c}_i :=\left\{Y_i(a):a\in \text{supp}(A) \right\}$, are $n$ i.i.d. copies of a random vector $(U,L,Z,W,A,Y^{c})$, and for each subject $i$, we observe $O_i=(L_i, Z_i, W_i, A_i,Y_i)$ where $Y_i = Y_i(A_i)$, as stated in Assumption~\ref{ass:consist}. That is, we never observe $U$ and, of the vector of potential outcomes under all possible treatment doses, we only observe the outcome corresponding to the dose actually received. 
We make the following assumptions:

\begin{assumption}[Consistency]
$Y\left( A\right) =Y$.\label{ass:consist}    
\end{assumption}

\begin{assumption}[Randomization of policy-related populations]  For all $(a,l,u)\in\supp(A,L,U)$ and $a'=q(a,l)$, it holds that
$\text{law}\left\{Y(a')|A=a, L=l, U=u\right\}=\text{law}\left\{Y(a')|A=a', L=l, U=u\right\}$.\label{ass:rand}
\end{assumption}

\begin{assumption}[Common support] For all $\left(l,u\right)\in\supp{(L,U)}$, $\supp{(A|L=l, U=u)}\equiv\supp{(A|L=l)}$.\label{ass:pos_mtp}
\end{assumption}

\begin{assumption}[Negative controls]
(i) $Z\bot Y\mid A,L,U$ and (ii)  $\left( A,Z\right) \bot W\mid U,L$.\label{ass:nct}
\end{assumption}

Above and hereafter, $V\bot R|S$ stands for $V$ conditionally independent of $R$ given $S$.

The first assumption is standard in causal inference. The second assumption is tantamount to exchangeability between the population with the given observed exposure and the population whose observed exposure matches the exposure the first population would have under the policy. This condition must hold for every possible observed exposure level. Assumption~\ref{ass:pos_mtp} is analogous to the positivity assumption of the propensity score in the proximal causal inference literature for binary treatments, which similarly requires positivity conditional on unmeasured confounders \citep{miao2018identifying, tchetgen2024introduction}. This assumption cannot be empirically tested because \(U\) is unobserved. Assumption~\ref{ass:nct} (i) defines $Z$ as a negative control treatment and (ii) $W$ as a negative control outcome. In our motivating example, anti-spike binding antibody concentration acts as a negative control treatment that can influence the outcome solely through neutralizing antibody titer. Meanwhile, the burden of non-COVID-19 respiratory symptoms serves as a negative control outcome, as they should not be affected by SARS-CoV-2 antibodies. This negative control outcome is defined by meeting the protocol's respiratory disease symptom criteria for a primary endpoint and testing negative for SARS-CoV-2 infection based on a nucleic acid amplification test.

We also impose the following assumption on the policy:

\begin{assumption}[Monotone and smooth policy] For each $l\in\supp(L)$, $q(\cdot,l)$ is strictly monotone and differentiable almost everywhere with respect to the distribution of $A\mid L=l$.\label{ass:q}
\end{assumption}

Assumption~\ref{ass:q} ensures that, for each $l$, the map $q: a\mapsto q(a,l)$ has an inverse that is almost everywhere differentiable with respect to the distribution of $A$ given $L=l$. The shift-intervention policy and its tapered version discussed in Example~\ref{ex2} satisfy this assumption.

Lastly, following \citet{miao2018identifying} and \citet{cui2024semiparametric}, we make assumptions that allow us to leverage the negative controls to address the unobserved confounding. In particular, we require that the negative controls are sufficiently informative about the unobserved confounder $U$---a property often referred to as \emph{$U$-relevance} in the proximal inference literature \citep{ghassami2022minimax, tchetgen2024introduction}. At a high level, $U$-relevance ensures that the negative controls carry enough information about $U$ to enable identification of causal effects. As established in Theorem~\ref{theo:identification} below, identifiability of the counterfactual MTP mean is then possible if, in addition to Assumptions~\ref{ass:consist}--\ref{ass:q}, one of Assumptions~\ref{ass:lat_out_br} or~\ref{ass:lat_tr_br} holds.

Hereafter, for any random vector $V$, $\mathcal{L}^2(V)$ denotes the space of functions of $V$ with finite second moment, and for any set $B\subset\mathbb{R}$, $I\{a\in B\}=1$ if $a\in B$ and $0$ otherwise. Additionally, we assume that, for every $l\in\text{supp}(L)$, $\text{supp}(A|L=l)$ is an interval, denoted as $[c(l),d(l)]$.

\begin{assumption}[Existence of latent outcome and observed treatment bridge functions]
\begin{enumerate}
    \item[]
    \item[i)] There exists $h_{0}\in\mathcal{L}^2(A,L,W)$, hereafter called the ``latent outcome bridge function", solving, for all $(a,l,u)\in\supp{(A,L,U)}$, the integral equation%
\begin{equation}
\mathbb{E}\left( Y\mid A=a,L=l,U=u\right) =\int h\left( a,l,w\right) p_{W|A,L,U}\left(
w\mid a,l,u\right) dw.  \label{lat_out_eqn}
\end{equation}
\item[ii)] 
There exists $g_{0}\in\mathcal{L}^2(A,L,Z)$, hereafter called the ``observed treatment bridge function", such that, $g_0(a,l,z)=0$ if $a\notin q([c(l),d(l)],l)$ and if $a\in q([c(l),d(l)],l)$ then, for any $(a,l,w)\in\supp{(A,L,W)}$ satisfying that the map $a'\mapsto q^{-1}(a',l)$ is differentiable at $a'=a$, $g_{0}$ solves the integral equation
\begin{equation}
\alpha_{0}\left( a,l,w\right) =\int g\left( a,l,z\right) p_{Z|A,L,W}\left(
z\mid a,l,w\right) dz,  \label{obs_tx_eqn}
\end{equation}
where
\[
\alpha_{0}\left( a,l,w\right) :=I_q(a,l) \frac{dq^{-1}(a,l)}{da}\frac{p_{A|L,W}\left\{q^{-1}\left( a,l\right)
\mid l,w\right\} }{p_{A|L,W}\left( a\mid l,w\right) }
\]
and $I_q(a,l):=I\left\{ a\in q\left([c(l),d(l)]  ,l\right) \right\}$.
\end{enumerate}
\label{ass:lat_out_br}
\end{assumption}

\begin{assumption}[Existence of latent treatment and observed outcome bridge functions] 
\begin{enumerate}
\item[]
\item[i)] There exists $g_{0}\in\mathcal{L}^2(A,L,Z)$, hereafter called the ``latent treatment bridge function", such that, $g_0(a,l,z)=0$ if $a\notin q([c(l),d(l)],l)$ and if $a\in q([c(l),d(l)],l)$ then, for any $(a,l,u)\in\supp{(A,L,U)}$ satisfying that the map $a'\mapsto q^{-1}(a',l)$ is differentiable at $a'=a$, $g_{0}$ solves the integral equation $\widetilde\alpha_{0}\left( a,l,u\right) =\int g\left( a,l,z\right) p_{Z|A,L,U}\left(
z\mid a,l,u\right) dz$ where
\[
\widetilde\alpha_{0}\left( a,l,u\right) :=I_q(a,l) \frac{dq^{-1}(a,l)}{da}\frac{p_{A|L,U}\left\{q^{-1}\left( a,l\right)
\mid l,u\right\} }{p_{A|L,U}\left( a\mid l,u\right) }.
\]
    \item[ii)] There exists $h_{0}\in\mathcal{L}^2(A,L,W)$, hereafter called the ``observed outcome bridge function", solving, for all $(x,l,z)\in\supp{(A,L,Z)}$, the integral equation%
\begin{equation}
\mathbb{E}\left( Y\mid A=a,L=l,Z=z\right) =\int h\left( a,l,w\right) p_{W|A,L,Z}\left(
w\mid a,l,z\right) dw.  \label{obs_out_eqn}
\end{equation}
\end{enumerate}
\label{ass:lat_tr_br}
\end{assumption}

The existence of the bridge functions postulated in Assumptions~\ref{ass:lat_out_br} or~\ref{ass:lat_tr_br}, encodes the precise sense in which the negative control variables suffice to capture the variability of the unmeasured $U$ in order to control for unmeasured confounding. To gain some intuition, consider Assumption~\ref{ass:lat_out_br} (a similar interpretation applies to Assumption~\ref{ass:lat_tr_br}, but with the roles of $W$ and $Z$ reversed). Following \citet{miao2018identifying} and \citet{cui2024semiparametric}, and using Piccard's theorem (see Theorem 15.18 in \cite{kress2014}), we show in Appendix~\ref{sec:appA_existence} that equation~\eqref{lat_out_eqn} has a solution if, for all $(a,l)$:
\begin{equation}
\eta(a,l,U=u)=0 \quad \forall  u\quad \text{whenever} \quad \mathbb{E}\{\eta(A,L,U)\mid A=a,L=l,W=w\}=0 \quad \forall w.
\label{cc1}
\end{equation}
This condition is sufficient when $W$ and $U$ are discrete and finitely valued; otherwise, it must be combined with a regularity condition provided in Appendix~\ref{sec:appA_existence}. Condition~\eqref{cc1} is often referred to as a completeness property. It asserts that, within any fixed level of \((A, L)\), no function of \(U\) can be  orthogonal to the information provided by \(W\). In other words, \(W\) must be sufficiently informative about \(U\) such that any transformation of \(U\) that averages to a constant when conditioned on \(W\) must itself be constant. In particular, this implies that, within any fixed level of $(A,L)$, the first, second, and all higher moments of $U$ vary with \(W\). Put differently, $W$ has predictive value for every direction of variation in \(U\).
Similarly, it can be shown that, under regularity conditions, equation~\eqref{obs_tx_eqn} has a solution if, for all $(a,l)$
\begin{equation*}
\eta(a,l,W=w)=0 \quad \forall  w\quad \text{whenever} \quad \mathbb{E}\{\eta(A,L,W)\mid A=a,L=l,Z=z\}=0 \quad \forall z.
\end{equation*}
This indicates that, conditional on $(A,L)$, $Z$ must have predictive value for every possible direction of variation in $W$. Under the conditional independence postulated by Assumption~\ref{ass:nct}(ii), this is only possible due to dependence of $U$ and $Z$. For additional insights, see \citet{miao2018identifying, tchetgen2024introduction}; and \citet{cui2024semiparametric}.

 The following result establishes that, for negative control variables, latent bridge functions are also observed bridge functions. Part (i) was established by \citet{kallus2022causal}, while part (ii) is new and specific to the MTP setting.

\begin{theorem}
\begin{enumerate}
    \item[]
    \item[i)][\cite*{kallus2022causal}] Under Assumption~\ref{ass:nct}, any latent outcome bridge function $h_0$, i.e., satisfying the conditions in Assumption~\ref{ass:lat_out_br}(i), is also an observed outcome bridge function, i.e. it satisfies the conditions in Assumption~\ref{ass:lat_tr_br}(ii).
    \item[ii)] Under Assumption~\ref{ass:nct}(ii), any latent treatment bridge function $g_0$ is also an observed treatment bridge function. 
\end{enumerate}
\label{theo:lat_imply_ob}
\end{theorem}

With these premises, and building on the approaches of \citet{miao2018identifying, kallus2022causal, ghassami2022minimax}; and \citet{cui2024semiparametric}, we are now in a position to derive the identification result for the MTP counterfactual mean. Hereafter, for any $h\in\mathcal{L}^2(A,L,W)$ and $g\in\mathcal{L}^2(A,L,Z)$, we define $\phi(O;h,g) := h\left\{q\left( A,L\right),L,W\right\}+ g(A,L,Z)\left\{Y-h(A,L,W)\right\}$.

\begin{theorem} Suppose Assumptions~\ref{ass:consist} --\ref{ass:q} hold. If, in addition, either Assumption~\ref{ass:lat_out_br} or Assumption~\ref{ass:lat_tr_br} holds, then $\psi_0$ is identified by the observed data and satisfies all of the following:
\begin{align*}
     \psi_0=\mathbb{E}\left[ h_0\left\{ q\left( A,L\right),L,W\right\}\right]= \mathbb{E}\left\{ Yg_0\left( A,L,Z\right)\right\}= \mathbb{E}\left\{ 
\phi(O; h^\dag,g^\dag)\right\},
\end{align*}
where $h_0$ is an observed outcome bridge function, $g_0$ is an observed treatment bridge function, and either $h^\dag$ is an observed outcome bridge function or $g^\dag$ is an observed treatment bridge function, but not necessarily both functions are observed bridge functions.

    \label{theo:identification}
\end{theorem}

The identifying expressions in Theorem~\ref{theo:identification} parallel those that hold when $U$ is observed. In that case, $\psi_0$ equals any of the three expressions in Theorem~\ref{theo:identification}, with $h_0(a,l,w)$ replaced by $\mathbb{E}(Y\mid A=a,L=l,U=u)$, $g_0(a,l,z)$ replaced by $\widetilde\alpha_0(a,l,u)$ as defined in Assumption~\ref{ass:lat_tr_br} and $h^\dag$ and $g^\dag$ by functions of $e(a,l,u)$ and $\alpha(a,l,u)$ such that $\alpha=\widetilde\alpha_0$ or $e(a,l,u)=\mathbb{E}(Y\mid A=a,L=l,U=u)$ \citep{diaz2012population,haneuse2013estimation}. In fact, when $Z=W$, the representations of the counterfactual MTP mean in Theorem~\ref{theo:identification} coincide exactly with their counterparts under no unmeasured confounding. We refer to these representations as outcome regression (OR), density-quotient-weighted (DQW), and doubly robust (DR), respectively.

The results in Theorem~\ref{theo:identification} align with those in the proximal causal inference literature for other target parameters. For example, \cite{miao2018identifying,kallus2022causal}; and \cite{cui2024semiparametric} established identifiability of the counterfactual mean in a hypothetical scenario where everyone receives the same treatment, under assumptions analogous to those in Theorem~\ref{theo:identification}, but with $\widetilde\alpha_0$ replaced by the inverse of the propensity score. Notably, as in their work, the identifying expressions in Theorem~\ref{theo:identification} remain valid even if equations~\eqref{obs_tx_eqn} and/or~\eqref{obs_out_eqn} have multiple solutions, provided that $g_0$ and $h_0$ are any solutions to these equations, respectively.

\section{Estimation}
\label{sec:est}
Theorem~\ref{theo:identification} suggests three estimators of $\psi_0$, namely
\begin{align*}
    \hat\psi^{OR}&=\mathbb{E}_n\big[\hat h\{q(A,L), L, W\}\big],\\
    \hat\psi^{DQW}&= \mathbb{E}_n\big[Y\hat g(A,L,Z)\big],\shortintertext{and}
\hat{\psi}^{DR}&=\mathbb{E}_n\big[\phi(O; \hat h, \hat g)\big],
\end{align*}
where $\hat{g}$ and $\hat{h}$ are estimators of some solutions $g_0$ and $h_0$ of equations~\eqref{obs_tx_eqn} and~\eqref{obs_out_eqn}, respectively. Here and throughout, $\mathbb{E}_n[\cdot]$ denotes the empirical mean operator, i.e., $ \mathbb{E}_n\left[t(O)\right]=\frac{1}{n}\sum_{i=1}^nt(O_i)$ for $n$ i.i.d. copies $O_1,\dots,O_n$ of $O$.

The estimators $\hat{\psi}^{OR}$ and $\hat{\psi}^{DQW}$ generally fail to achieve $\sqrt{n}$--consistency unless $\hat{h}$ and $\hat{g}$ converge to some respective observed bridge functions at parametric rates. However, this requires correctly specified parametric models for the bridge functions—an often unrealistic assumption. In contrast, as established in Theorem~\ref{theo:asymp_distr} below, the estimator $\hat{\psi}^{DR}$ can achieve $\sqrt{n}$--consistency and asymptotic normality under regularity conditions, even when the bridge functions are estimated non-parametrically. For these reasons, we focus hereafter exclusively on $\hat{\psi}^{DR}$ with bridge functions estimated non-parametrically, using recently proposed methods tailored to our problem \citep{dikkala2020minimax, ghassami2022minimax,olivas2025source}. Nevertheless, for completeness, in Appendix~\ref{sec:parametic_est} we provide unbiased estimating equations for parametric bridge function estimation. We also describe the construction of doubly-robust confidence intervals, i.e., intervals whose asymptotic coverage probability equals the nominal level as long as one parametric bridge estimator converges in probability to a solution of the corresponding observed bridge equation, regardless of which one. 

Following the modern theory of debiased machine learning \citep{robins2008higher,chernozhukov2018double}, we describe a cross-fitted version of the estimator \( \hat{\psi}^{\text{DR}} \). For both \( \hat{\psi}^{\text{DR}} \) and its cross-fitted version to converge to a normal distribution with mean zero at a \(\sqrt{n}\)--rate, a certain random quantity—defined later—must be of order \( o_p\left(n^{-1/2}\right) \). The estimator \( \hat{\psi}^{\text{DR}} \) additionally requires that the estimators $\hat{h}$ and $\hat{g}$, along with the observed bridge functions \( h_0\) and \( g_0 \) to which they converge, lie within function classes subject to specific size constraints, such as Donsker classes. In contrast, this condition is not necessary for the cross-fitted version of \( \hat{\psi}^{\text{DR}} \) to achieve \(\sqrt{n}\) convergence. We focus on the cross-fitted estimator because this relaxation broadens the set of data-generating processes under which the estimator of the target parameter is asymptotically normal.

To construct the cross-fitted DR estimator, we partition the sample into $K$ approximately equal-sized subsets, denoted $\mc I_1, \dots, \mc I_K$, where $K$ is a small number, such as $K = 5$.
For each $k\in\{1,\dots,K\}$, using data from all subsamples except $\mc I_k$, we obtain estimators $\hat h^{(-k)}$ and $\hat g^{(-k)}$ for the bridge functions $h_0$ and $g_0$. The cross-fitted DR estimator is then calculated as 
\[
\hat\psi_{CF}^{DR} =\frac{1}{n}\sum_{k=1}^K\sum_{i:O_i\in \mc I_k}\phi\left\{O_i;\hat h^{(-k)},\hat g^{(-k)}\right\}.
\]

The next theorem establishes results on the asymptotic behavior of $\hat\psi_{CF}^{DR}$. Hereafter, for any function $f$ of the random vector $O$, we define $\Vert f\Vert_\infty:=\sup_{o\in\supp{(O)}}|f(o)|$ and $\Vert f\Vert_2:=\left[\mathbb{E}\{f(O)^2\}\right]^{1/2}$. Note that $\Vert \cdot \Vert_2$ is the 2-norm in the space $\mathcal{L}^2(O)$.

\begin{theorem}
    Suppose that the assumptions of Theorem~\ref{theo:identification} hold. Let $h_0$ and $g_0$ denote observed outcome and treatment bridge functions, respectively. If, for some $B>0$, $|Y|\le B$, $|\alpha_0(A,L,W)|\le B$, and for each $k$, $\Vert\hat h^{(-k)}-h_0\Vert_2=o_p(1)$, $\Vert\hat g^{(-k)}-g_0\Vert_2=o_p(1)$ and either $\Vert h_0\Vert_\infty+\Vert \hat g^{(-k)}\Vert_\infty\le B$ or $\Vert \hat h^{(-k)} \Vert_\infty\ + \Vert g_0\Vert_\infty\le B$, then 
    \begin{equation*}
\sqrt{n}\left\{ \hat{\psi}_{CF}^{DR}-\psi _{0}\right\} =\sqrt{n}\mathbb{E}_{n}%
\left\{ \phi (O;h_{0},g_{0})%
-\psi _{0} \right\} +\sqrt{n}R_{n}+o_{p}(1),
\end{equation*}
    where $R_n=\sum_{k=1}^K\frac{n_k}{n} \:\int\big\{\hat h^{(-k)}(a,l,w)-h_0(a,l,w)\big\}\big\{g_0(a,l,z)-\hat g^{(-k)}(a,l,z)\big\}dP(a,l,w,z)$ and $n_k$ denotes the cardinality of $\mc I_k$. In particular, if $R_n=o_p\left(n^{-1/2}\right)$ and $\mathbb{E}\left\{\phi(O;h_0,g_0)^2\right\}<\infty$, then $ \sqrt{n}\left\{\psi_{CF}^{DR}-\psi_0\right\}\stackrel{\text{d}}{\longrightarrow}\mc{N}(0,\tau^2)$ where $\tau ^{2}=\mathbb{E}\left[\left\{\phi(O;h_0,g_0)-\psi_0\right\}^2\right] $.
\label{theo:asymp_distr}
\end{theorem}

In the next section, we introduce non-parametric estimators $\hat h^{(-k)}$ and $\hat g^{(-k)}$ that, under suitable regularity conditions, converge in $\mc{L}^2$ to the observed bridge functions that minimize the norm associated with the function class used for estimation, among all observed bridge functions contained within that class. These estimators yield $R_n=o_p\left(n^{-1/2}\right)$, which in turn ensures that $\hat\psi_{CF}^{DR}$ is asymptotically normal. The expression for $\tau^2$ suggests estimating it with
\[
\hat \tau^2=\frac{1}{n}\sum_{k=1}^K\sum_{i:O_i\in I_k}\left[\phi\left\{O_i;\hat h^{(-k)},\hat g^{(-k)}\right\}-\hat \psi_{CF}^{DR}\right]^2.
\]
Under regularity conditions, $\hat \tau^2$  is a consistent estimator of $\tau^2$, and consequently, the Wald interval $\psi_{CF}^{DR}\:\pm \:n^{-1/2} \:\Phi^{-1}(1-\xi/2) \:\hat{\tau} $, where $\Phi^{-1}(1-\xi/2) $ is the $1-\xi/2$ quantile of the standard normal distribution, covers $\psi_0 $ with probability converging to $1-\xi$. 

\subsection{Estimation of the Bridge Functions}
\label{sec:est_nuisances}
To estimate the observed bridge functions, we employ kernel regularized adversarial stabilized estimators, a class of adversarial methods recently introduced by \cite{dikkala2020minimax} and \cite{ghassami2022minimax}, and further studied by \cite{olivas2025source}.

The estimation procedure is as follows. Define 
\begin{align*}
    \Psi^{(-k)}_{H,n}(h,g) &:= \mathbb{E}_n^{(-k)}\big[g(A,L,Z)\left\{Y-h(A,L,W)\right\}-g(A,L,Z)^2\big],\\
    \Psi^{(-k)}_{G,n}(g, h) &:= \mathbb{E}_n^{(-k)}\big[h\left\{q(A,L),L,W\right\}-I_q(A,L)h(A,L,W)g(A,L,Z)-I_q(A,L)h(A,L,W)^2\big],
\end{align*} 
where $\mathbb{E}_n^{(-k)}$ denotes the empirical mean over all the observations excluding those in fold $\mc I_k$, and, recall, $I_q(a,l):=I\{a\in q([c(l),d(l)],l)\}$. Let $\mc{H}, \mc{H}' \subseteq \mathcal{L}^2(A,L,W)$ and $\mc{G}, \mc{G}' \subseteq \mathcal{L}^2(A,L,Z)$ denote function classes, each contained in possibly distinct reproducing kernel Hilbert spaces (RKHSs) specified by the data analyst. Let $\Vert \cdot \Vert_{\mc{H}}$, $\Vert \cdot \Vert_{\mc{H}'}$, $\Vert \cdot \Vert_{\mc{G}}$, and $\Vert \cdot \Vert_{\mc{G}'}$ denote the norms of the RKHSs in which $\mc{H}$, $\mc{H}'$, $\mc{G}$, and $\mc{G}'$ are included. For each $k\in\{1,\dots,K\}$, we compute the estimators $\hat h^{(-k)}$ and $\hat g^{(-k)}$ of the bridge functions, using all observations except those in fold $\mc I_k$, as follows:
\begin{align}
    \hat h^{(-k)} & =\arg\min_{h\in\mathcal{H}}\max_{g\in\mathcal{G}'}\left\{ \Psi^{(-k)}_{H,n}(h,g)-\lambda_{\mathcal{G'}}\Vert g\Vert_{\mathcal{G'}}^2+\lambda_{\mc{H}}\Vert h\Vert_{\mc{H}}^2\right\},\label{eq:minmax_h}\\
    \hat g^{(-k)} &=I_q(A,L)\cdot \arg\min_{g\in\mathcal{G}}\max_{h\in\mathcal{H}'}\left\{\Psi^{(-k)}_{G,n}(g, h)-\lambda_{\mathcal{H'}}\Vert h\Vert_{\mathcal{H'}}^2+\lambda_{\mc{G}}\Vert g\Vert_{\mc{G}}^2\right\},\label{eq:minmax_g}
\end{align}
where $\lambda_{\mc{H}}$, $\lambda_{\mc{H}'}$, $\lambda_{\mc{G}}$, and $\lambda_{\mc{G}'}$ are positive tuning parameters.
By restricting attention to function classes within RKHSs, the estimators $\hat{h}^{(-k)}$ and $\hat{g}^{(-k)}$ admit closed-form expressions. Before presenting these expressions, a few remarks are in order.

\begin{remark}
The min-max optimization problems~\eqref{eq:minmax_h} and~\eqref{eq:minmax_g} include penalization terms in both the outer and inner optimization components. The outer terms, $\lambda_{\mathcal{H}}\Vert h\Vert_{\mathcal{H}}^2$ and $\lambda_{\mathcal{G}}\Vert g\Vert_{\mathcal{G}}^2$, known as Tikhonov regularization terms, primarily serve to control the complexity of the solutions and enhance their stability. Tikhonov regularization is standard for solving integral equations like~\eqref{obs_tx_eqn} and~\eqref{obs_out_eqn}, which are ill-posed because their solutions are highly sensitive to small perturbations in the equations' left-hand sides. The inner terms, $\lambda_{\mathcal{G'}}\Vert g\Vert_{\mathcal{G'}}^2$ and $\lambda_{\mathcal{H'}}\Vert h\Vert_{\mathcal{H'}}^2$, are included solely for computational tractability. Penalizing with RKHS norms—rather than, for example, empirical $\mc{L}^2(\mathbb{P}_n)$ norms—avoids the inversion of nearly ill-posed matrices and ensures numerically stable computations \citep{olivas2025source}.
\end{remark}

\begin{remark}
    The estimators $\hat h^{(-k)}$ and $\hat g^{(-k)}$ converge to the observed bridge functions with minimum RKHS norm, denoted $h_0$ and $g_0$, within their respective function classes. Theoretical bounds on their convergence rates rely on a ``source condition'' linking the smoothness of these solutions to the smoothing properties of the associated conditional expectation operators. Under regularity conditions and with $\lambda_{\mathcal{H}}$ and $\lambda_{\mathcal{G}}$ chosen to converge to zero at appropriate rates, \cite{olivas2025source} showed that the remainder term $R_{n}$ is $o_{p}(n^{-1/2})$, ensuring asymptotic normality of the MTP mean estimator. Importantly, this analysis remains valid even if multiple solutions to the observed bridge equations exist, as the estimators consistently target the solutions with minimum RKHS norm.
\end{remark}

\begin{remark}
    The selection of the tuning parameters in practice remains an open problem, the investigation of which is beyond the scope of this work. While a natural approach is to select them by cross-validation, a key challenge lies in the choice of the risk estimator to minimize. See Section~\ref{sec:exp} for the strategy we used in our simulations and data analysis.
\end{remark}

We now provide the expressions for $\hat h^{(-k)}$ and $\hat g^{(-k)}$. To simplify notation, we omit the superscript $\left(-k\right) $ from the bridge function estimators and denote with $n$ the size of the sample from which they are computed. Let $K_{\mc{H}}$, $K_{\mc{H}'}$, $K_{\mc{G}}$, and $K_{\mc{G}'}$ denote the kernel functions associated with the RKHSs, and let $K_{\mc{H},n}$, $K_{\mc{H}',n}$, $K_{\mc{G},n}$, and $K_{\mc{G}',n}$ be the $n\times n$ matrices whose $(i,j)$-th entry corresponds to the value of the respective kernel function evaluated at observations $i$ and $j$. By the ``Representer Theorem'' \citep{scholkopf2001generalized}, $\hat h$ and $\hat g$ admit the following closed form expressions $\hat h(a,l,w)= \sum_{i=1}^n\gamma_iK_{\mc{H}}\left\{(a,l,w),(A_i,L_i,W_i)\right\}$ and $\hat g(a,l,z)=I_q(a,l)\sum_{i=1}^n\theta_iK_{\mc{G}}\left\{ (a,l,z),(A_i,L_i,Z_i)\right\}$ for some constants $\gamma_i$ and $\theta_i$, $i=1,\dots,n$. In Appendix~\ref{sec:rkhs_est}, we show that $\gamma:=\left(\gamma_{1},\dots,\gamma_{n}\right)$ and $\theta:=\left(\theta_{1},\dots,\theta_{n}\right)$ are given by: 
\begin{align*}
\gamma=&\left(K_{\mc{H},n}\Gamma_{\mc{G}'}K_{\mc{G}',n}K_{\mc{H},n}+n^2\lambda_{\mc{H}}K_{\mc{H},n}\right)^{-1} K_{\mc{H},n}\Gamma_{\mc{G}'}K_{\mc{G}',n} \widetilde{Y}_n,\\
\theta=&\left(K_{\mc{G},n}J_n\Gamma_{\mc{H}'}K_{\mc{H}',n}J_nK_{\mc{G},n}+n^2\lambda_GK_{\mc{G},n}\right)^{-1} K_{\mc{G},n}J_n\Gamma_{\mc{H}'} \widetilde{K}_{\mc{H},n}^\top1_{n}.
\end{align*}
 Here, $\widetilde{Y}_n=(Y_1,\dots,Y_n)^\top$ is the vector of outcomes, $1_n=(1,\dots,1)^\top$ the vector of ones, $J_n$ the $n\times n$ diagonal matrix with entries $I_q(A_1,L_1),\dots,I_q(A_n,L_n)$, and $\widetilde{K}_{\mc{H},n}$ is the $n\times n$ matrix with $(i,j)$-entry $K_{\mc{H}'}\left\{(q(A_i,L_i),L_i,W_i),(A_j,L_j,W_j)\right\}$. Moreover, $\Gamma_{\mc{G}'}=4^{-1}\left(n^{-1}K_{\mc{G}',n}+\lambda_{\mc{G}'}Id_{n}\right)^{-1}$
and $\Gamma_{\mc{H}'}=4^{-1}\left(n^{-1}K_{\mc{H}',n}J_n+\lambda_{\mc{H}'}Id_{n}\right)^{-1}$ where $Id_n$ is the $n\times n$ identity matrix. 
 
\section{Numerical Experiments}
\label{sec:exp}
We conducted simulation experiments to assess the finite-sample performance of \(\hat\psi_{CF}^{DR}\) under varying levels of correlation between the negative controls (\(Z\) and \(W\)) and the unmeasured confounder (\(U\)), conditional on observed confounders (\(L\)). For comparison, we evaluated two estimators of the MTP mean that assume no unmeasured confounding: augmented inverse probability weighted estimator (Non-proximal AIPW) and targeted minimum loss-based estimator (Non-proximal TMLE) \citep{diaz2012population, diaz2023nonparametric}. Each experiment included 500 replications, and we considered sample sizes $n=750,1500,$ $ 3000,\text{and} \:6000$. 

Let $\mc{TN}_{[a,b]}(\mu,\sigma^2)$ denote the $\mc{N}(\mu,\sigma^2)$ distribution truncated to the interval $[a,b]$, and let $\mc{TN}_c(\mu,\sigma^2)$ denote  $\mc{TN}_{[a,b]}(\mu,\sigma^2)$ with $a=\mu-3\sigma$ and $b=\mu+3\sigma$. For each $i=1,\dots,n$, we generated  $L_i\sim\mc{TN}_c(0,1)$, $U_i\sim\mc{TN}_c(0.5L_i, 1)$, $Z_i\sim\mc{TN}_c(0.2L_i + \beta_z U_i,1)$, $W_i\sim\mc{TN}_c(0.5L_i + \beta_w U_i,1)$, $A_i\sim\mc{TN}_{[-2,2]}(0.3L_i + U_i,1)$, and $Y_i\sim\text{Bernoulli}\left(p_i\right)$, where $p_i=\big\{1+\exp(1-0.5L_i+U_i+1.5A_i+0.75A_i^2)\big\}^{-1}$. Data were generated across nine scenarios by combining $\beta_z\in\{2, 1, 0.5\}$ (corresponding to $\text{Cor}(Z, U|L) = 0.894, 0.707$ and $ 0.447$) and $\beta_w\in\{-2, -1, -0.5\}$ (corresponding to $\text{Cor}(W, U|L) = -0.894, -0.707$ and $-0.447$). The target parameter was $\mathbb{E}\left[Y\left\{q(A)\right\}\right]$, with $q(a)=a+0.4\cdot I\left\{-2\le a\le 0.6\right\}+\frac{0.4(2-a)}{1.4}\cdot I\left\{0.6< a\le2\right\}$, the $\varepsilon$--upper-tail tapering shift intervention with $\delta=0.4$ and $\varepsilon=1$. Since $(U,L)$ suffices to control for confounding, $\mathbb{E}\left[Y\left\{q(A)\right\}\right]=0.2512$ under all considered scenarios. Theorem \ref{theo:identification}'s assumptions---including the existence of bridge functions---are verified in Appendix~\ref{sec:appC_validating_assumptions}.

We computed \(\hat\psi_{CF}^{DR}\) with bridge functions estimated as in~\eqref{eq:minmax_h} and~\eqref{eq:minmax_g}, using Gaussian RKHSs. We standardized \(A\), \(L\), \(Z\), and \(W\) to have empirical mean 0 and empirical standard deviation 1 before estimating the bridge functions to prevent any of these variables from disproportionately influencing the estimation due to differences in scale. We used tuning parameters $\lambda_{\mc{H}}=c_1\{\log(n')/n'\}^{1/2}$, $\lambda_{\mc{G}'}=c_2\log(n')/n'$, $\lambda_{\mc{G}}=c_3\{\log(n')/n'\}^{1/2}$, and $\lambda_{\mc{H}'}=c_4\log(n')/n'$, where $n'=n/3$, $c_1, c_3\in\{10^{-5},10^{-4}, 10^{-3}, 10^{-2}, 10^{-1}\}$ and $c_2, c_4\in\{0.1, 1, 10,100\}$, and fixed kernel bandwidths $(\sigma^2_{\mc{H}},\sigma^2_{\mc{G}'},\sigma^2_{\mc{G}},\sigma^2_{\mc{H}'})$. The bandwidth $\sigma_{\mc{G}'}^2$ ($\sigma_{\mc{H}'}^2$) was fixed at 1/4 times the median of the pairwise euclidean distance between the observed vectors $(A_i,L_i,Z_i)$ ($(A_i,L_i,W_i)$) \citep{garreau2017large}. The bandwidths $\sigma^2_{\mc{H}}$ and $\sigma^2_{\mc{G}}$ were also computed from the median of the pairwise Euclidean distances, but multiplied by $c_5\in\{1/4,1/2,1,2,4\}$. We used 3-fold cross-fitting, i.e. we partitioned the sample into three folds $\mc I_1,\mc I_2$, and $\mc I_3$ and proceeded as explained in Section $\ref{sec:est}$. 
To compute \(\hat{h}^{(-1)}\), we first used fold 2 data to estimate the minimum-norm solution \(h_0\) using the various combinations of hyperparameters and bandwidths. In fold 3, we selected the configuration that minimized a specific empirical loss function, detailed in Appendix~\ref{sec:cv}. Finally, \(\hat{h}^{(-1)}\) was set as the estimator from fold 2 corresponding to the selected configuration. This procedure was applied similarly to compute \(\hat{h}^{(-2)}\), \(\hat{h}^{(-3)}\), and \(\hat{g}^{(-k)}\) for \(k = 1, 2, 3\).

We implemented the non-proximal MTP estimators using the R package \texttt{lmtp} and adjusting for the observed confounder $L$, and the negative controls $Z$ and $W$. Adjusting for the negative controls ensured a fair comparison with the proximal estimators.

\begin{figure}
\centerline{\includegraphics[width=6.5in]{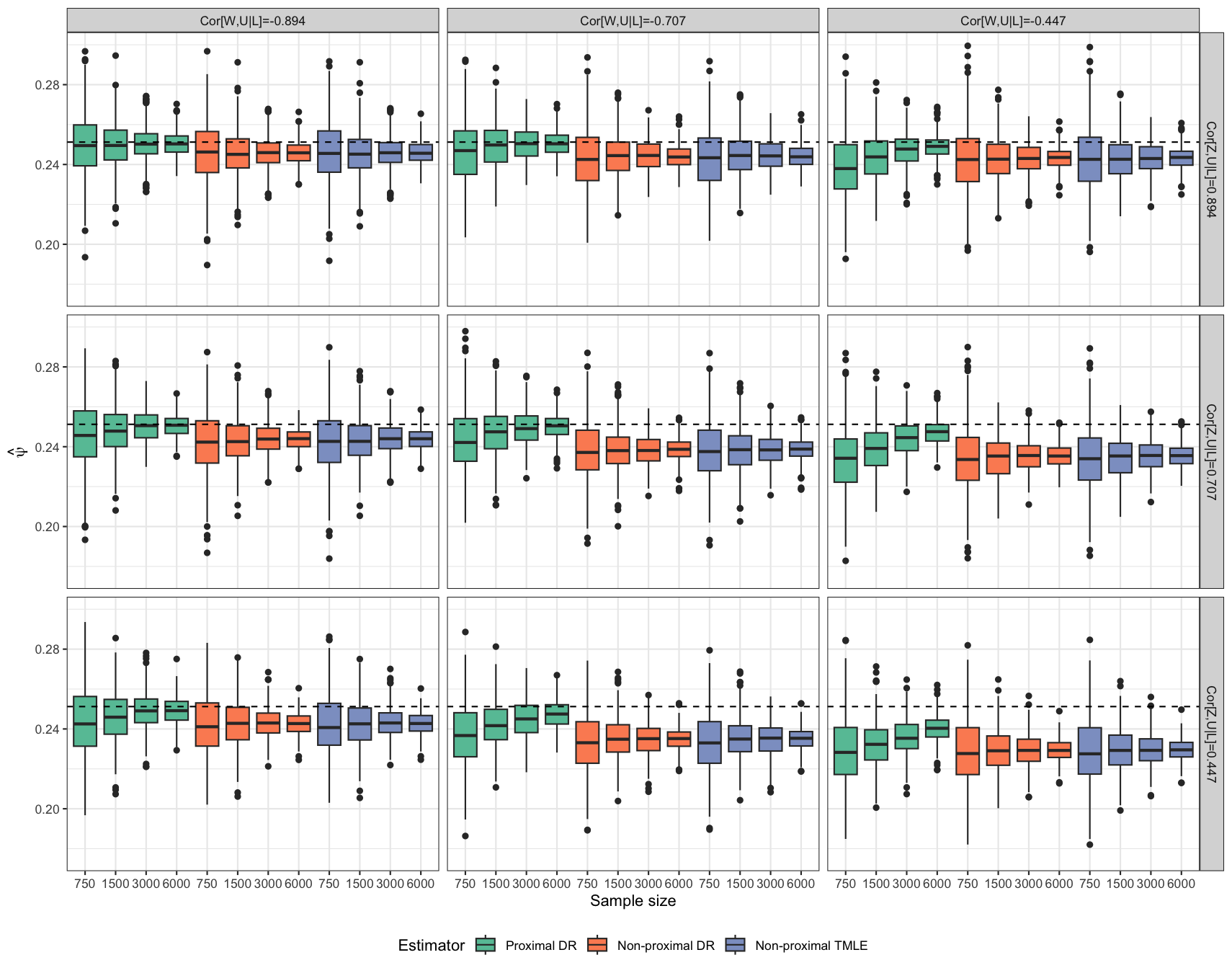}}
    \caption{Boxplots of the estimators from numerical experiments with data generated under varying conditional correlations between the proxies $Z$ and $W$ and the unmeasured confounder $U$, with sample sizes $n = 750$, $1500$, $3000$, and $6000$. The dashed line represents the true parameter value. Non-proximal estimators rely on the (incorrect) assumption of no unmeasured confounding.}\label{fig:sims1}
\end{figure}

\begin{figure}
\centerline{\includegraphics[width=6.5in]{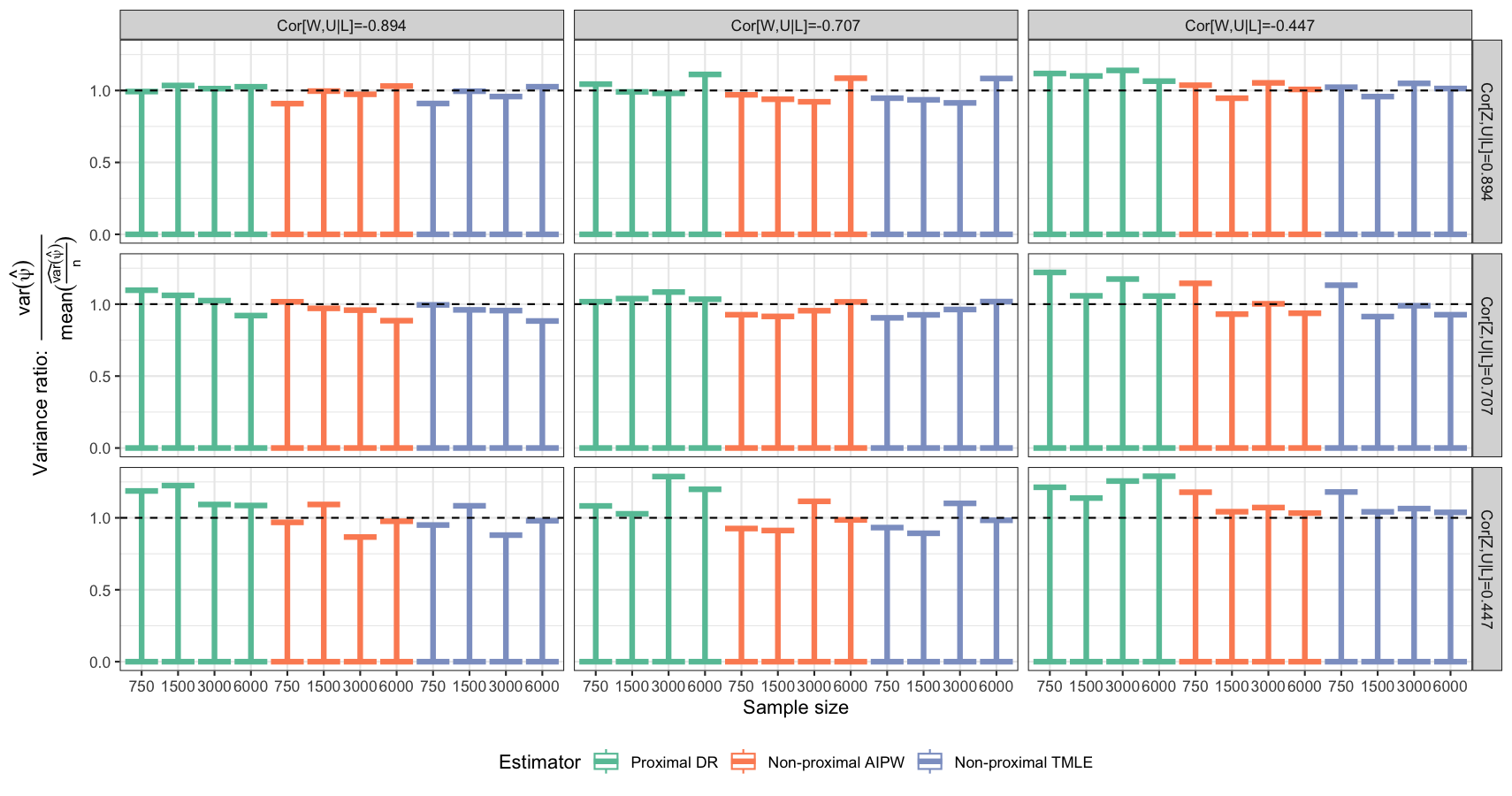}}
    \caption{Ratio of empirical variance to the mean estimated asymptotic variance for each estimator, from numerical experiments with data generated under varying conditional correlations between the proxies $Z$ and $W$ and the unmeasured confounder $U$, across sample sizes $n = 750$, $1500$, $3000$, and $6000$. The dashed line indicates a variance ratio of 1.}\label{fig:sims2}
\end{figure}

\begin{figure}
\centerline{\includegraphics[width=6.5in]{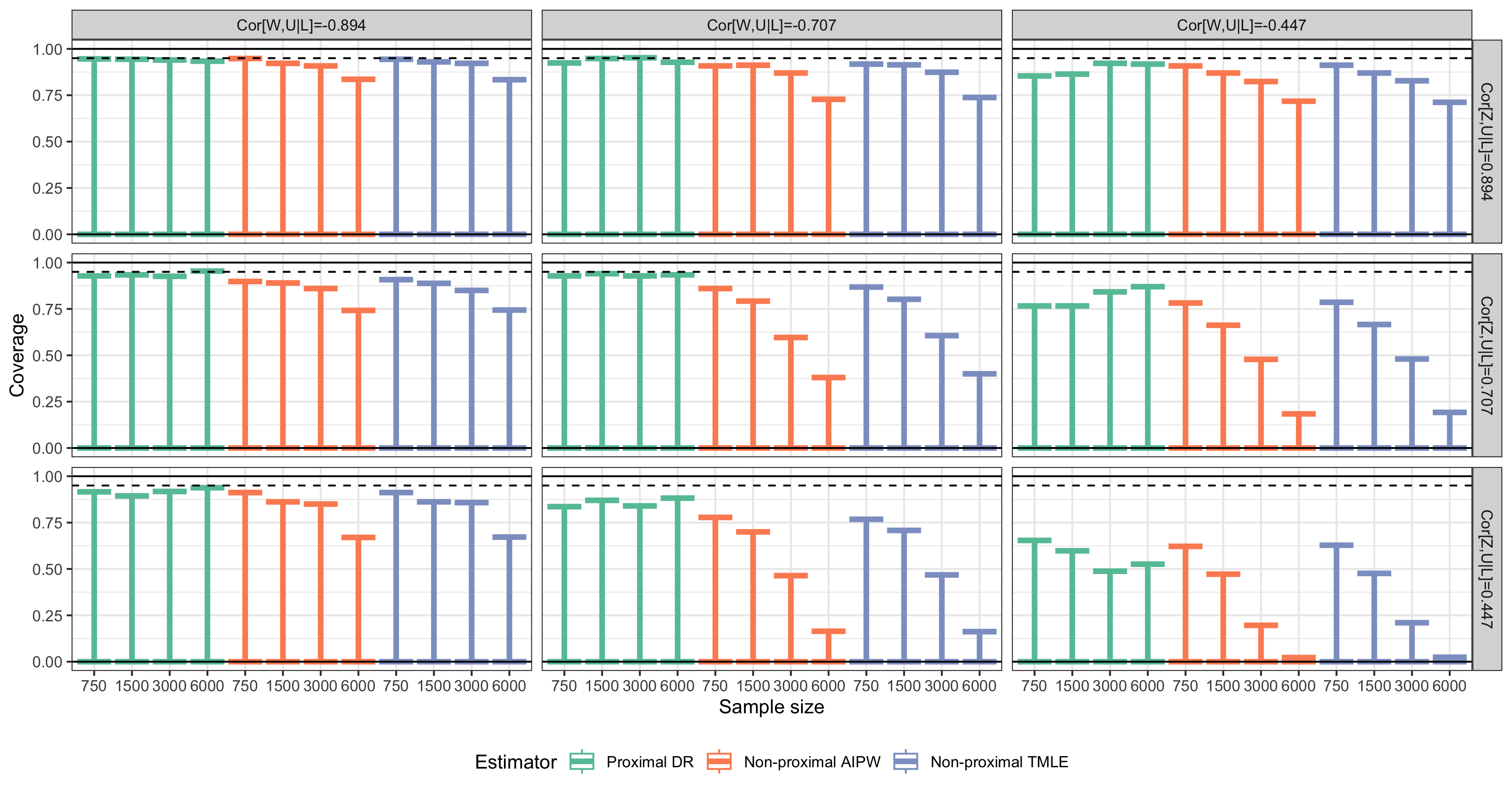}}
    \caption{Coverage probabilities of the confidence intervals from numerical experiments with data generated under varying conditional correlations between the proxies $Z$ and $W$ and the unmeasured confounder $U$, across sample sizes $n = 750$, $1500$, $3000$, and $6000$. The dashed line indicates the nominal coverage level of 0.95. Non-proximal estimators rely on the (incorrect) assumption of no unmeasured confounding.}\label{fig:sims3}
\end{figure}

The performance of our estimator (Proximal DR) and its competitors is summarized in Figures~\ref{fig:sims1}--\ref{fig:sims3}. In most scenarios, the estimator $\hat\psi_{CF}^{DR}$ exhibits low bias, and the average estimated asymptotic variance closely matches its empirical variance. The coverage of the Wald confidence intervals centered at $\hat\psi_{CF}^{DR}$ is close to the nominal level of 0.95. Overall, the proposed estimator performed as predicted by theory across scenarios with varying correlations between the negative controls and the unmeasured confounder, with performance deteriorating as these correlations decrease. As expected, it outperformed conventional non-proximal methods that cannot account for unmeasured confounding.

In Appendix~\ref{sec:parametic_est}, we also present results for a doubly-robust estimator that employs parametric estimators of the bridge functions. These simulations illustrate that estimation of the counterfactual MTP mean also deteriorates as the conditional correlation between the negative controls and the unmeasured confounder weakens, even when the estimation is based on correctly specified parametric models for the bridge functions.

\section{Application to the ENSEMBLE Trial}
\label{sec:appl}
We applied our methodology to investigate log10 neutralizing antibody titer against the ancestral strain at the Day 29 (D29) visit post-vaccination as a correlate of protection against a new COVID-19 diagnosis in the ENSEMBLE trial. The analysis included 1,164 vaccine recipients who were SARS-CoV-2 seronegative at baseline, had D29 antibody measurements, and came from the six participating countries (out of eight) with at least 50 subjects having antibody measurements. These subjects
were selected using a two-phase case-cohort sampling design, as described in \cite{gilbert2022immune}. To account for this sampling design, we employed a weighted version of our estimator, as detailed in Appendix~\ref{sec:rkhs_wt}. The outcome of interest was the binary indicator of developing virologically confirmed COVID-19 with onset between 7 and 115 days after D29. During this follow-up period, 295 COVID-19 cases were observed among vaccine recipients with antibody measurements. The remaining 869 vaccine recipients were treated as non-cases, regardless of follow-up length.

Our analysis included age, geographic region (Latin America: Argentina, Brazil, Colombia, and Peru; South Africa; or the United States), and a pre-specified risk score for COVID-19 as baseline covariates. Log10 anti-spike binding antibody concentration against the ancestral strain served as a negative control treatment, and the burden of non–SARS-CoV-2 respiratory symptoms during follow-up as a negative control outcome. The burden of respiratory symptoms was calculated as the average number of symptoms per day with negative SARS-CoV-2 antigen test results, normalized by the maximum value within country to account for seasonal variation across locations. Descriptive information on baseline covariates, antibody titers, and the negative control outcome by case status is provided in Appendix~\ref{sec:details_app}.

\begin{figure}
\centerline{ \includegraphics[width=6in]{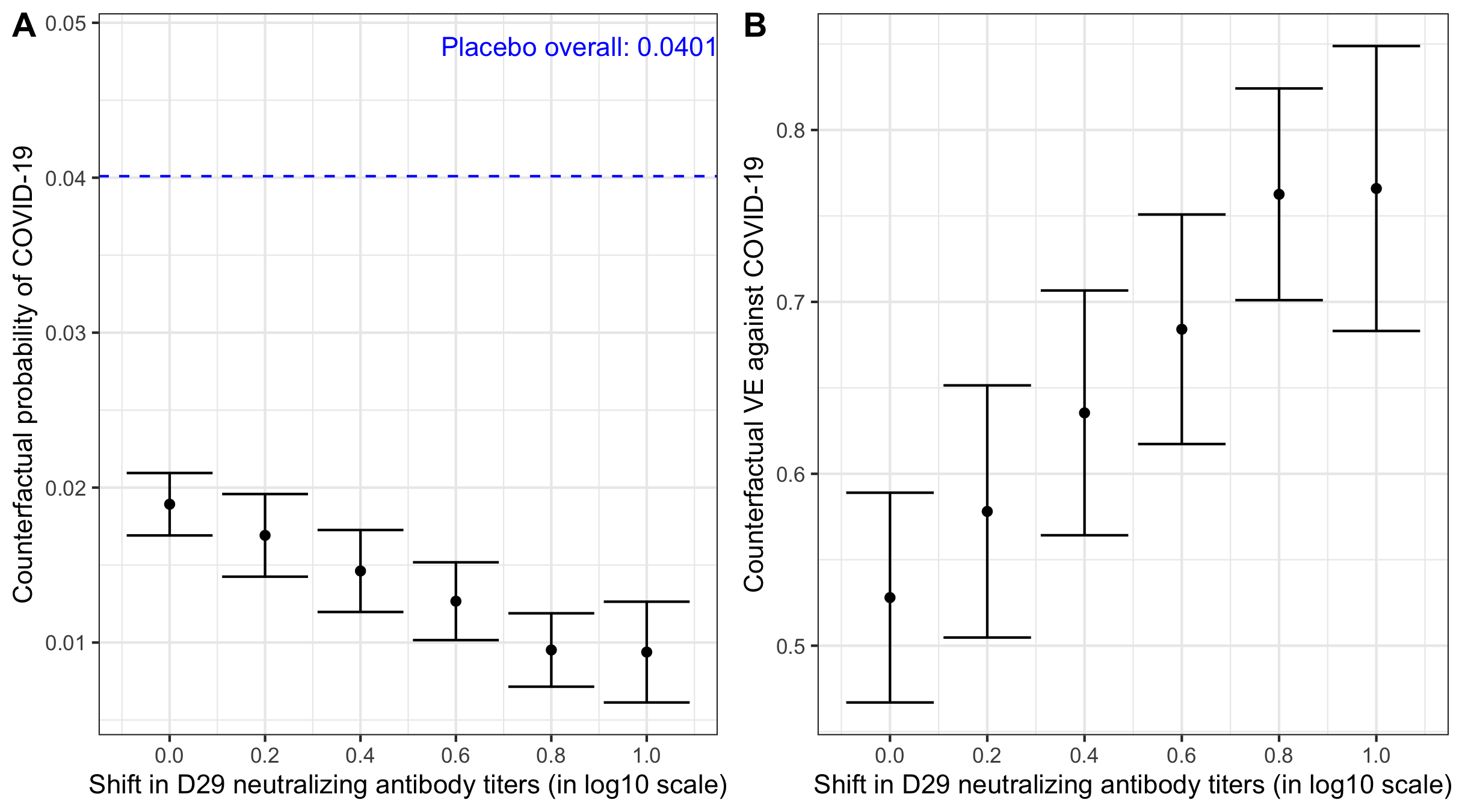}}
   \caption{Results from the application to the ENSEMBLE trial data. Panel A shows observed and counterfactual probabilities of new COVID-19 diagnosis, and Panel B shows  observed and counterfactual vaccine efficacy (VE) against new COVID-19 for $\varepsilon$--upper-tail tapered shift intervention policies with $\varepsilon=1$ and shifts $\delta=0$ (no shift), 0.2, 0.4, 0.6, 0.8, and 1 logarithm, as described in Example~\ref{ex2}. These policies make shifts that are linearly tapered to zero for individuals whose observed titers are within $\delta+\varepsilon$ units of the upper limit of the observed distribution, and apply the full shift to all other individuals. The bars indicate the $95\%$ confidence interval at each shift without any correction.}\label{fig:covid}
\end{figure}

We evaluated five $\varepsilon$--upper-tail tapered shift intervention policies with \(\delta \in \{0.2, 0.4, 0.6, 0.8, 1\}\), \(\varepsilon \equiv 1\), and constant \(c(\cdot)\) and \(d(\cdot)\). Neutralizing antibody titers at D29 (log10 scale) ranged from 0.3889 to 4.2175. The percentages of vaccine recipients with observed antibody titers within the range of each policy were 42.5\%, 38.2\%, 27.3\%, 18.8\%, and 11.3\%, respectively. We applied our doubly-robust cross-fitted estimator using the same folds, function classes, and tuning parameters as in the numerical experiments in Section~\ref{sec:exp}. The estimated probability of new COVID-19 diagnosis between 7 and 115 days after D29 for the analyzed sample was 0.0189 (95$\%$ CI 0.0169 - 0.0209), and decreased to 0.0094 (95$\%$ CI 0.0061 - 0.0126) under the policy with $\delta = 1$. Figure~\ref{fig:covid} presents the results for the policies we considered. 

Additionally, for each policy $q$, we estimated  $VE(q) := 1 - \mathbb{E}[Y\{X = 1, q(A)\}]/\mathbb{E}[Y(X = 0)]$ as well as $VE:=VE(Id)$. Here, \(Y(X = 0)\) is the outcome under placebo, \(Y\{X = 1, q(A)\}\) is the outcome under a vaccine formulation that induces an antibody titer \(q(A)\) instead of the observed titer \(A\), and $Id$ is the no-shift MTP, so that $VE$ corresponds to the usual vaccine efficacy measure. $VE(q)$ quantifies the relative reduction in the probability of COVID-19 between a hypothetical world where everyone receives the vaccine and develops neutralizing antibody titers according to policy $q$, and a world where everyone receives placebo.

We estimated $\mathbb{E}[Y(X = 0)]$ using the sample mean of the outcomes of the 17,108 placebo recipients from the same six countries as the analyzed vaccine recipients. Among these, 686 COVID-19 cases occurred during the same follow-up period, yielding an observed probability of COVID-19 of 0.0401 (95$\%$ CI 0.0372 - 0.0430). Combining this with data from the analyzed vaccine recipients, the observed VE in the trial was 0.528 (95$\%$ CI 0.467 - 0.589). Under the $\varepsilon$--upper-tail tapering shift intervention policy that increases each participant's antibody titers by one log10 unit, VE improved to 0.766 (95$\%$ CI 0.683 - 0.849), with a monotone pattern consistent with the expected protective effect of higher titers (Figure~\ref{fig:covid}). Non-proximal TMLE VE estimates, provided in Appendix~\ref{sec:details_app}, were generally lower and non-monotone across shift values, likely due to sampling variability and residual unmeasured confounding.

\section{Discussion}
\label{sec:discussion}
In this work, we established identification conditions for causal estimands involving modified treatment policies in the presence of unmeasured confounding, assuming the availability of negative control treatment and outcome variables. Unmeasured confounding is a significant concern in the immunobridging studies motivating our methodology, as individual immunity capacity levels are likely important but unobserved confounders.  

We described a cross-fitted estimator based on nonparametric estimation of the observed bridge functions using recently developed kernel-regularized adversarial methods, which, under regularity conditions, is asymptotically normal.

The numerical experiments demonstrated that our estimator performed as predicted by theory across various scenarios with differing correlations between the negative controls and the unmeasured confounder. As expected, it outperformed conventional non-proximal methods that do not account for unmeasured confounding, in scenarios where we have access to suitable negative controls. Its utility was further illustrated in the ENSEMBLE trial, where it yielded vaccine efficacy estimates that increased monotonically with larger tapered shifts, unlike non-proximal estimates, which were generally lower and non-monotone.

Future research directions include extending this work to accommodate censored time-to-event outcomes and developing proximal methods for longitudinal modified treatment policies. These extensions would enable the optimization of sequential treatment decisions while addressing time-varying unmeasured confounding.

\section*{Acknowledgements}

The authors thank the ENSEMBLE study participants and investigators and Janssen Pharmaceuticals and the Department of Health and Human Services for generating and providing the data.

Antonio Olivas and Peter Gilbert were supported by the National Institute Of Allergy and Infectious Diseases of the National Institutes of Health (R37AI054165 to P.B.G.). Antonio Olivas was also supported by the National Institute Of Allergy and Infectious Diseases grant UM1AI068635. Andrea Rotnitzky was supported by the National Heart, Lung, and Blood Institute grant R01-HL137808, and by the National Institute Of Allergy and Infectious Diseases grants UM1-AI068635 and R37-AI029168. The content is solely the responsibility of the authors and does not necessarily represent the official views of the National Institutes of Health.

\appendix
\section*{Appendices}

\section{Proximal Identification and Estimation in a Potentially Restricted Population}
\label{sec:webappA}
\label{sec:app_mtp_identification}

\renewcommand{\theequation}{A.\arabic{equation}}
\setcounter{equation}{0}
\renewcommand{\theassumption}{A.\arabic{assumption}}
\setcounter{assumption}{0}
\renewcommand{\thetheorem}{A.\arabic{theorem}}
\setcounter{theorem}{0}

Throughout we let $P$ denote the distribution of $O=(L,Z,W,A,Y)$. The symbol $\mathbb{E}$ stands for expectation under $P$. In this Appendix we extend the identifiability results and estimators in main text to target estimands of the form 
\begin{equation}
\psi_0 =\mathbb{E}\left[ Y\left\{ q\left( A,L\right) \right\} |(A,L)\in \mathcal{S}\right],\label{target_p}
\end{equation}
where $q:\supp{(A,L)}\mapsto\mathbb{R}$  is a specified map and $\mathcal{S}$ is a specified subset of $\supp{(A,L)}$ such that $P\left\{(A,L)\in\mathcal{S}\right\}>0$ and that for every $(a,l)\in\mathcal{S}$, $\left(q(a,l),l\right)$ belongs to $\supp(A,L)$. The estimands studied in the main text are of the form~\eqref{target_p} with $\mathcal{S}=\supp{(A,L)}$.

In Example~\ref{ex1} of Section~\ref{sec:ident}, the target estimand is also of the form~\eqref{target_p} with $q(a,l)=a+\delta$. The maximal choice of $\mathcal{S}$ in that example is $\mathcal{S}=\{(a,l):l\in\supp(L),\:a\in[c(l),d(l)-\delta]\}$. However, subpopulations $\left\{O:(A,L)\in\mathcal{S'}\right\}$, where $\mathcal{S}'$ is a strict subset of $\mathcal{S}$, may also be of interest, particularly for comparing the counterfactual MTP mean across different policies within a common population.

\subsection{Proximal Identifying Assumptions}

For each $\l\in\supp{(L)}$, define $\mathcal{S}(l):=\{a\in\supp{(A)}:(a,l)\in\mathcal{S}\}$. 
\begin{assumption}[Consistency]
$Y\left( A\right) =Y$.\label{ass:consist0}    
\end{assumption}

\begin{assumption}[Randomization of policy related populations]   For all $(a,l,u)\in\supp(A,L,U)$ such that $a\in\mathcal{S}(l)$ and $a'=q(a,l)$, the conditional distribution of $Y(a')|A=a, L=l, U=u$ is equal to the conditional distribution of $Y(a')|A=a', L=l, U=u$.\label{ass:rand0}
\end{assumption}

\begin{assumption}[Common support] For all $\left(l,u\right)\in\supp{(L,U)}$, $\supp{(A|L=l, U=u)} = \supp{(A|L=l)}$.\label{ass:pos_mtp0}
\end{assumption}

\begin{assumption}[Negative controls]
(i) $Z\bot Y|A,L,U$ and (ii) $\left( A,Z\right) \bot W|U,L$.\label{ass:nct0}
\end{assumption}

\begin{assumption}[Monotone and smooth policy] For each $l\in\supp{(L)}$, the map $q(\cdot,l):\mc{S}(l)\mapsto\supp{(A)}$ is strictly monotone
and differentiable almost everywhere with respect to the distribution of $A$ given $L=l$.\label{ass:q0}
\end{assumption}

 Assumption~\ref{ass:q0} ensures that, for each $l\in\supp{(L)}$, the policy map $q:a\mapsto q(a,l)$ has an inverse $q^{-1}:q\left(\mc{S}(l),l\right)\mapsto\mc{S}(l)$ that is well-defined and almost everywhere differentiable with respect to the distribution of $A$ given $L=l$.

\begin{assumption}[Existence of latent outcome bridge and observed treatment bridge functions]
\begin{enumerate}
    \item[]
    \item[i)] There exists $h_{0}\in\mathcal{L}^2(A,L,W)$, hereafter called ``latent outcome bridge function", solving for all $(a,l,u)\in\supp{(A,L,U)}$ such that $a\in\mathcal{S}(l)\cup q\left(\mathcal{S}(l),l\right)$, the integral equation%
\begin{equation}
\mathbb{E}\left( Y|A=a,L=l,U=u\right) =\int h\left( a,l,w\right) p_{W|A,L,U}\left(
w|a,l,u\right) dw.  \label{lat_out_eqn0}
\end{equation}
\item[ii)] 
There exists $g_{0}\in\mathcal{L}^2(A,L,Z)$, hereafter called ``observed treatment bridge function", satisfying $g_{0}(a,l,z) = 0$ if $a \in \mathcal{S}(l) - q\left(\mathcal{S}(l),l\right)$ and such that at every $a \in q\left(\mathcal{S}(l),l\right) $ at which the map $a'\mapsto q^{-1}(a',l)$ is differentiable at $a'=a$, $g_0$ solves the integral equation
\begin{equation}
\alpha_{0}\left( a,l,w\right) =\int g\left( a,l,z\right) p_{Z|A,L,W}\left(
z|a,l,w\right) dz,  \label{obs_tx_eqn0}
\end{equation}
where
\[
\alpha_{0}\left( a,l,w\right) :=I_q(a,l)\cdot \frac{dq^{-1}(a,l)}{da}\frac{p_{A|L,W}\left\{q^{-1}\left( a,l\right)
|l,w\right\} }{p_{A|L,W}\left( a|l,w\right) }
\]
and $I_q(a,l):=\left\{ a\in q\left(\mc{S}(l),l\right) \right\}$.
\end{enumerate}
\label{ass:assumption7_0}
\end{assumption}

\begin{assumption}[Existence of latent treatment bridge and observed outcome bridge functions]
\begin{enumerate}
    \item[]
    \item[i)] 
There exists $g_{0}\in\mathcal{L}^2(A,L,Z)$, hereafter called ``latent treatment bridge function", satisfying $g_{0}(a,l,z) = 0$ if $a \in \mathcal{S}(l) - q\left(\mathcal{S}(l),l\right)$ and such that at every $a \in q\left(\mathcal{S}(l),l\right) $ at which the map $a'\mapsto q^{-1}(a',l)$ is differentiable at $a'=a$, $g_0$ solves the integral equation
\begin{equation}
\widetilde\alpha_{0}\left( a,l,u\right) =\int g\left( a,l,z\right) p_{Z|A,L,U}\left(
z|a,l,u\right) dz,  \label{lat_tx_eqn0}
\end{equation}
where
\[
\widetilde\alpha_{0}\left( a,l,u\right) :=I_q(a,l)\frac{dq^{-1}(a,l)}{da}\frac{p_{A|L,U}\left\{q^{-1}\left( a,l\right)
|l,u\right\} }{p_{A|L,U}\left( a|l,u\right) }
\]
      \item[ii)] There exists $h_{0}\in\mathcal{L}^2(A,L,W)$, hereafter called ``observed outcome bridge function", solving for all $(a,l,w)\in\supp{(A,L,W)}$ such that $a\in\mathcal{S}(l)\cup q\left(\mathcal{S}(l),l\right)$, the integral equation%
\begin{equation}
\mathbb{E}\left( Y|A=a,L=l,Z=z\right) =\int h\left( a,l,w\right) p_{W|A,L,Z}\left(
w|a,l,z\right) dw.  \label{obs_out_eqn0}
\end{equation}

\end{enumerate}

\label{ass:assumption8_0}
\end{assumption}

\subsection{Conditions for Existence of the Latent Bridge Functions}\label{sec:appA_existence}

We consider first sufficient conditions for the existence of solutions to equations~\eqref{lat_out_eqn0} and~\eqref{lat_tx_eqn0} when the variables $Z$, $W$, and $U$ are discrete and take values in finite sets. For equation~\eqref{lat_out_eqn0}, sufficient conditions were discussed in \cite{miao2018identifying, shi2020multiply}; and \cite{tchetgen2024introduction}. For equation~\eqref{lat_tx_eqn0}, conditions can be derived by mimicking the analysis given in \cite{kallus2022causal} and \cite{cui2024semiparametric} for the existence of solutions to the treatment bridge equations corresponding to the average treatment effect (ATE) and generalized ATE respectively. Here, we review these analyses. 

Suppose that $W$, $Z$, and $U$ take a finite number of values, denoted by $w_i$, $z_j$, and $u_k$ for $i=1,\dots,n_W$, $j=1,\dots,n_Z$, and $k=1,\dots,n_U$, respectively. For any $(a,l,u)\in\supp{(A,L,U)}$ such that $a\in\mathcal{S}(l)\cup q(\mathcal{S}(l),l)$, let $\mathbb{P}\left({\bf W}|{\bf U}, a,l\right)$ denote the $n_W\times n_U$ matrix whose $(i,k)$-entry is $P(W=w_i|U=u_k, A=a, L=l)$. Similarly, let $\mathbb{P}\left({\bf Z}|{\bf U}, a,l\right)$ be the $n_Z\times n_U$ matrix with $(j,k)$-entry $P(Z=z_j|U=u_k, A=a, L=l)$. Define $r_0(a,l, {\bf U})$ and $a_0(a,l,{\bf U})$ as the $n_U$-dimensional column vectors whose $k$-th entries are $\mathbb{E}\left(Y|A=a, L=l, U=u_k\right)$ and $\widetilde\alpha_0(a,l,u_k)$, respectively. With this notation, the outcome bridge equation~\eqref{lat_out_eqn0} is:
\[
r_0(a,l, {\bf U})= \mathbb{P}\left({\bf W}|{\bf U}, a,l\right)^\top h(a,l,{\bf W}),
\]
and the treatment bridge equation~\eqref{lat_tx_eqn0} is
\[
a_0(a,l,{\bf U})=\mathbb{P}\left({\bf Z}|{\bf U}, a,l\right)^\top g(a,l,{\bf Z}).
\]

It follows that a solution to equation~\eqref{lat_out_eqn0} exists if and only if $r_0(a,l, {\bf U})$ lies in the vector space spanned by the row vectors of $\mathbb{P}\left({\bf W}|{\bf U}, a,l\right)$. Likewise, a solution to equation~\eqref{lat_tx_eqn0} exists if and only if $a_0(a,l,{\bf U})$ lies in the vector space spanned by the rows of $\mathbb{P}\left({\bf Z}|{\bf U}, a,l\right)$. If $n_W\ge n_U$ ($ n_Z \ge n_U$) and the matrix $\mathbb{P}\left({\bf W}|{\bf U}, a,l\right)$ ($\mathbb{P}\left({\bf Z}|{\bf U}, a,l\right)$) is of full column order, then the solution to the outcome bridge equation~\eqref{lat_out_eqn0} (treatment bridge equation~\eqref{lat_tx_eqn0}) always exists. If $n_W>n_U$ then equation~\eqref{lat_out_eqn0} may have many solutions. Likewise, if $n_Z>n_U$ then equation~\eqref{lat_tx_eqn0} may have many solutions.

When $W$, $Z$, and $U$ are continuous, as noted by \cite{miao2018identifying} and \cite{cui2024semiparametric}, we can deduce sufficient conditions for the existence of solutions to equations~\eqref{lat_out_eqn0} by invoking Picard's theorem  (Theorem 15.18 in \cite{kress2014}). Specifically, equation~\eqref{lat_out_eqn0} can be written~as
\begin{equation}
    \mathcal{T}^W_{a,l}h(a,l,U)=\mathbb{E}(Y|A=a,L=l,U),\label{eq:out_picard}
\end{equation}
where for each $(a,l)\in\supp{(A,L)}$, we let $\mc{T}_{a,l}^W:\mathcal{L}^2(W|A=a,L=l)\mapsto\mathcal{L}^2(U|A=a, L=l)$ denote the conditional expectation operator defined by $\mathcal{T}^W_{a,l}h(a,l,U):=\mathbb{E}\{h(A,L,W)|A=a,L=l,U\}$ and throughout we assume that $\mathbb{E}(Y|A=a,L=l,U)\in\mc{L}^2(U|A=a,L=l)$. Then, equation~\eqref{eq:out_picard} has a solution if and only if $\mathbb{E}(Y|A=a,L=l,U)$ lies in the range of the, continuous and linear, operator $\mc{T}_{a,l}^W$. Now, the closure of the range of a continuous and linear operator is equal to the orthogonal complement of the null space of the adjoint of the operator. For $\mc{T}_{a,l}^W$, its adjoint is the operator $\mc{T}_{a,l}^U:\mathcal{L}^2(U|A=a,L=l)\mapsto\mathcal{L}^2(W|A=a, L=l)$ defined as $\mathcal{T}^U_{a,l}g(a,l,w):=\mathbb{E}\{g(A,L,U)|A=a,L=l,W=w\}$, whose kernel is $\left\{\eta(a,l,U):\mathbb{E}[\eta(A,L,U)|A=a,L=l,W]=0\right\}$. Thus, in particular, the closure of the range of the operator $\mc{T}_{a,l}^W$ equals $\mc{L}^2(U|A=a,L=l)$ if and only if the kernel of $\mathcal{T}^U_{a,l}$ is the set $\{0\}$, equivalently, if and only if
\begin{equation}
\eta(a,l,U=u)=0 \:\forall u \quad \text{whenever} \quad \mathbb{E}\:\{\eta(A,L,U)|A=a,L=l,W=w\}=0 \:\forall w.
\label{eq:complete_W}
\end{equation}

Condition~\eqref{eq:complete_W} is the completeness condition discussed in the main text. So, we conclude that the completeness condition ensures that the closure of the range of the operator $\mc{T}_{a,l}^W$ is the entire space $\mc{L}^2(U|A=a,L=l)$. However, this is not sufficient to ensure that $\mathbb{E}\:\{Y|A=a,L=l,U\}$ is in the range of $\mc{T}_{a,l}^W$, as it may fall in the boundary of the range. Picard's theorem provides a sufficient condition for a given element of $\mc{L}^2(U|A=a,L=l)$, in our case $\mathbb{E}\:\{Y|A=a,L=l,U\}$, to not fall in the boundary, provided the operator $\mc{T}_{a,l}^W$ is compact. Compactness of $\mc{T}_{a,l}^W$ holds under the regularity condition that (see Examples 2.2 and 2.3 in \cite{carrasco2007linear}):
\begin{equation}
    \iint p_{W|A,L,U}(w|a,l,u)p_{U|A,L,W}(u|a,l,w)dwdu<\infty.
    \label{eq:compactness_op}
\end{equation}

Under~\eqref{eq:compactness_op}, $\mc{T}_{a,l}^W$ admits a singular value decomposition $(\sigma_{a,l,j}, \varphi_{a,l,j},\phi_{a,l,j})_{j=1}^\infty$ (see Theorem 15.16 in \cite{kress2014}). Under completeness and compactness, Piccard's sufficient condition for $\mathbb{E}\:\{Y|A=a,L=l,U\}$ to fall outside the boundary of the range of $\mc{T}_{a,l}^W$ is 
\begin{equation}
\sum_{j=1}^\infty\sigma^{-2}_{a,l,j}\left\{\int \mathbb{E}(Y|A=a, L=l, U=u)\phi_{a,l,j}(a,l,u)p_{U|A,L}(u|a,l)du\right\}^2<\infty.
    \label{eq:reg_cond0}
\end{equation}

Thus, we conclude that the completeness condition~\eqref{eq:complete_W} together with the regularity conditions~\eqref{eq:compactness_op} and~\eqref{eq:reg_cond0} suffice for equation~\eqref{eq:out_picard} to have a solution.

Likewise, we can invoke Picard’s theorem to derive sufficient conditions for the existence of a solution to equation~\eqref{lat_tx_eqn0}. These are the completeness condition $\eta(a,l,U=u)=0$ for all $u$ whenever $\mathbb{E}\:\{\eta(A,L,U)|A=a,L=l,Z=z\}=0$ for all $z$, and the regularity conditions that $\iint p_{Z|A,L,U}(Z|a,l,u)p_{U|A,L,Z}(u|a,l,z)dzdu$ and $\sum_{j=1}^\infty\tilde\sigma^{-2}_{a,l,j}\{\int \widetilde\alpha_0(a,l,u)\tilde\phi_{a,l,j}(a,l,u) p_{U|A,L}(u|a,l)du\}^2$ are both finite, where $(\tilde\sigma_{a,l,j}, \tilde\varphi_{a,l,j}, \tilde\phi_{a,l,j})_{j=1}^\infty$ is the singular value decomposition of the operator $\mathcal{T}_{a,l}^Z: \mc{L}^2(Z|A=a,L=l) \mapsto \mc{L}^2(U|A=a,L=l)$ defined as $\mathcal{T}^Z_{a,l}g(a,l,U):=\mathbb{E}\left\{g(A,L,Z)|A=a,L=l,U\right\}$.

\subsection{Proximal Identification Results}
\label{sec:app_identif_results}
Next, we extend the identification result for the counterfactual MTP mean over $\{O: (A,L) \in \mathcal{S}\}$. Define $\phi(O;h,g) := h\{q\left( A,L\right),L,W\}\cdot I\{(A,L)\in\mathcal{S}\}+ g(A,L,Z)\{Y-h(A,L,W)\}$. The following Theorem and the subsequent Theorem~\ref{theo:asymp_distr0} are proved in Appendix~\ref{sec:proofs_theorems}.

\begin{theorem} Suppose Assumptions~\ref{ass:consist0}--\ref{ass:q0} hold. If, in addition, either Assumption~\ref{ass:assumption7_0} or Assumption~\ref{ass:assumption8_0} holds, then $\psi_0$ is identified by the observed data and it satisfies:
\begin{align}
\label{out_reg0}
     \psi_0&= \frac{\mathbb{E}\left[ h_0
\left\{ q\left( A,L\right),L,W\right\}\cdot I\left\{(A,L)\in\mathcal{S}\right\}\right]}{P\left\{(A,L)\in\mathcal{S}\right\}}
\\
\label{ipw0}
           &= \frac{\mathbb{E}\left\{ Yg_0\left( A,L,Z\right)\right\}}{P\left\{(A,L)\in\mathcal{S}\right\}}\\
           &= \frac{\mathbb{E}\left\{\phi(O; h^\dag,g^\dag)\right\}}{P\left\{(A,L)\in\mathcal{S}\right\}}
\nonumber
\end{align}
where $h_0$ is an observed outcome bridge function, $g_0$ is an observed treatment bridge function and, either $h^\dag$ is an observed outcome bridge function or $g^\dag$ is an observed treatment bridge function but not necessarily both functions are observed bridge functions.

    \label{theo:identification0}
\end{theorem}

Theorem~\ref{theo:identification0} offers a slight weakening of the conditions for the outcome regression representation~\eqref{out_reg0} of the preceding theorem that would follow if one were to apply the identification strategy in \cite{kallus2022causal} to our target parameter $\psi_0$. Specifically, following the identification analysis of \cite{kallus2022causal} we arrive at the conclusion that the representation holds under Assumption~\ref{ass:assumption7_0}. Theorem~\ref{theo:identification0} establishes that this representation is valid even when Assumption~\ref{ass:assumption7_0} fails as long as Assumption~\ref{ass:assumption8_0} holds. Likewise, for the representation~\eqref{ipw0}, Theorem~\ref{theo:identification0} establishes its validity even if Assumption~\ref{ass:assumption8_0} fails, provided that Assumption~\ref{ass:assumption7_0} holds. 

\subsection{Estimation}
\label{sec:estimation}

 The estimator $\hat{\psi}_{CF}^{DR}$ of the main text extends to the following estimator for the target parameter defined in~\eqref{target_p} of this Appendix:
 \begin{equation*}
\hat{\psi}_{CF}^{DR}:=\frac{\sum_{k=1}^{K}\sum_{i:O_{i}\in \mc I_{k}}\phi
\left\{ O_{i};\hat{h}^{(-k)},\hat{g}^{(-k)}\right\} }{\sum_{i=1}^{n}I\left\{
(A_{i},L_{i})\in \mathcal{S}\right\} }.
\end{equation*}
where $\hat{h}^{(-k)}$ and $\hat{g}^{(-k)}$
are defined as in the main text except that their computation is restricted to the units in all the sample but sample $\mc I_k$ and satisfying $A\in \mathcal{S}(L)\cup q(\mc{S}(L),L)$. 

Note that when $\mathcal{S}=\supp{(A,L)}$, i.e., when $I\left\{(A,L)\in%
\mathcal{S}\right\}\equiv1$, this estimator $\hat\psi_{CF}^{DR}$ coincides with the estimator proposed in Section~\ref{sec:est}.

The following Theorem extends the Theorem~3 in Section~\ref{sec:est}. 

\begin{theorem}
    Suppose that the assumptions of Theorem \ref{theo:identification0} hold. Let $h_0$ and $g_0$ denote observed outcome and treatment bridge functions. If for some $B>0$, $|\alpha_0(A,L,W)|\le B$, $|Y|\le B$ and for each $k$, $\Vert\hat h^{(-k)}-h_0\Vert_2=o_p(1)$, $\Vert\hat g^{(-k)}-g_0\Vert_2=o_p(1)$ and either $\Vert h_0\Vert_\infty+\Vert \hat g^{(-k)}\Vert_\infty\le B$ or $\Vert \hat h{(-k)} \Vert_\infty\ + \Vert g_0\Vert_\infty\le B$ for some $B$, then the estimator $\hat\psi_{CF}^{DR}$ satisfies
    \begin{equation*}
\sqrt{n}\left\{ \hat{\psi}_{CF}^{DR}-\psi _{0}\right\} =\sqrt{n}\mathbb{E}_{n}%
\left[ p_0^{-1} \phi (O;h_{0},g_{0})%
-\psi _{0} p_0^{-1}I\left\{ (A,L)\in \mathcal{S}\right\} \right] +\sqrt{n}R_{n}+o_{p}(1).
\end{equation*}
    where $R_n=p_0^{-1}\sum_{k=1}^K\frac{n_k}{n} \:\int\{\hat h^{(-k)}(a,l,w)-h_0(a,l,w)\}\{g_0(a,l,z)-\hat g^{(-k)}(a,l,z)\}dP(a,l,w,z)$
with $p_0:=P\left\{ (A,L)\in 
\mathcal{S}\right\}$ and $n_k$ denoting the cardinality of $\mc I_k$. In particular, if $R_n=o_p\left(n^{-1/2}\right)$ and $\mathbb{E}\left\{\phi(O;h_0,g_0)^2\right\}<\infty$, then $ \sqrt{n}\left\{\hat\psi_{CF}^{DR}-\psi_0\right\}\stackrel{\text{d}}{\longrightarrow}\mc{N}(0,\tau^2)$, where $\tau ^{2}=\mathbb{E}\Big\{ \Big[ \frac{\phi (O;h_{0},g_{0})}{%
P\left\{ (A,L)\in \mathcal{S}\right\} }-\psi _{0}\frac{I\{ (A,L)\in 
\mathcal{S}\} }{P\{ (A,L)\in \mathcal{S}\} }\Big]
^{2}\Big\} $.
\label{theo:asymp_distr0}
\end{theorem}

\section{Additional Details on the Estimation of the Bridge Functions}
\label{sec:webappB}

\renewcommand{\theequation}{B.\arabic{equation}}
\setcounter{equation}{0}
\renewcommand{\theassumption}{B.\arabic{assumption}}
\setcounter{assumption}{0}
\renewcommand{\thetheorem}{B.\arabic{theorem}}
\setcounter{theorem}{0}

To establish convergence guarantees for the estimators of the bridge functions, we assume that equations~\eqref{obs_out_eqn0} and~\eqref{obs_tx_eqn0} admit solutions within the RKHSs used for estimation, and let $h_0$ and $g_0$ denote the corresponding observed bridge functions with minimum RKHS norm in these spaces. Under this and further assumptions outlined below, Theorems~4 and~5 of \cite{olivas2025source}---with $(A,L,Z,W,Y)$ identified with $O$ in that paper, $(A,L,Z)$ with $Z$, and $(A,L,W)$ with $W$---provide upper bounds on the root-mean-squared errors $\Vert\hat h-h_0\Vert_2$ and $\Vert \hat g-g_0\Vert_2$, as well as on the weak errors $\Vert \mathbb{E}\{\hat h(A,L,W)-h_0(A,L,W)|A,L,Z\}\Vert_2$ and $\Vert \mathbb{E}\{\hat g(A,L,Z)-g_0(A,L,Z)|A,L,W\}\Vert_2$. These bounds are expressed in terms of the regularization parameters and the critical radii of the RKHSs involved, and rely on a source condition imposing smoothness constraints on $h_0$ and $g_0$ tailored to the geometry of the RKHSs $\mc{H}$ and $\mc{G}$, a local closedness and boundedness condition on the RKHSs involved in the inner and outer optimization problems (ensuring regular behavior of the conditional expectation operators on bounded subsets), and additional regularity conditions satisfied by the RKHSs we use. See \cite{olivas2025source} for full details.

In particular, when $h_0$ and $g_0$ belong to Gaussian RKHSs and satisfy source conditions of orders $\beta_h$ and $\beta_g$, and the regularization parameters are chosen as in Section~\ref{sec:exp}---$\lambda_\mc{H}$ and $\lambda_\mc{G}$ of order equal to the critical radius of $\mc{H}$ and $\mc G$, respectively, and $\lambda_{\mc{H}'}$ and $\lambda_{\mc{G}'}$ of order equal to the square of the critical radius---Theorems 4 and 5 of \cite{olivas2025source} guarantee that $\hat h$ and $\hat g$ are norm-consistent. When $\beta_h+\beta_g>1$, these choices ensure that the product of the weak error of one estimator and the root-mean-squared error of the other is $o_p(n^{-1/2})$, thereby satisfying the conditions of Theorem~\ref{theo:asymp_distr0} for asymptotic normality of the MTP mean estimator $\psi_{CF}^{DR}$.

\subsection{Estimators Using Reproducing Kernel Hilbert Spaces}
\label{sec:rkhs_est}

In this section, we derive closed-form expressions for the estimators of the bridge functions. To apply Theorems~4 and~5 from \cite{olivas2025source}, we work within bounded RKHS balls, choosing $B$ large enough so that the minimum RKHS-norm bridge functions satisfy $\Vert h_0\Vert_\mc{H}\le B$ and $\Vert g_0\Vert_\mc{G}\le B$. These bounds were omitted from the main text to avoid unnecessary technical detail, but are included here because the theoretical results require both the true functions and their estimators to lie within RKHS balls of finite radius.

Accordingly, define $\mc{H}_B := \{h \in \mc{H} : \Vert h \Vert_\mc{H} \le B\}$. The estimator for $h_0$ is
\[
\hat h = \arg\min_{h\in\mc{H}_B}\max_{g\in\mc{G}'}\mathbb{E}_n\!\left[g(A,L,Z)\left\{Y-h(A,L,W)\right\}-g(A,L,Z)^2\right]-\lambda_{\mc{G}'}\Vert g\Vert_{\mc{G}'}^2+\lambda_{\mc{H}}\Vert h\Vert_{\mc{H}}^2.
\]

We first analyze the inner maximization. For any $h\in\mc{H}_B$, we show that
\begin{align}
\max_{g\in\mathcal{G}'}\mathbb{E}_n&\left[g(A,L,Z)\left\{Y-h(A,L,W)\right\}-g(A,L,Z)^2\right]-\lambda_{\mc{G}'}\Vert g\Vert_{\mc{G}'}^2\label{eq:inner_h}\\
&=\frac{1}{4}\left\{\xi_n(h)\right\}^\top\left(\frac{1}{n}K_{\mc{G}',n}+\lambda_{\mc{G}'}Id_n\right)^{-1}K_{\mc{G}',n}\:\xi_n(h)\nonumber,
\end{align}
where $K_{\mc{G}',n}$ is the $n\times n$ matrix with $(i,j)$-th entry $K_{\mc{G}'}\{(A_i,L_i,Z_i),(A_j,L_j,Z_j)\}$, $\xi_n(h):=\frac{1}{n}\big(\widetilde{Y}_n-\tilde h_n\big)$, $\widetilde{Y}_n=\left(Y_1,\dots,Y_n\right)^\top$, and $\tilde h_n=\left[h(A_1,L_1,W_1),\dots, h(A_n,L_n,W_n)\right]^\top$.

By the generalized representer theorem (see \cite{scholkopf2001generalized}), the solution to this the maximization problem has the form
\[
g(a,l,z)=\sum_{j=1}^n\alpha_jK_{\mc{G}'}\{(a,l,z), (A_j,L_j,Z_j)\},\]
where $\alpha =(\alpha_1,\dots,\alpha_n)^\top\in\mathbb{R}^n$. This ensures that the solution is fully characterized by the observed data through the kernel evaluations. For $i=1,\dots,n$, we can write $g(A_i,L_i,Z_i)$ as $g(A_i,L_i,Z_i) = e_i^\top   K_{\mc{G},n} \alpha$
where $e_i$ is the $i$-th standard basis vector in $\mathbb{R}^n$. Hence, the terms in~\eqref{eq:inner_h} evaluated at functions $g$ as in the last display, can be written as follows:
\begin{align*}
    \mathbb{E}_n\left[Yg(A,L,Z)\right] &=\frac{1}{n} \sum_{i=1}^nY_ie_i^\top K_{\mc{G}',n} \alpha= \frac{1}{n}\widetilde{Y}_n^\top K_{\mc{G}',n}\alpha,\\
    \mathbb{E}_n\left[g(A,L,Z)h(A,L,W)\right]&= \frac{1}{n}\sum_{i=1}^nh(X_i,L_i,W_i)e_i^\top   K_{\mc{G}',n} \alpha =  \frac{1}{n}\tilde h_n^\top K_{\mc{G}',n}\alpha,\\
    \mathbb{E}_n\left[g(A,L,Z)^2\right] &= \frac{1}{n}\sum_{i=1}^n\alpha^\top K_{\mc{G}',n}e_ie_i^\top   K_{\mc{G}',n} \alpha=\frac{1}{n}\alpha^\top K_{\mc{G}',n}^2\alpha,\\
    \Vert g\Vert_{\mc{G}'}^2&=\sum_{i=1}^n\sum_{j=1}^n\alpha_i\alpha_jK_{\mc{G}'}\{(A_i,L_i,Z_i), (A_j,L_j,Z_j)\}=\alpha^\top K_{\mc{G}',n}\alpha.
\end{align*}
Then, $\mathbb{E}_n\left[g(A,L,Z)\left\{Y-h(A,L,W)\right\}-g(A,L,Z)^2\right]-\lambda_{\mc{G}'}\Vert g\Vert_{\mc{G}'}^2$ reduces to the following concave quadratic function of $\alpha$:
\[
\frac{1}{n}\alpha^\top K_{\mc{G}',n}\big(\widetilde{Y}_n-\tilde h_n\big) -\alpha^\top\left(\frac{1}{n}K_{\mc{G}',n}^2+\lambda_{\mc{G}'}K_{\mc{G}',n}\right)\alpha,
\]
with unique maximizer $\alpha^*=\frac{1}{2n}(n^{-1}K_{\mc{G}',n}+\lambda_{\mc{G}'}Id_n)^{-1}\big(\widetilde{Y}_n-\tilde h_n\big)$. This can be obtained by taking the derivative of the quadratic function with respect to $\alpha$
and setting it to zero. Hence,
\begin{align*}
    \max_{g\in\mathcal{G}'}\mathbb{E}_n&\left[g(A,L,Z)\left\{Y-h(A,L,W)\right\}-g(A,L,Z)^2\right]-\lambda_{\mc{G}'}\Vert g\Vert_{\mc{G}'}^2
    =\left\{\xi_n(h)\right\}^\top\Gamma_{\mc{G}'} K_{\mc{G}',n} \left\{\xi_n(h)\right\},
\end{align*}
with  $\Gamma_{\mc{G}'} = 4^{-1}(n^{-1}K_{\mc{G}',n}+\lambda_{\mc{G}'}Id_n)^{-1}$, and the original optimization problem in $h$ reduces to:
\[
\hat h = \arg\min_{h\in\mc{H}_B}\left\{\xi_n(h)\right\}^\top\Gamma_{\mc{G}'} K_{\mc{G}',n} \left\{\xi_n(h)\right\}+\lambda_{\mc{H}}\Vert h\Vert^2_{\mc{H}}.
\]
By the representer theorem, the solution to this minimization problem takes the form
\[
\hat h(a,l,w) = \sum_{j=1}^n\gamma_jK_{\mc{H}}\{(a,l,w),(A_j,L_j,W_j)\},
\]
where $\gamma=(\gamma_1,\dots,\gamma_n)^\top$ is the coefficient vector to be determined. Substituting this representation into the objective function, we obtain:
\[
\frac{1}{n^2}\widetilde{Y}_n^\top\Gamma_{\mc{G}'} K_{\mc{G}',n} \widetilde{Y}_n-\frac{2}{n^2}\gamma^\top K_{\mc{H},n}\Gamma_{\mc{G}'} K_{\mc{G}',n} \widetilde{Y}_n  +\gamma^\top\left(\frac{1}{n^2}K_{\mc{H},n}\Gamma_{\mc{G}'} K_{\mc{G}',n}K_{\mc{H},n}+\lambda_{\mc{H}}K_{\mc{H},n}\right)\gamma.
\]
where $K_{\mc{H},n}$ is the $n\times n$ matrix with $(i,j)$-th entry given by $K_{\mc{H}}\left\{(A_i,L_i,W_i),(A_j,L_j,W_j)\right\}$. This is a convex quadratic function of $\gamma$, whose unique minimizer---found by solving the first order equation---is $\gamma^*=\left(\Gamma_{\mc{G}'} K_{\mc{G}',n}K_{\mc{H},n}+n^2\lambda_{\mc{H}}Id_n\right)^{-1} \Gamma_{\mc{G}'} K_{\mc{G}',n} \widetilde{Y}_n$.

It remains to ensure that $\Vert \hat h\Vert_{\mc{H}}\le B$ for some pre-specified $B$. If $\left(\gamma^*\right)^\top K_{\mc{H},n}\gamma^*\le B^2$, the condition is satisfied. Otherwise, by Exercise 4.22 of \cite{boyd2004convex}, the optimal $\gamma^*$ is given by $\gamma^*=\left(P+B^{-2}\tilde uK_{\mc{H},n}\right)^{-1}\tilde q$ where $P=\frac{1}{n^2}K_{\mc{H},n}\Gamma_{\mc{G}'} K_{\mc{G}',n}K_{\mc{H},n}+\lambda_{\mc{H}}K_{\mc{H},n}$, $\tilde q = \frac{1}{n^2}K_{\mc{H},n}\Gamma_{\mc{G}'} K_{\mc{G}',n} \widetilde{Y}_n$, and $\tilde u$ is the largest solution to the nonlinear equation:
\[
\tilde q^\top K_{\mc{H},n}^{-1/2}\left(K_{\mc{H},n}^{-1/2}PK_{\mc{H},n}^{-1/2}+\frac{u}{B^2} Id_n\right)^{-2}K_{\mc{H},n}^{-1/2}\tilde q =B^2.
\]

We now derive the closed-form expression of the estimator $\hat g$ for $g_0$. Define $\mc{G}_B := \{g \in \mc{G} : \Vert g \Vert_\mc{G} \le B\}$. The estimator for $g_0$ is $\hat g(a,l,z)=I_q(a,l)\cdot \tilde g(a,l,z)$ where $\tilde g$ is
\begin{align*}
    \tilde g = \arg\min_{g\in\mc{G}_B}\max_{h\in\mc{H}'}&\mathbb{E}_n\Big[h\left\{q(A,L),L,W\right\}I\left\{(A,L)\in\mc{S}\right\}-I_q(A,L)h(A,L,W)g(A,L,Z)\\
    &\quad\quad-I_q(A,L)h(A,L,W)^2\Big]-\lambda_{\mc{H}'}\Vert h\Vert_{\mc{H}'}+\lambda_{\mc{G}}\Vert g\Vert_{\mc{G}}^2.
\end{align*}
We first show that for any $g\in\mc{G}_B$, the inner maximization satisfies
\begin{align}
\max_{h\in\mc{H}'}&\mathbb{E}_n\Big[h\left\{q(A,L),L,W\right\}I\left\{(A,L)\in\mc{S}\right\}-I_q(X,L)h(A,L,W)g(A,L,Z)\nonumber\\
    &\quad\quad-I_q(A,L)h(A,L,W)^2\Big]-\lambda_{\mc{H}'}\Vert h\Vert_{\mc{H}'}\label{eq_gn}\\
    &\quad=\frac{1}{4}\left\{\zeta_n(g)\right\}^\top\left(\frac{1}{n}K_{\mc{H}',n}R_nK_{\mc{H}',n}+\lambda_{\mc{H}'}K_{\mc{H}',n}\right)^{-1}\left\{\zeta_n(g)\right\},\nonumber
\end{align}
where $K_{\mc{H}',n}$ is the $n\times n$ matrix with $(i,j)$-th entry $K_{\mc{H}'}\left\{(A_i,L_i,W_i),(A_j,L_j,W_j)\right\}$, $\zeta_n(g):=\frac{1}{n}\left[\widetilde{K}_{\mc{H},n}^\top\tilde d_n-K_{\mc{H}',n}R_n\tilde g_n\right]$, $\tilde d_n$ is the vector $\big[I\left\{(A_1,L_1)\in\mc{S}\right\},\cdots,I\left\{(A_n,L_n)\in\mc{S}\right\}\big]^\top$, $\tilde g_n$ is the vector $\big[g(A_1,L_1,Z_1),\dots, g(A_n,L_n,Z_n)\big]^\top$, $R_n$ is the diagonal matrix with values $I_q(A_1,L_1),\dots,I_q(A_n,L_n)$, and $\widetilde{K}_{\mc{H}',n}$ is the  $n\times n$ matrix with $(i,j)$-th entry $K_{\mc{H}'}\{(q(A_i,L_i),L_i,W_i),(A_j,L_j,W_j)\}$.

By the generalized representer theorem, the maximizer of problem~\eqref{eq_gn} has the form
\[
h(a,l,w)=\sum_{j=1}^n\alpha_jK_{\mc{H}'}\{(a,l,w), (A_j,L_j,W_j)\},\]
where $\alpha =(\alpha_1,\dots,\alpha_n)^\top\in\mathbb{R}^n$. Then, the terms in~\eqref{eq_gn} can be expressed as follows:
\begin{align*}
    \mathbb{E}_n\left[h\left\{q(A,L),L,W\right\} I\left\{(A,L)\in\mc{S}\right\}\right] &=\frac{1}{n} \sum_{i=1}^nI\left\{(A_i,L_i)\in\mc{S}\right\} \widetilde{K}_{\mc{H}',n} \alpha= \frac{1}{n}(\tilde d_n)^\top \widetilde{K}_{\mc{H}',n}\alpha,\\
    \mathbb{E}_n\left[I_q(A,L) h(A,L,W)g(A,L,Z)\right]&= \frac{1}{n}\sum_{i=1}^ng(A_i,L_i,A_i)I_q(A_i,L_i)e_i^\top   K_{\mc{H}',n} \alpha= \frac{1}{n}\tilde g_n^\top R_nK_{\mc{H}',n}\alpha,\\
    \mathbb{E}_n\left[I_q(A,L) h(A,L,W)^2\right] &= \frac{1}{n}\sum_{i=1}^n\alpha^\top K_{\mc{H}',n}e_iI_q(A_i,L_i)e_i^\top   K_{\mc{H}',n} \alpha=\frac{1}{n}\alpha^\top K_{\mc{H}',n}R_nK_{\mc{H}',n}\alpha,\\
    \Vert h\Vert_{\mc{H}'}^2&=\alpha^\top K_{\mc{H}',n}\alpha.
\end{align*}
Then, $\mathbb{E}_n\big[h\{q(A,L),L,W\}I\{(A,L)\in\mc{S}\}-I_q(A,L)h(A,L,W)g(A,L,Z)-I_q(A,L)h(A,L,W)^2\big]-\lambda_{\mc{H}'}\Vert h\Vert_{\mc{H}'}$ is a quadratic function of $\alpha$:
\begin{align*}
\frac{1}{n}\alpha^\top\left(\widetilde{K}_{\mc{H}',n}^\top\tilde d_n-K_{\mc{H}',n}R_n\tilde g_n\right) -\alpha^\top\left(\frac{1}{n}K_{\mc{H}',n}R_nK_{\mc{H}',n}+\lambda_{\mc{H}'}K_{\mc{H}',n}\right)\alpha,
\end{align*}
with maximizer $\alpha^*=\frac{1}{2n}(n^{-1}K_{\mc{H}',n}R_nK_{\mc{H}',n}+\lambda_{\mc{H}'}K_{\mc{H}',n})^{-1}\big(\widetilde{K}_{\mc{H}',n}^\top \tilde d_n-\widetilde{K}_{\mc{H}',n}R_n \tilde g_n\big)$. Hence,
\begin{align*}
\max_{h\in\mc{H}'}&\mathbb{E}_n\Big[h\left\{q(A,L),L,W\right\}I\left\{(A,L)\in\mc{S}\right\}-I_q(A,L)h(A,L,W)g(A,L,Z)\\
    &\quad\quad-I_q(A,L)h(A,L,W)^2\Big]-\lambda_{\mc{H}'}\Vert h\Vert_{\mc{H}'}\\
    &\quad=\frac{1}{4}\left\{\zeta_n(g)\right\}^\top\left(\frac{1}{n}K_{\mc{H}',n}R_nK_{\mc{H}',n}+\lambda_{\mc{H}'}K_{\mc{H}',n}\right)^{-1}\left\{\zeta_n(g)\right\},
\end{align*}
and the original optimization problem simplifies to:
\[
\tilde g = \arg\min_{g\in\mc{G}_B}\frac{1}{4}\left\{\zeta_n(g)\right\}^\top\left(\frac{1}{n}K_{\mc{H}',n}R_nK_{\mc{H}',n}+\lambda_{\mc{H}'}K_{\mc{H}',n}\right)^{-1}\left\{\zeta_n(g)\right\}+\lambda_{\mc{H}}\Vert h\Vert^2_{\mc{H}}.
\]
By the representer theorem, the solution to this optimization problem has the form
\[
\tilde g(a,l,z) = \sum_{j=1}^n\theta_jK_{\mc{G}}((a,l,z),(A_j,L_j,Z_j)),
\]
where $\theta=(\theta_1,\dots,\theta_n)^\top\in\mathbb{R}^n$ is the vector of coefficients to be determined. Let $\Gamma_{\mc{H}'}=4^{-1}(n^{-1}K_{\mc{H}',n}R_n+\lambda_{\mc{H}'}Id_n)^{-1}$. Then, the objective becomes
\begin{align*}
    \frac{1}{4n^2}(\tilde d_n)^\top\widetilde{K}_{\mc{H}',n}&\left(\frac{1}{n}K_{\mc{H}',n}R_nK_{\mc{H}',n}+\lambda_{\mc{H}'}K_{\mc{H}',n}\right)^{-1}K_{\mc{H}',n}^\top\tilde d_n\\
    &-\frac{2}{n^2}\theta^\top K_{\mc{G},n}R_n\Gamma_{\mc{H}'} \widetilde{K}_{\mc{H}',n}^\top \tilde d_n  +\theta^\top\left(\frac{1}{n^2}K_{\mc{G},n}R_n\Gamma_{\mc{H}'} K_{\mc{H}',n}R_nK_{\mc{G},n}+\lambda_{\mc{G}}K_{\mc{G},n}\right)\theta,
\end{align*}
where $K_{\mc{G},n}$ is the $n\times n$ matrix with $(i,j)$-th entry equal to $K_{\mc{G}}\left\{(A_i,L_i,Z_i),(A_j,L_j,Z_j)\right\}$.
The unique solution to this convex quadratic function of $\theta$ is
\[
\theta^*=\left(R_n\Gamma_{\mc{H}'} K_{\mc{H}',n}R_nK_{\mc{G},n}+n^2\lambda_{\mc{G}}Id_n\right)^{-1} R_n\Gamma_{\mc{H}'} \widetilde{K}_{\mc{H}',n}^\top\tilde d_n.
\]

If $\left(\theta^*\right)^\top K_{\mc{G},n}\theta^*\le B^2$, it follows that $\left\Vert \tilde g\right\Vert_{\mc{G}}\le B$. If $\left(\theta^*\right)^\top K_{\mc{G},n}\theta^*> B^2$, the optimal $\theta^*$ ensuring $\left\Vert \tilde g\right\Vert_{\mc{G}}\le B$ is given by $\theta^*=\left(P+B^{-2}\tilde u K_{\mc{G},n}\right)^{-1}\tilde q$,
where $P=\frac{1}{n^2}K_{\mc{G},n}R_n\Gamma_{\mc{H}'} K_{\mc{H}',n}R_nK_{\mc{G},n}+\lambda_{\mc{G}}K_{\mc{G},n}$, $\tilde q = \frac{1}{n^2}K_{\mc{G},n}R_n\Gamma_{\mc{H}'} \widetilde{K}_{\mc{H}',n}^\top \tilde d_n$, and $\tilde u$ is the largest solution to the nonlinear equation
\[
\tilde q^\top K_{\mc{G},n}^{-1/2}\left(K_{\mc{G},n}^{-1/2}PK_{\mc{G},n}^{-1/2}+\frac{u}{B^2} Id_n\right)^{-2}K_{\mc{G},n}^{-1/2}\tilde q =B^2.
\]

\subsection{Estimators Accounting for Two-Phase Sampling}\label{sec:rkhs_wt}
In the ENSEMBLE trial, the exposure---neutralizing antibody levels ($A$)---was measured only on a subset of study participants. For these participants, $A$ was obtained by thawing and analyzing, at the end of follow-up, previously frozen blood samples. Subjects were selected to have $A$ measured based on specified probabilities that depended on the outcome $Y$ and a vector of measured baseline covariates $V$ that included some of the variables in $L$.

Letting $\Delta_i$ denote the indicator that subject $i$ has $A_i$ measured, the observed data are $n$ i.i.d. copies of the vector $\widetilde{O}=(\Delta,\Delta A, L,V,Z,W,Y)$. By design, the missingness probabilities satisfy
\begin{equation}
P(\Delta=1\vert A,V,L,Y,Z,W)=P(\Delta=1\vert V,Y).
\label{MAR}
\end{equation}
Then, letting $S :=P(\Delta=1\vert V,Y)^{-1}$, and with $\phi(O;h,g)$ as defined in Section~\ref{sec:app_identif_results} for $O:=(A,L,Z,W,Y)$, under the assumptions of Theorem~\ref{theo:identification0} and equality~\eqref{MAR}, it holds
\[
\psi_0 := \frac{\mathbb{E} \left\{\Delta S\phi(O;h^\dag,g^\dag)\right\}}{\mathbb{E}\left[\Delta S\cdot I\left\{(A,L)\in\mc{S}\right\}\right]},
\]
where either $h^{\dag}$ is an observed outcome bridge function or $g^{\dag}$ is an observed treatment bridge function, but not necessarily both. The equality, which follows by taking conditional expectation given $(V,L,Z,W,Y)$ inside the outer expectations in the numerator and denominator, suggests the following doubly-robust cross-fitted estimator
\[
\hat\psi_{CF}^{DR} =\frac{\sum_{k=1}^K\sum_{i:O_i\in \mc I_k}\Delta_iS_i\phi\left\{O_i;\hat h^{(-k)},\hat g^{(-k)}\right\}}{\sum_{i=1}^n\Delta_iS_iI\left\{(A_i,L_i)\in\mc{S}\right\}},
\]
where $\mc I_1,\dots,\mc I_K$ is a partition of the entire sample into roughly equal size subsamples and $\hat{h}^{(-k)}$ and $\hat{g}^{(-k)}$ are estimators of the observed negative control treatment and outcome functions computed using units from the entire sample but the subsample $\mc I_k$. We emphasize that all expressions involving $\phi$ are multiplied by $\Delta$, so $\phi$ is only evaluated for units with $\Delta=1$. For these units, $A$ is observed and equals $\Delta A$, ensuring that the original definition of $\phi$ applies without modification.

The next Theorem extends Theorem~\ref{theo:asymp_distr0} to the present setting with missing covariates.
\begin{theorem}
    Suppose the observed data are $n$ i.i.d. copies of $\widetilde{O}=(\Delta,\Delta A, L,V,Z,W,Y)$, that~\eqref{MAR} holds, and that there exists $\sigma>0$ such that $S:=P(\Delta=1\vert V,Y)^{-1} \ < \sigma \: (a.e.)$. 
    Suppose that the Assumptions of Theorem~\ref{theo:asymp_distr0} hold for some observed outcome and treatment bridge functions $h_0$ and $g_0$. Then the estimator $\hat\psi_{CF}^{DR}$ satisfies
    \begin{equation*}
\sqrt{n}\left\{ \hat{\psi}_{CF}^{DR}-\psi _{0}\right\} =\sqrt{n}\mathbb{E}_{n}%
\left[\Delta S \left\{ \frac{\phi (O;h_{0},g_{0})}{P\left\{ (A,L)\in \mathcal{S}\right\} }%
-\psi _{0}\frac{I\left\{ (A,L)\in \mathcal{S}\right\} }{P\left\{ (A,L)\in 
\mathcal{S}\right\} }\right\}\right] +\sqrt{n}R_{n}+o_{p}(1).
\end{equation*}
    where $ R_n=\sum_{k=1}^K\frac{n_k}{n} \:\int\big\{\hat h^{(-k)}(a,l,w)-h_0(a,l,w)\big\}\big\{g_0(a,l,z)-\hat g^{(-k)}(a,l,z)\big\}dP(a,l,w,z)$ with $n_k$ denoting the cardinality of $\mc I_k$, and recall that $O=(A,L,Z,W,Y)$ is used inside $\phi$. In particular, if $\mathbb{E}\left\{\phi(O;h_0,g_0)^2\right\}<\infty$ and $R_n=o_p\left(n^{-1/2}\right)$, then $ \sqrt{n}\left\{\psi_{CF}^{DR}-\psi_0\right\}\stackrel{\text{d}}{\longrightarrow}\mc{N}(0,\tau^2)$ where $\tau ^{2}=\mathbb{E}\Big\{ S \:\Big[ \frac{\phi (O;h_{0},g_{0})}{%
P\left\{ (A,L)\in \mathcal{S}\right\} }-\psi _{0}\frac{I\left\{ (A,L)\in 
\mathcal{S}\right\} }{P\left\{ (A,L)\in \mathcal{S}\right\} }\Big]
^{2}\Big\} $.
\label{theo:asymp_distr_delta}
\end{theorem}
Now, note that any observed outcome bridge function $h_0$ satisfies
\[
h_0 \in \arg\min_{h \in \mc{L}^2(A,L,W)}\max_{{g}\in \mc{L}^2(A,L,Z)} \mathbb{E}\left[\Delta\cdot S\left\{g(A,L,Z)\left[Y-h(A,L,W)\right]-g(A,L,Z)^2\right\}\right].
\]
Thus, in analogy with the case in which $A$ is always observed, we propose the estimator  
\[
\hat h^{(-k)} = \arg\min_{h\in\mc{H}_B}\max_{g\in\mc{G}'}\mathbb{E}_n^{(-k)}\left[\Delta S\left\{g(A,L,Z)\left[Y-h(A,L,W)\right]-g(A,L,Z)^2\right\}\right]-\lambda_{\mc{G}'}\Vert g\Vert_{\mc{G}'}^2+\lambda_{\mc{H}}\Vert h\Vert_{\mc{H}}^2,
\]
where $\mc{H}$ and $\mc{G'}$ are RKHSs. It can be shown that the bounds on the root-mean-squared error $\Vert\hat h-h_0\Vert_2$ and on the weak error $\Vert\mathbb{E}\{\hat h(A,L,W)-h_0(A,L,W)|A,L,Z\}\Vert_2$ established by \cite{olivas2025source} continue to hold up to constants depending on $\sigma$.

Likewise, any treatment bridge function can be written as $I_q(a,l)g_0(a,l,z)$, where
\begin{align*}
g_0\in\arg\min_{g\in\mc{L}^2(A,L,Z)}\max_{h\in\mc{L}^2(A,L,w)}\mathbb{E}&\big[\Delta\cdot S\big\{h\left\{q(A,L),L,W\right\}\cdot I\left\{(A,L)\in\mc{S}\right\}\\
    &-I_q(A,L)h(A,L,W)g(A,L,Z)-I_q(A,L)h(A,L,W)^2\big\}\big].
\end{align*}
Hence, letting $\mc{G}$ and $\mc{H}'$ be RKHSs, we propose the estimator $\hat g=I_q(A,L)\cdot \tilde g(A,L,Z)$, where 
\begin{align*}
    \tilde g =\arg\min_{g\in\mc{G}_B}\max_{h\in\mc{H}'}\mathbb{E}_n^{(-k)}&\big[\Delta S\big\{h\left\{q(A,L),L,W\right\} I\left\{(A,L)\in\mc{S}\right\}-I_q(A,L)h(A,L,W)g(A,L,Z)\\
    &-I_q(A,L)h(A,L,W)^2\big\}\big]-\lambda_{\mc{H}'}\Vert h\Vert_{\mc{H}'}^2+\lambda_{\mc{G}}\Vert g\Vert_{\mc{G}}^2.
\end{align*}

Finally, we derive the closed-form expressions for $\hat h^{(-k)}$ and $\hat g^{(-k)}$. As in the main text, we drop the superscript $(-k)$, and let $n$ denote the size of the full sample excluding the subsample $\mc I_k$. Let $m=\sum_{i=1}^n\Delta_i$ be the number of observations for which $A$ is observed, and let $\ell_1,\dots,\ell_m$ be their corresponding indices. Using an approach similar to the one described in Appendix~\ref{sec:rkhs_est}, the closed-form solution for $\hat h$ is given by
\[
\hat h(a,l,w) = \sum_{j=1}^m\gamma_j^*K_{\mc{H}}((a,l,w),(A_{\ell_j},L_{\ell_j},W_{\ell_j})),
\]
where $\gamma^*=\left(S_m\Gamma_{\mc{G}'} K_{\mc{G}',m}S_mK_{\mc{H},m}+n^2\lambda_{\mc{H}}Id_m\right)^{-1}S_m\Gamma_{\mc{G}'} K_{\mc{G}',m}S_m \widetilde{Y}_m$. Here, $K_{\mc{H},m}$ and  $K_{\mc{G}',m}$ are the $m\times m$ empirical kernel matrices evaluated at the observations $\ell_1,\dots,\ell_m$, $S_m$ is an $m\times m$ diagonal matrix containing the weights $S_{\ell_1},\dots,S_{\ell_m}$, $\Gamma_{\mc{G}'} = 4^{-1}(n^{-1}S_mK_{\mc{G}',m}+\lambda_{\mc{G}'}Id_m)^{-1}$, and $\widetilde Y_m=\left(Y_{\ell_1},\dots,Y_{\ell_m}\right)^\top$. If $(\gamma^*)^\top K_{\mc{H},n}\gamma^*>B$ for the pre-specified $B$, we obtain a constrained solution as outlined in the previous section.

The estimator $\hat g$ for $g_0$ is $\hat g(a,l,z)=I\left\{a\in q(\mc{S}(l),l)\right\}\cdot\tilde g(a,l,z)$, where $\tilde g$ is given by
\[
\tilde g(a,l,z) = \sum_{j=1}^m\theta^*_jK_{\mc{G}}((a,l,z),(A_{\ell_j},L_{\ell_j},Z_{\ell_j})),
\]
with $\theta^*=\left(S_mR_m\Gamma_{\mc{H}'} K_{\mc{H}',m}R_mS_mK_{\mc{G},m}+n^2\lambda_{\mc{G}}Id_m\right)^{-1}S_mR_m\Gamma_{\mc{H}'} \widetilde{K}_{\mc{H}',m}^\top S_m\tilde d_m$. Here, $K_{\mc{G},m}$ and  $K_{\mc{H}',m}$ are the $m\times m$ empirical kernel matrices evaluated at the observations $\ell_1,\dots,\ell_m$, $\widetilde{K}_{\mc{H}',m}$ is the $m\times m$ matrix with $(i,j)$--entry 
$\widetilde{K}_{\mc{H}',m}=
         K_{\mc{H}'}\{(q(A_{\ell_i},L_{\ell_i}),L_{\ell_i},W_{\ell_i}),(A_{\ell_j},L_{\ell_j},W_{\ell_j})\}$, $R_m$ is an $m\times m$ diagonal matrix with values $I_q(A_{\ell_1},L_{\ell_1}),\dots,I_q(A_{\ell_m},L_{\ell_m})$ in its diagonal, $\Gamma_{\mc{H}'}=4^{-1}(n^{-1}K_{\mc{H}',m}S_mR_m+\lambda_{\mc{H}'}Id_m)^{-1}$, and $\tilde d_m$ is the vector $\big[I\left\{(A_{\ell_1},L_{\ell_1})\in\mc{S}\right\},\dots,I\left\{(A_{\ell_m},L_{\ell_m})\in\mc{S}\right\}\big]^\top$. If $\left(\theta^*\right)^\top K_{\mc{G},n}\theta^*>B$, a constrained solution can be obtained as described previously.

\section{Extended Numerical Experiments and Implementation} \label{sec:webapp_c}

\renewcommand{\theequation}{C.\arabic{equation}}
\setcounter{equation}{0}

\renewcommand{\thefigure}{C.\arabic{figure}}
\setcounter{figure}{0}

In this section, we expand upon the numerical experiments presented in Section~\ref{sec:exp} and provide further details on both the implementation of our methodology and its application to the ENSEMBLE trial data.

\subsection{Data-Generating Process and Policy}\label{sec:dgp}

Recall that $\mc{TN}_{[a,b]}(\mu,\sigma^2)$ denotes a normal distribution $\mc{N}(\mu,\sigma^2)$ truncated to the interval $[a,b]$, and that $\mc{TN}_c(\mu,\sigma^2)$ denotes the specific case where $a=\mu-3\sigma$ and $b=\mu+3\sigma$. For each $i=1,\dots,n$, we generate: the observed confounder $L_i\sim\mc{TN}_c(0,1)$, the unobserved confounder $U_i\sim\mc{TN}_c(\beta_1L_i, 1)$, the negative controls $Z_i\sim\mc{TN}_c(\beta_2L_i + \beta_3 U_i,1)$ and $W_i\sim\mc{TN}_c(\beta_4L_i + \beta_5 U_i,1)$, the exposure $A_i\sim\mc{TN}_{[c,d]}(\beta_6L_i + \beta_7U_i+\beta_8Z_i,1)$, and the outcome $Y_i\sim\text{Bernoulli}\left(p_i\right)$ where $p_i=\left[1+\exp\left\{-\left(\mu+\beta_9L_i+\beta_{10}U_i+\beta_{11}A_i+\beta_{12}W_i+\gamma A_i^2\right)\right\}\right]^{-1}$. This data-generating process includes the parameter $\beta_8$, allowing the negative control treatment to have a direct effect on the primary treatment, and the parameter $\beta_{12}$, allowing the negative control outcome to directly influence the outcome of interest. 

For parameters $\delta>0$ and $\varepsilon\ge0$ satisfying $c+\delta<d-\varepsilon$, and for $r\in\{0,1\}$, we define the policy $q^{\delta,\varepsilon,r}:[c,d-r(\varepsilon+\delta)]\rightarrow[c+\delta,d-r\varepsilon]$ as
 \begin{equation}
      q^{\delta,\varepsilon,r}(a):=a+\delta I\{c\le a\le  d-\varepsilon-\delta\}+\frac{\delta}{\delta+\varepsilon}(d-a) I\{d-\varepsilon-\delta< a\le d-r(\varepsilon+\delta)\}.\label{q_general}
 \end{equation}
 When $r=0$, this corresponds to the $\varepsilon$--upper-tail tapering shift intervention policy discussed in Example~\ref{ex2} of Section~\ref{sec:ident}. When $r=1$, it reduces to the standard shift intervention policy restricted to the subpopulation $\{O_i\,:\,A_i\in[c,d-\delta-\varepsilon]\}$, since the second term in \eqref{q_general} vanishes (the set $\{a: d-\varepsilon-\delta < a \le d-r(\varepsilon+\delta)\}$ is empty). The parameter $r$ allows the policy to accommodate interventions applicable to the entire population (Strategy~2 in Example~\ref{ex2}) as well as those restricted to a subpopulation (Strategy~1 in Example~\ref{ex1}).
    
    In the numerical experiments reported in Section~\ref{sec:exp}, we used the following parameter setting $(\beta_1,\dots,\beta_{12})=(0.5, 0.2,\beta_z,0.5, \beta_w,0.3, 1, 0, 0.5,-1,-1.5,0)$, $c=-2$, $d=2$, $\mu=-1$, and $\gamma=-0.75$, and targeted the counterfactua MTP mean under the policy $q^{\delta=0.4,\varepsilon=1,r=0}$.

\subsection{Verifying Assumption~\ref{ass:consist0}--\ref{ass:assumption8_0} for the Data-Generating Process and Policy}\label{sec:appC_validating_assumptions}

By design, the data-generating process satisfies Assumptions~\ref{ass:consist0}--\ref{ass:nct0}, and the policy~\ref{q_general} satisfies Assumption~\ref{ass:q0}. In fact, the inverse of $q^{\delta,\varepsilon,r}$, defined over $[c+\delta,d-r\varepsilon]$, is given by $(q^{\delta,\varepsilon,r})^{-1}(a)=a-\delta V(a)$,
where
\begin{equation}
    V(a):=I\left\{c+\delta\le a\le d-\varepsilon\right\}+\frac{d-a}{\varepsilon} I\left\{d-\varepsilon< a\le d-r\varepsilon\right\}.\label{eq:def_va}
\end{equation}

To verify Assumptions \ref{ass:assumption7_0} and \ref{ass:assumption8_0}, we provide explicit or accurate approximate solutions to the bridge equations~\eqref{lat_out_eqn0} and~\eqref{lat_tx_eqn0}. When $\beta_{10}/\beta_5$ and $\beta_{12}$ are small, in Appendix~\ref{sec:deriv_out_br} we show that an accurate approximate latent (and therefore observed) outcome bridge function is
    \begin{align*}
        h(a,l,w;\varphi)\approx\Big(1+\frac{0.9733}{2}\varphi_3^2\Big)\Big[1+\exp\Big\{-\varphi_0-\beta_{11}\:a-\Big(\beta_9-\frac{\beta_4\beta_{10}}{\beta_5}\Big)l-\Big(\beta_{12}+\frac{\beta_{10}}{\beta_5}\Big)w-\gamma\: a^2\Big\}\Big]^{-1},
    \end{align*}
    where
    \[
\varphi_0\approx\mu+\log\Big(1+\frac{0.9733}{2}\beta_{12}^2\Big)-\log\Big\{1+0.9733\,\Big(\beta_{12}+\frac{\beta_{10}}{\beta_5}\Big)^2\Big\}.
\]
As shown in Appendix~\ref{sec:parametic_est}, employing this general parametric form in estimation yields good finite-sample performance.

An exact latent (and observed) treatment bridge function when $\beta_8 = 0$ (i.e., there is no direct effect of $Z$ on $A$) is:
    \begin{align*}
        g(a,l,z)&=\rho(a)\exp\Big[\delta\Big\{a-\Big(\beta_6-\frac{\beta_2\beta_7}{\beta_3}\Big)l-\frac{\beta_7}{\beta_3}z-\frac{\delta}{2}\Big(1+\frac{\beta_7^2}{\beta_3^2}\Big)V(a)\Big\}V(a)\Big],
    \end{align*}
where
\[
\rho (a):=\frac{\Phi(3)-\Phi(-3)}{\Phi\big\{\frac{\delta\beta_7}{\beta_3} V(a)+3\big\}-\Phi\big\{\frac{\delta\beta_7}{\beta_3} V(a)-3\big\}}\Big[I\left\{c+\delta\le a\le d-r\varepsilon\right\}+\frac{\delta}{\varepsilon}\,I\left\{d-\varepsilon< a\le d-r\varepsilon\right\}\Big]
\]
and $\Phi(\cdot)$ denotes the standard normal CDF. An approximate latent (and observed) treatment bridge function for small nonzero $\beta_8$ is derived in Appendix~\ref{sec:deriv_tr_br}.

\subsection{Proximal Parametric Estimators}\label{sec:parametic_est}

The data-generating process motivates the following parametric model for the outcome bridge function $h_0$ (its derivation is provided in Appendix~\ref{sec:deriv_out_br}):
\[
h(a,l,w;\varphi)=\left(1+\frac{0.9733}{2}\varphi_3^2\right)\left[1+\exp\left\{-(\varphi_0+\varphi_1a+\varphi_2l+\varphi_3w+\varphi_4a^2)\right\}\right]^{-1},
\]
where $\varphi=(\varphi_0,\varphi_1,\varphi_2,\varphi_3,\varphi_4)^\top$.

For the treatment bridge function $g_0$ under the policy defined in~\eqref{q_general}, and in the case $\beta_8=0$, the data-generating process suggests the following parametric model (derived in Appendix~\ref{sec:deriv_tr_br})
\[
g(a,l,z;\eta)=\rho(a;\eta)\exp\left[\left\{\eta_0a+\eta_1l+\eta_2z+\eta_3V(a)\right\}V(a)\right],
\]
where $\eta=(\eta_0,\eta_1,\eta_2,\eta_3)^\top$ $V(a)$ is defined in~\eqref{eq:def_va} and
\[
\rho(a;\eta):=\frac{\Phi(3)-\Phi(-3)}{\Phi\left(3-\eta_2V(a)\right)-\Phi\left(-3-\eta_2V(a)\right)}\Big[I\left\{c+\delta\le a\le d-r\varepsilon\right\}+\frac{\delta}{\varepsilon}\, I\left\{d-\varepsilon< a\le d-r\varepsilon\right\}\Big].
\]

The integral equations~\eqref{obs_out_eqn0} and~\eqref{obs_tx_eqn0} motivate parametric estimators $\hat\varphi$ and $\hat\eta$, obtained by solving the following estimating equations:
\begin{align*}
    \mathbb{E}_n\left[\left\{Y-h(A,L,W;\varphi)\right\}(1,A,L,Z,A^2)^\top\right]&=(0,0,0,0,0)^\top,\\
    \mathbb{E}_n\left[I\left\{(A,L)\in\mc{S}\right\}\cdot\left(1,q^{\delta,\varepsilon,r}(A),\,L,W\right)^\top -g(A,L,Z;\eta )(1,A,L,W)^\top\right]&=(0,0,0,0)^\top.
\end{align*}
Using $\hat \varphi$ and $\hat \eta$, we compute the following three parametric estimators for $\psi_0$:
\begin{align*}
\hat \psi_{p,OR}&=\mathbb{E}_n\left[h\{q^{\delta,\varepsilon,r}(A,L), W,L;\hat \varphi\}\right],\\
    \hat\psi_{p,DQW} &= \mathbb{E}_n\left[Y\cdot g(A,Z,L;\hat\eta)\right],\\
    \hat\psi_{p,DR}&=\mathbb{E}_n\left[h\{q^{\delta,\varepsilon,r}(A,L), W,L;\hat\varphi\}\cdot I\{(A,L)\in\mc{S}\}+ g(A,Z,L;\hat\eta)\cdot\left\{Y- h(A,W,L;\hat\eta)\right\}\right].
\end{align*}

For each estimator, we construct 95\% confidence intervals using influence-function-based standard errors that account for the estimation of the nuisance functions. Let $\Psi(O)$ denote the population influence function of the parametric estimator under consideration, and let $\widehat\Psi_i$ denote the corresponding estimated influence function for observation $i$. The population influence function $\Psi(O)$ is derived using standard results for Z-estimators (see Chapter 5 of \cite{van2000asymptotic}), and the variance of the limiting mean zero normal distribution of the scaled error is estimated via the empirical variance of the estimated influence functions:
\[
\frac{1}{n}\sum_{i=1}^n(\widehat\Psi_i-\hat\psi)^2.
\]
Wald-type 95\% confidence intervals are then computed as $\hat\psi\pm1.96\sqrt{\widehat{\text{var}}(\hat\psi)}$. These intervals are asymptotically valid under standard regularity conditions, provided the nuisance models are correctly specified. For the doubly-robust parametric estimator $\hat\psi_{p,DR}$, consistency and valid inference require only one of the two parametric models for the nuisance functions to be correctly specified.

We assessed the finite-sample performance of the proximal parametric estimators using the same data-generating process and target parameter as in Section~\ref{sec:exp}. Because the parametric approach is less computationally demanding, we considered larger sample sizes $n\in\{750, 3000, 12000, 48000\}$ and a broader set of 16 scenarios formed by combining $\beta_z,-\beta_w\in\{2,1,0.5,0.25\}$. The values $\beta_z=0.25$ and $\beta_w=-0.25$ correspond to conditional correlations $\text{Cor}(Z,U|L)=0.235$ and $\text{Cor}(W,U|L)=-0.235$. For each scenario and sample size, we performed 500 simulation replications. For each nuisance function, we considered both a correctly specified and a misspecified parametric model: the misspecified model for $h$ omitted the quadratic $a^2$ term, and the misspecified model for $g$ replaced the term $\eta_3 V(a)$ in the exponential with the constant $\eta_3$. Combining these specifications results in a total of eight parametric estimators: two outcome regression estimators (based on the correctly specified or misspecified parametric model for the outcome bridge function), two density-quotient-weighted estimators (based on the correctly specified or misspecified parametric model for the treatment bridge function), and four doubly-robust estimators (combining all possible correct/misspecified parametric models for both bridge functions).
  
\begin{sidewaysfigure}[h!]
\centerline{\includegraphics[width=9in]{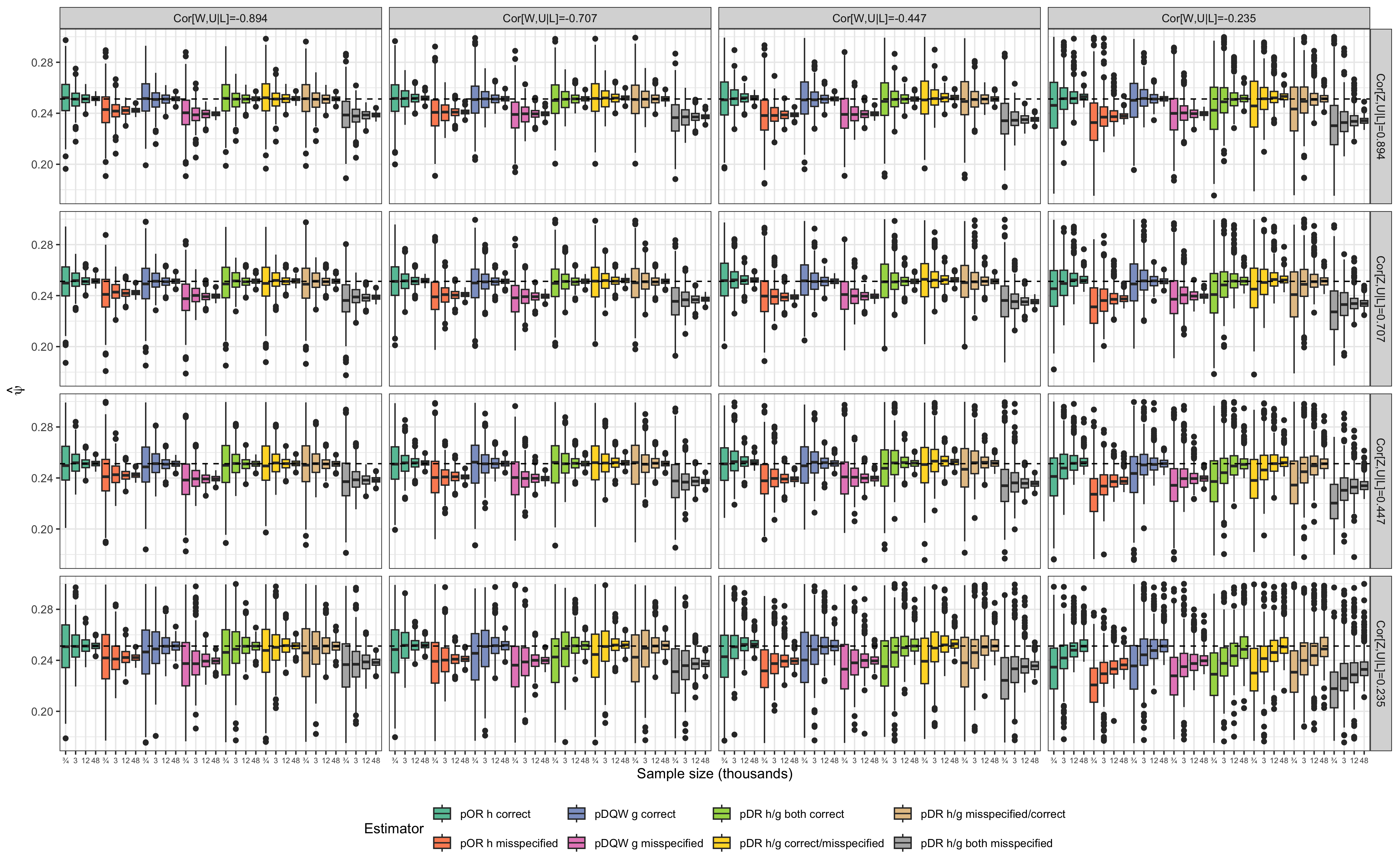}}
    \caption{Boxplots of the parametric estimators from numerical experiments with data generated under varying conditional correlations of the proxies $Z$ and $W$ with the unmeasured confounder $U$ and sample sizes $n=750, 3000, 12000$, and 48000. Dashed lines indicates the true parameter value.}\label{fig:num_exp_param_bp}
\end{sidewaysfigure}

\begin{sidewaysfigure}
\centerline{\includegraphics[width=9in]{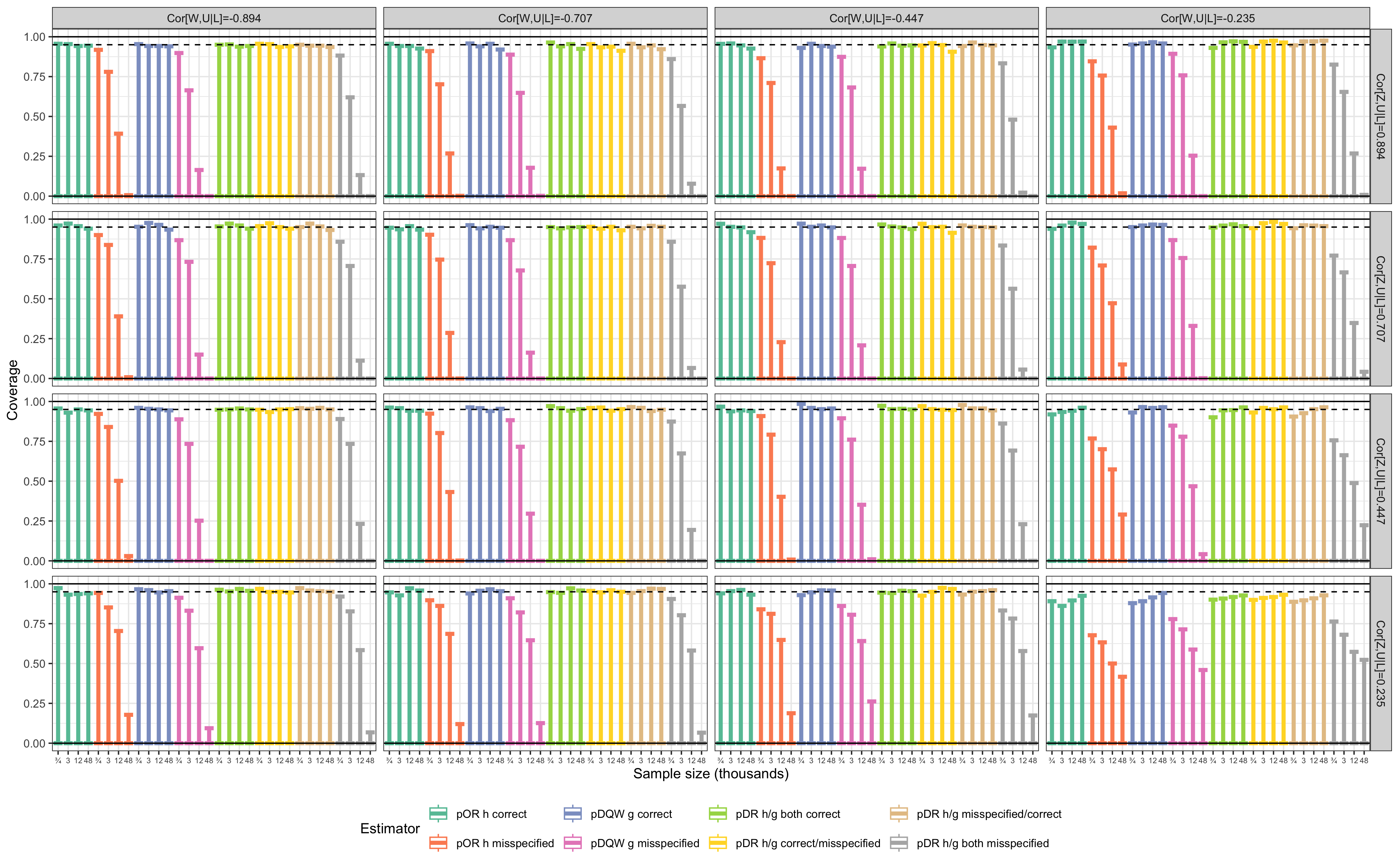}}
    \caption{Empirical coverage of 95\% confidence intervals for the parametric estimators from numerical experiments with data generated under varying conditional correlations of the proxies $Z$ and $W$ with the unmeasured confounder $U$ and sample sizes $n=750, 3000$, 12000, and 48000. The dashed line indicates the nominal coverage level of 0.95.}\label{fig:num_exp_param_cov}
\end{sidewaysfigure}

  Figures~\ref{fig:num_exp_param_bp} and~\ref{fig:num_exp_param_cov} display boxplots and coverage probabilities of the confidence intervals for the eight parametric estimators described above. Across the considered scenarios, estimators based on at least one correctly specified model exhibit low bias and achieve coverage probabilities close to the nominal level. As expected, estimators relying solely on misspecified models are biased and show poor coverage. Furthermore, consistent with the findings for nonparametric estimators in the main text, we observe that the challenge of estimating the bridge functions under weak conditional correlation between the negative controls and the unmeasured confounder diminishes as the sample size increases.
  
\subsection{Additional Details on the Cross-Validation Procedure}\label{sec:cv}

 To implement the cross-validation procedure described in the main text, we evaluate candidate bridge functions using the following empirical risk functions, defined for any $h\in\mathcal{L}^2(A,L,W)$ and $g\in\mathcal{L}^2(A,L,Z)$:
\[
\mathcal{R}_n^1(h):=\sup_{g\in\mathcal{G}'}\mathbb{E}_n\left[g(A,L,Z)\left\{Y-h(A,L,W)\right\}-g(A,L,Z)^2\right]-\lambda_{\mc{G}'}\Vert g\Vert_{\mc{G}'}^2,
\]
and
\begin{align*}
    \mathcal{R}_n^2(g):=\sup_{h\in\mathcal{H}'}\mathbb{E}_n&\big[h\left\{q(A,L),L,W\right\}\cdot I\left\{(A,L)\in\mc{S}\right\}-I_q(A,L) h(A,L,W)g(A,L,Z)\\
    &-I_q(A,L)h(A,L,W)^2\big]-\lambda_{\mc{H}'}\Vert h\Vert_{\mc{H}'}^2.
\end{align*}
When the function classes $\mc{G}'$ and $\mc{H}'$ are RKHSs generated by Gaussian or Mat\'ern kernels, the solutions to the observed bridge equations~\eqref{obs_out_eqn0}and~\eqref{obs_tx_eqn0} minimize the population version of these risk functions, making them suitable for cross-validation.

We now provide the closed-form expressions of the empirical risk functions when candidate bridge functions are evaluated in fold 3 (testing set). Let $n_3$ denote the number of observations in fold 3, and $\ell_1^{(3)},\dots,\ell_{n_3}^{(3)}$ their corresponding indices. When $\mc{G}'$ and $\mc{H}'$ are RKHSs, these empirical risks take the forms:
\[
\mathcal{R}_n^1(h)=\frac{1}{4}\left\{\xi_{n_3}(h)\right\}^\top\left(\frac{1}{n_3}K_{\mc{G}',n_3}^0+\lambda_{\mc{H}'}Id_{n_3}\right)^{-1}K_{\mc{G}',n_3}^0\left\{\xi_{n_3}(h)\right\}
\]
and
\[
\mathcal{R}_n^2(g)=\frac{1}{4}\left\{\zeta_{n_3}(g)\right\}^\top\left(\frac{1}{n_3}K_{\mc{H}',n_3}^0R_{n_3}K_{\mc{H}',n_3}^0+\lambda_{\mc{H}'}K_{\mc{H}',n_3}^0\right)^{-1}K_{\mc{H}',n_3}^0\left\{\zeta_{n_3}(g)\right\}.
\]
Here, $K_{\mc{G}',n_3}^0$ and $K_{\mc{H}',n_3}^0$ are $n_3\times n_3$ matrices whose $(i,j)$-th entries correspond to the kernel functions $K_{\mc{G}'}^0$ and $K_{\mc{H}'}^0$ evaluated at observations $\ell_i^{(3)}$ and $\ell_j^{(3)}$, and $R_{n_3}$ is the diagonal matrix with diagonal entries $I_q(A_{\ell_1^{(3)}},L_{\ell_1^{(3)}}),\dots,I_q(A_{\ell_{n_3}^{(3)}},L_{\ell_{n_3}^{(3)}})$. The vectors $\xi_{n_3}(h)$ and $\zeta_{n_3}(g)$ are defined as
\[
\xi_{n_3}(h):=\frac{1}{n_3}\big(\widetilde{Y}_{n_3}-\tilde h_{n_3}\big)\quad\text{and}\quad\zeta_{n_3}(h):=\frac{1}{n_3}\big\{\big(\widetilde{K}_{\mc{H}',n_3}^0\big)^\top\tilde d_{n_3}-K_{\mc{H}',n_3}^0R_{n_3}\tilde g_{n_3}\big\},
\]
where $\widetilde{Y}_{n_3}=(Y_{\ell_1^{(3)}},\dots,Y_{\ell_{n_3}^{(3)}})^\top$, $\tilde h_n=\big[h(A_{\ell_1^{(3)}},L_{\ell_1^{(3)}},W_{\ell_1^{(3)}}),\dots, h(A_{\ell_{n_3}^{(3)}},L_{\ell_{n_3}^{(3)}},W_{\ell_{n_3}^{(3)}})\big]^\top$ and $\tilde g_n=\big[g(A_{\ell_1^{(3)}},L_{\ell_1^{(3)}},Z_{\ell_1^{(3)}}),\dots, g(A_{\ell_{n_3}^{(3)}},L_{\ell_{n_3}^{(3)}},Z_{\ell_{n_3}^{(3)}})\big]^\top$. The matrix $\widetilde{K}_{\mc{H}',n_3}^0$ is the $n_3\times n_3$ matrix whose $(i,j)$-entry is given by $K_{\mc{H}'}^0\big\{(q(A_{\ell_i^{(3)}},L_{\ell_i^{(3)}}),L_{\ell_i^{(3)}},W_{\ell_i^{(3)}}),(A_{\ell_j^{(3)}},L_{\ell_j^{(3)}},W_{\ell_j^{(3)}})\big\}$. The kernel bandwidth for $K_{\mc{G}',n_3}^0$ is set to 1/4 times the median of the pairwise Euclidean distances between the observed vectors $(A_{\ell_i^{(3)}},L_{\ell_i^{(3)}},Z_{\ell_i^{(3)}})$, and the bandwidth for $K_{\mc{H}',n_3}^0$ is chosen analogously using $(A_{\ell_i^{(3)}},L_{\ell_i^{(3)}},W_{\ell_i^{(3)}})$. Finally, the regularization parameters $\lambda_{\mc{G}'}$ and $\lambda_{\mc{H}'}$ are both set to $\log(n_3)/n_3$, corresponding to the order of the squared critical radius for Gaussian RKHSs.

\label{sec:cv}

\subsection{Performance of the Proximal Estimator Under No Unmeasured Confounding}

To evaluate the performance of our estimator in the absence of unmeasured confounding, we generated data using the general mechanism described in Appendix~\ref{sec:dgp}, with the same parameter specification and policy as in Section~\ref{sec:exp}, except that the effect of the unobserved variable $U$ on the exposure was set to zero ($\beta_7=0$), i.e., $U$ is no longer a confounder. We focused on the scenario $\beta_z=2$ and $\beta_w=-2$, in which both $Z$ and $W$ are highly correlated with $U$. Under this configuration, $Z$ and $W$ act as precision variables. For this scenario, the counterfactual MTP mean under the policy $q^{\delta=0.4,\varepsilon=1,r=0}$ is 0.2081. Data were generated for sample sizes $n\in\{750, 1500, 3000\}$, and for each sample size, 500 repetitions of the data-generating and estimation procedures were performed.

\begin{figure}
\centerline{\includegraphics[width=5in]{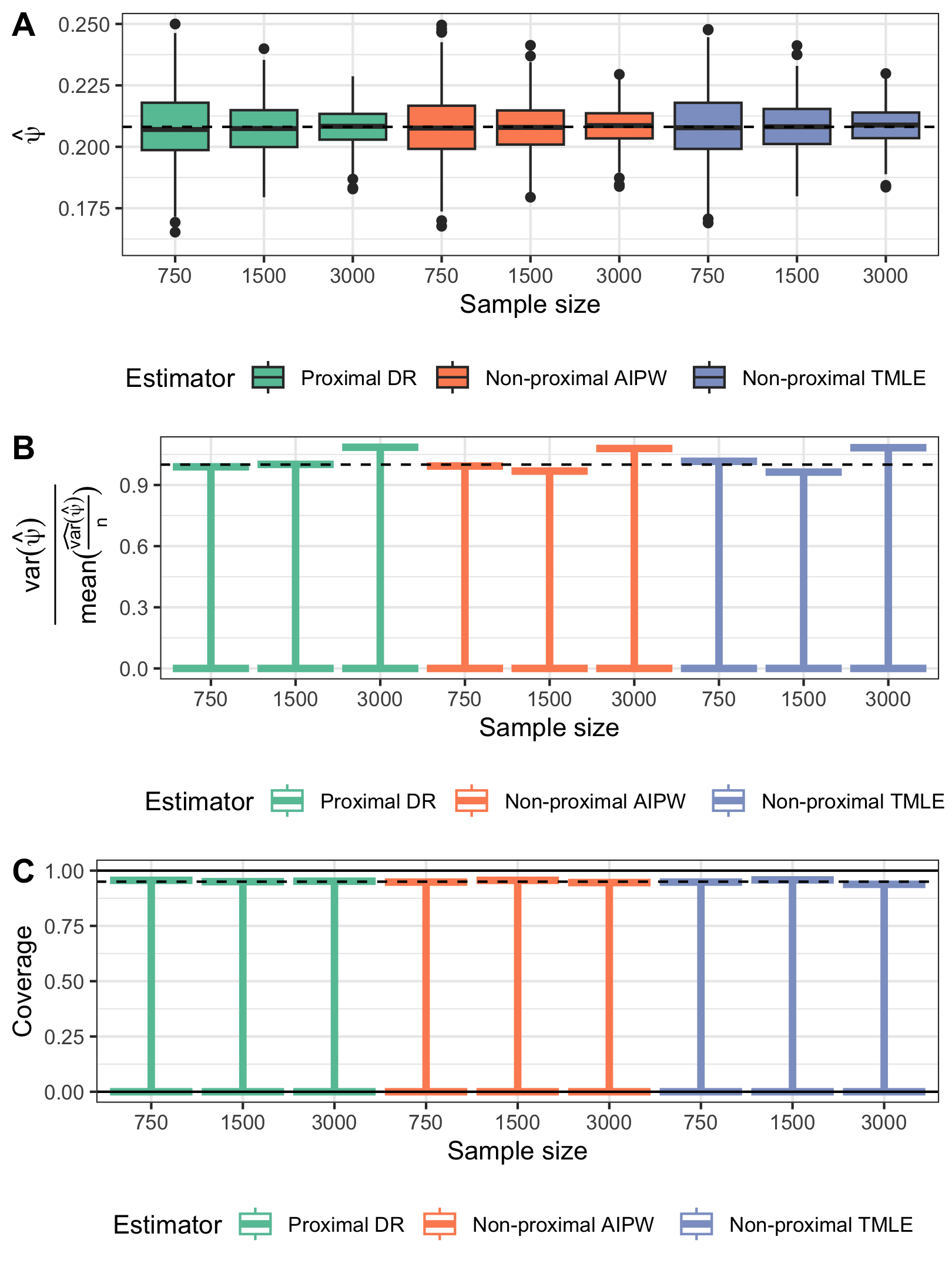}}
   \caption{Finite-sample performance of the doubly-robust cross-fitted estimator $\hat\psi_{CF}^{DR}$ and competitors based on numerical experiments \textit{without} unmeasured confounding. \textbf{Panel A}: boxplots of the estimator across simulations. \textbf{Panel B}: ratio of the empirical variance to the mean estimated asymptotic variance. \textbf{Panel C}: empirical coverage of 95\% confidence intervals. The dashed line indicates the true parameter value (Panel A), a variance ratio of 1 (Panel B), and nominal 95\% coverage (Panel C).}\label{fig:num_exp_nuc}
\end{figure}

Figure~\ref{fig:num_exp_nuc} presents boxplots, variance ratios, and coverage probabilities for the proximal doubly-robust cross-fitted estimator and its competitors. In this setting, both proximal and non-proximal estimators exhibit low bias, estimated asymptotic variances closely match the empirical variances, and coverage probabilities are at the nominal level.

\subsection{Proximal Estimation of the Counterfactual MTP Mean in a Restricted Population}

\begin{figure}
\centerline{\includegraphics[width=5in]{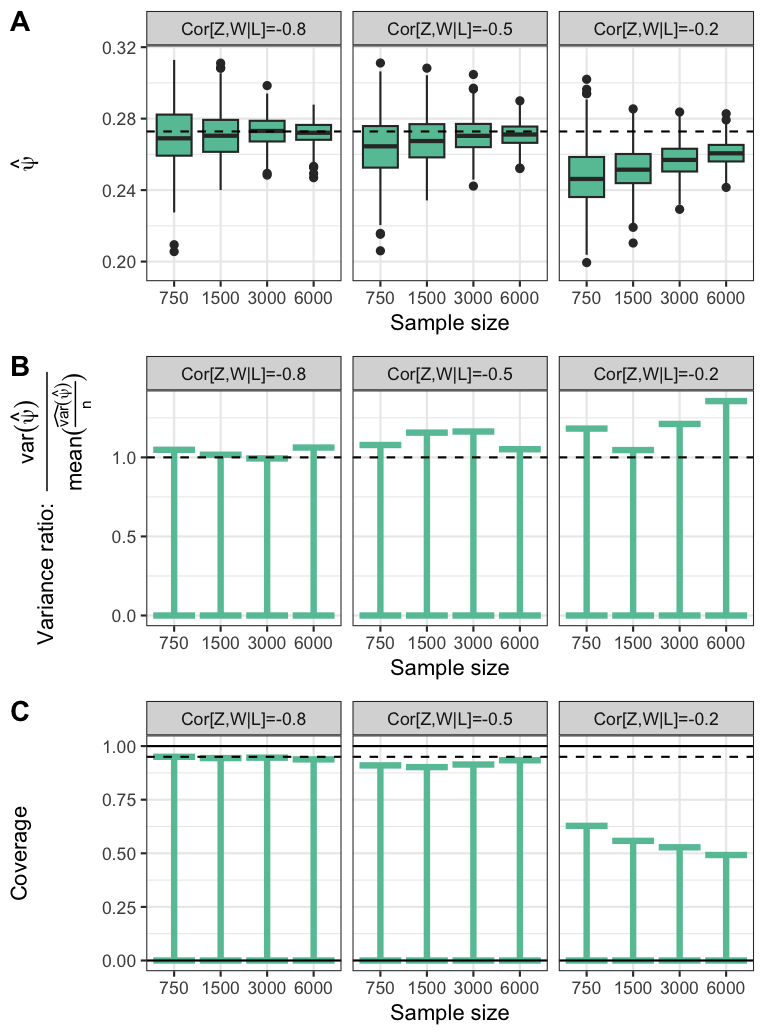}}
   \caption{Finite-sample performance of the doubly-robust cross-fitted estimator $\hat\psi_{CF}^{DR}$ for estimating the counterfactual MTP mean in a \textit{restricted population}, based on numerical experiments with data
generated under unmeasured confounding. \textbf{Panel A}: boxplots of the estimator across simulations. \textbf{Panel B}: ratio of the empirical variance to the mean estimated asymptotic variance. \textbf{Panel C}: empirical coverage of 95\% confidence intervals. The dashed line indicates the true parameter value (Panel A), a variance ratio of 1 (Panel B), and nominal 95\% coverage (Panel C).}\label{fig:num_exp_rest}
\end{figure}

To illustrate the performance of our proximal estimator for a target parameter under Strategy~1—outlined in Example~\ref{ex1} of Section~\ref{sec:ident}—we conducted numerical experiments using the same data-generating processes as in Section~\ref{sec:exp}, but restricted to the three scenarios with $\beta_z = -\beta_w \in \{2,1,0.5\}$. In this case, we estimate $\mathbb{E}\left\{Y(A+0.4)\mid A\in[-2,1.6]\right\}$, the counterfactual MTP mean under a shift intervention applied to the subpopulation $\{A_i: A_i\in[-2,1.6]\}$, which equals $0.2728$ under all considered scenarios. This policy corresponds to $q^{\delta=0.4,\varepsilon=0,r=1}(a)$ as defined in~\eqref{q_general}, for which Assumptions~\ref{ass:consist0}--\ref{ass:assumption8_0} have been shown to hold. For each setting, we performed 500 repetitions of the data-generating and estimation procedures. Figure~\ref{fig:num_exp_rest} presents boxplots, variance ratios, and coverage probabilities of the proximal doubly-robust cross-fitted estimator, whose performance is comparable to that observed for the estimand under Strategy~2, i.e., to the results reported in Section~\ref{sec:exp}.

\subsection{Performance of the Proximal Estimator under a Data-Generating Mechanism where $Z$ Affects $A$ and $W$ Affects $Y$}\label{sec:exp_case2}

We evaluated the performance of our estimator under a data-generating mechanism in which $Z$ directly influences $A$ and $W$ directly affects $Y$. Data were simulated using the process described in Appendix~\ref{sec:dgp}, with parameter values
\[
(\beta_1,\dots,\beta_{12},c,d,\mu,\gamma)=(0.5, 0.2,\beta_z,0.5, \beta_w,0.3, 1, 0.3, 0.5,-1,-1.5,-0.3,-2,2,-1,-0.75).
\]
We considered four scenarios with $\beta_z=-\beta_w\in\{3,1.5,1,0.75\}$, corresponding to conditional correlations $\text{cor}(W,U\mid L)=-0.90$, $-0.69$, $-0.50$, and $-0.36$, respectively. The target estimand was the same as in Section~\ref{sec:exp}, namely $\mathbb{E}\left[Y\left\{q^{\delta=0.4,\varepsilon=1,r=0}(A)\right\}\right]$, which equals 0.1943, 0.2297, 0.2400, and 0.2466 for $\beta_z=3, 1.5, 1$, and 0.75, respectively. For each combination of parameter configuration and sample size $n \in \{750, 1500, 3000, 6000\}$, 500 replications of the data-generating and estimation procedures were performed.

\begin{figure}
\centerline{\includegraphics[width=6in]{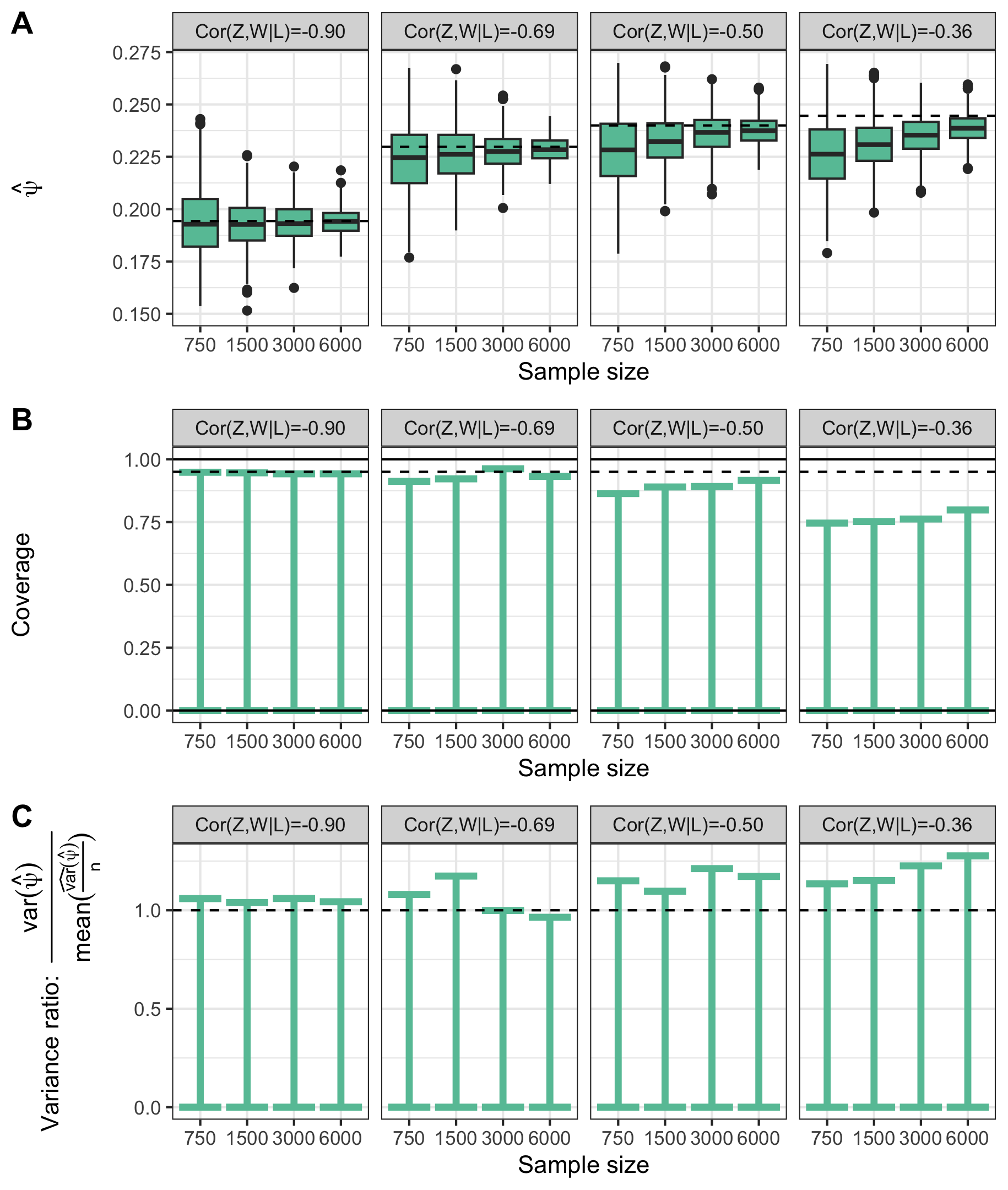}}
   \caption{Finite-sample performance of the doubly-robust cross-fitted estimator $\hat\psi_{CF}^{DR}$ based on numerical experiments under a more challenging data-generating mechanism, where $Z$ directly affects $X$ and $W$ directly affects $Y$.  Results are shown across varying conditional correlations of the proxies $Z$ and $W$ with the unmeasured confounder $U$. \textbf{Panel A}: boxplots of the estimator across simulations. \textbf{Panel B}: ratio of the empirical variance to the mean estimated asymptotic variance. \textbf{Panel C}: empirical coverage of 95\% confidence intervals. The dashed line indicates the true parameter value (Panel A), a variance ratio of 1 (Panel B), and nominal 95\% coverage (Panel C).}\label{fig:num_exp_arrows}
\end{figure}

Figure~\ref{fig:num_exp_arrows} presents boxplots, variance ratios, and confidence interval coverage of the proximal doubly-robust cross-fitted estimator. As expected, estimator performance depends on the strength of the conditional correlation between the negative controls and the unobserved confounder. For the highest- and intermediate-high-correlation scenarios, the estimator exhibits low bias, the estimated asymptotic variances closely match the empirical variances, and the confidence intervals achieve coverage near the nominal 0.95 level. In the intermediate- and lowest-correlation scenarios, bias and coverage improve with increasing sample size, but variance ratios slightly exceed 1, indicating that the estimated asymptotic variance underestimates the empirical variance. For the intermediate-correlation scenario, bias is low and coverage approaches 0.95 for the largest sample size; for the lowest-correlation scenario, bias is modest but coverage remains below 0.80.

In this particular setting, where $Z$ directly affects the exposure and $W$ directly affects the outcome, the estimation task appears somewhat more challenging than in the setting reported in Section~\ref{sec:exp}, likely because the direct effects reduce the effectiveness of the negative controls as proxies for the unobserved confounder. Nevertheless, the expected pattern of decreasing bias and improving coverage with increasing sample size remains apparent.

\subsection{Vignette Application to a Single Data Set}\label{sec:vignette}

We provide R code and a mock dataset simulating data from the ENSEMBLE trial in the GitHub repository \href{https://github.com/aolivasm/proxi-mtp}{\texttt{github.com/aolivasm/proxi-mtp}} to demonstrate the use of our \texttt{pmtp} R function for implementing the proximal doubly-robust cross-fitted estimator.

The mock dataset \texttt{sim\_trial\_data.csv} includes the following variables:
\begin{itemize}
    \item Outcome: \texttt{Y} (taking values 0 or 1).
    \item Covariates: \texttt{L1, L2, L3}.
    \item Biomarker: univariate biomarker \texttt{A}.
    \item Negative control treatment: \texttt{Z}.
    \item Negative control outcome: \texttt{W}.
    \item Sampling weights: \texttt{wt}.
\end{itemize}

The function \texttt{pmtp()} estimates the counterfactual MTP mean $\mathbb{E}\left[Y\left\{q(A)\right\}|A\in\mc{S}\right]$ for a user-specified policy $q(a)$ over a population of interest $\mc{S}\subset\supp{(A)}$, using the proximal doubly-robust cross-fitted estimator described in Section \ref{sec:estimation}. Bridge functions are estimated within RKHSs generated by Gaussian kernels, following the procedures outlined in Sections~\ref{sec:rkhs_est}, \ref{sec:rkhs_wt}, and \ref{sec:cv}. The function accommodates both binary and continuous outcomes but does not handle time-to-event (censored) outcomes. \texttt{pmtp()} uses only complete cases. For data collected under a two-phase sampling design, users can set treatment values to \texttt{NA} for observations outside the reduced sample and provide the corresponding sampling weights for observations within the reduced sample; this is sufficient for proper estimation.

The \texttt{pmtp()} function requires the following inputs:
\begin{itemize}
    \item \texttt{data}: A \texttt{data.frame} with the dataset to analyze.
    \item \texttt{trt}: String for the biomarker (treatment) variable name.
    \item \texttt{outcome}: String for outcome variable name.
    \item \texttt{covariates}: String or vector of strings for covariate(s) name(s).
    \item \texttt{nct}: String or vector of strings for negative control treatment variable(s) name(s).
    \item \texttt{nco}: String or vector of strings for negative control outcome variable(s) name(s).
    \item \texttt{policy}: A list of policy functions. Currently, \texttt{pmtp()} supports only policies that depend on the treatment variable; policies that also depend on covariates are not supported.
\end{itemize}

The following inputs are optional. If not provided, default values will be used:
\begin{itemize}
    \item \texttt{weights}: String for weights variable name.
    \item \texttt{ind\_S}: A numeric vector of 0s and 1s indicating membership in the target population $\mathcal{S}$. Default is a vector of 1s (i.e., all units are in $\mathcal{S}$).
    \item \texttt{theta}: Scaling factor $c_0$ for the regularization parameter $c_0\cdot\frac{\log(n)}{n}$ in the risk function used during cross-validation to select the optimal bridge functions. Default is \texttt{1}.
    \item \texttt{K\_folds}: Integer indicating the number of folds used for cross-fitting. Must be at least 3. Default is \texttt{3}.
    \item \texttt{lm\_H\_list}, \texttt{lm\_G\_list}: Lists of scaling factors $c_1$ for the regularization parameters $c_1\left(\frac{\log(n)}{n}\right)^{1/2}$ in the outer minimization step of the outcome and treatment bridge function estimation, respectively. Defaults are \texttt{10\^{}seq(-5, -1, by = 1)}.
    \item \texttt{lm\_Gh\_list}, \texttt{lm\_Hg\_list}: Lists of scaling factors $c_2$ for the regularization parameters $c_2\frac{\log(n)}{n}$ in the inner maximization step of the outcome and treatment bridge function estimation, respectively. Defaults are \texttt{10\^{}seq(-1, 2, by = 1)}.
     \item \texttt{bw0\_H}, \texttt{bw0\_G}: Optional initial bandwidths for the function classes $\mc{H}$ and $\mc{H}'$, and $\mc{G}$ and $\mc{G}'$, respectively. Default is \texttt{NULL}. If not provided, the initial bandwidths are set to the median (or weighted median, if weights are provided) of the pairwise Euclidean distances between the corresponding observed vectors.
    \item \texttt{bw\_int\_scale\_list}, \texttt{bw\_ext\_scale\_list}: Lists of scaling factors for the bandwidths of the Gaussian kernels used in the internal and external steps of the bridge function estimation procedure. Defaults are \texttt{1/4} and \texttt{c(1/4,1/2,1,2,4)}, respectively.
    \item \texttt{bw\_int\_fixed\_scale}: Scale factor used for the internal bandwidth of the Gaussian kernel used in the risk function for cross-validation. Default is \texttt{1/4}.
    \item \texttt{inf.fun}: Logical flag indicating whether to return the influence function evaluated at the sample observations. Default is \texttt{FALSE}.
    \item \texttt{show.prog}: Logical flag indicating whether to display progress messages. Default is \texttt{FALSE}.
    \item \texttt{control.folds}: Logical flag indicating whether fold assignment in cross-fitting should be stratified based on the distribution of the treatment variable. Default is \texttt{FALSE}.
\end{itemize}

The function outputs a \texttt{data.frame} with the proximal doubly-robust cross-fitted estimator for each specified policy and also returns a \texttt{proximal.mtp} object with detailed information about the estimation procedure. In addition, the  \texttt{summary.proximal.mtp()} function reports, for each specified policy, the proportion of observations with treatment values within the image of the policy, along with the estimated standard errors and 95\% confidence intervals for the corresponding estimators. 

\textit{Input to conduct analysis on the mock dataset}

We illustrate our estimator for the following two policies:

\begin{verbatim}
policy_q1 <- function(x) {
  (x+0.4)*(x+0.4<=3.5-1)+(0.4*3.5+1*x)/(0.4+1)*(x+0.4>3.5-1)
}
policy_q2 <- function(x) {
  (x+0.8)*(x+0.8<=3.5-1)+(0.8*3.5+1*x)/(0.8+1)*(x+0.8>3.5-1)
}
\end{verbatim}

\begin{verbatim}
obj.pmtp = pmtp(data = sim_trial_data,
                trt = "A",
                outcome = "Y",
                covariates = c("L1", "L2", "L3"),
                nct = "Z",
                nco = "W", 
                weights = "wt",
                policy = list(policy_q1, policy_q2),
                control.folds = TRUE)
\end{verbatim}

\textit{Output using the} \texttt{summary.proximal.mtp()} \textit{function}

\begin{verbatim}
summary(obj.pmtp)

Summary of Proximal MTP Estimation
-------------------------------------------------- 
Number of observations: 1000
Number in target population: 1000 (100%)
-Proportion in the image of Policy_q^1: 38%
-Proportion in the image of Policy_q^2: 27%

-------------------------------------------------- 
Proximal Estimators and 95% Confidence Intervals:
-------------------------------------------------- 
Estimator: Proximal doubly-robust cross-fitted
           Estimate Std.Error CI.Lower CI.Upper
Policy_q^1   0.0139     7e-04   0.0126   0.0152
Policy_q^2   0.0101     7e-04   0.0088   0.0114
\end{verbatim}

\subsection{Additional Details on the Application to the ENSEMBLE Trial}
\label{sec:details_app}

This section provides supplementary descriptive information and additional analyses to complement Section~\ref{sec:appl}.

The ENSEMBLE trial enrolled 17,444 vaccine recipients across the six countries included in our evaluation. Of these, 1,164 participants had both day 29 binding and neutralizing antibody measurements and were therefore included in the analytic sample. Among the 1,164 participants, 40\% were enrolled at sites in Latin America (Argentina, Brazil, Colombia, or Peru), 41\% in South Africa, and the remaining 19\% in the United States. At baseline, the median age was 54 years (interquartile range [IQR] 41–64), and the median COVID-19 risk score was 0.49 (IQR 0.34–0.69). Between 7 and 115 days after day 29, there were 295 COVID-19 cases.

\begin{figure}
\centerline{ \includegraphics[width=6in]{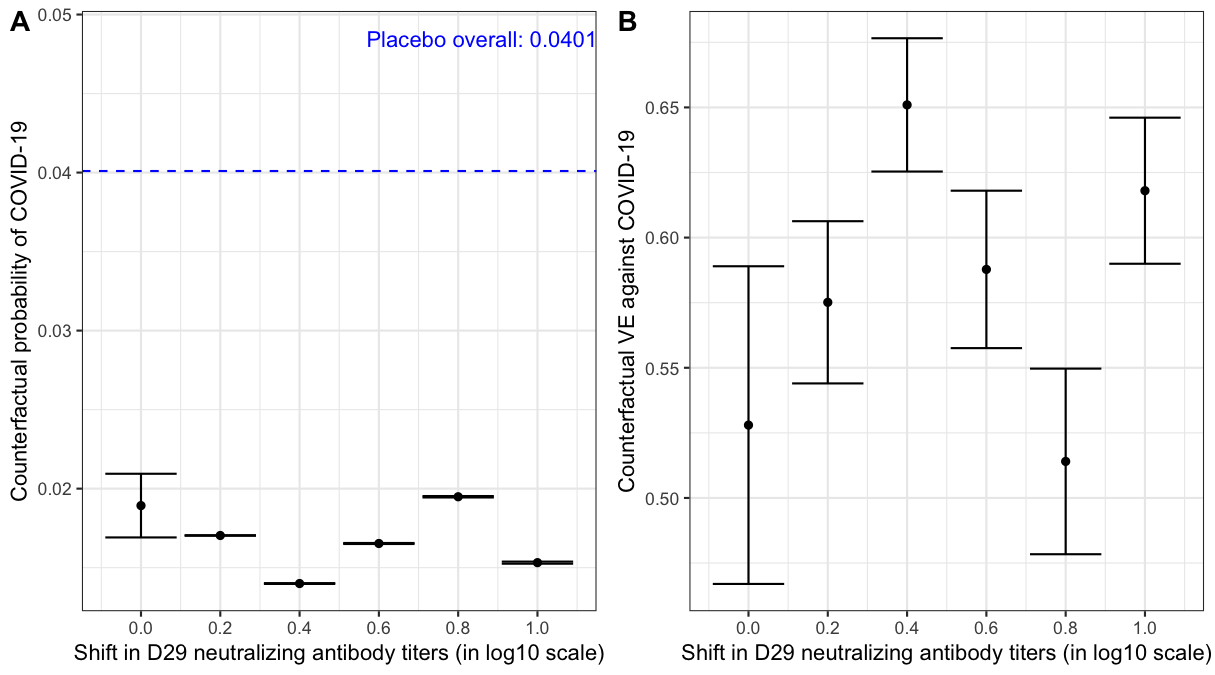}}
   \caption{Results from the application to the ENSEMBLE trial data using the non-proximal TMLE estimator. Panel A displays the observed and counterfactual probabilities of new COVID-19 diagnosis, and Panel B presents the corresponding observed and counterfactual vaccine efficacy for the $\varepsilon$--upper-tail tapered shift intervention policies with $\varepsilon=1$ and shifts $\delta=0$ (no shift), 0.2, 0.4, 0.6, 0.8, and 1 logarithm, as described in Example~\ref{ex2}. The bars indicate the $95\%$ confidence interval at each shift without any correction.}\label{fig:appl_tmle}
\end{figure}

Day 29 neutralizing antibody titers and binding antibody concentrations are summarized in the violin plots in Figure~\ref{fig:abs}. The median day 29 log\textsubscript{10} neutralizing antibody titer was 0.3889 in both cases and non-cases. The median day 29 log\textsubscript{10} binding antibody concentration was 1.4694 among cases and 1.5490 among non-cases. For the negative control outcome, 211 cases and 427 non-cases reported respiratory symptoms that tested negative for SARS-CoV-2; the mean burden of respiratory symptoms was 0.216 among cases and 0.128 among non-cases. Figure~\ref{fig:appl_tmle} presents the counterfactual probabilities of new COVID-19 diagnosis and the corresponding counterfactual vaccine efficacies under the 
$\varepsilon$-upper-tail–tapered shift interventions considered in Section~\ref{sec:appl}, estimated using the weighted version of the non-proximal TMLE estimator described in Section~\ref{sec:exp}. Importantly, in the implementation of the non-proximal TMLE estimator we did not adjust for the negative control treatment, as doing so could both alter the target estimand and induce collider bias. The resulting vaccine efficacy estimates are generally lower than those obtained using our proximal estimator and exhibit a non-monotone pattern across shift values.

\begin{figure}
\centerline{\includegraphics[width=6in]{FigS6_app_antibodies.png}}
   \caption{Day 29 antibody marker level by COVID-19 outcome status. Panel A shows violin plots of neutralizing antibody titers, and Panel B shows violin plots of binding anti-Spike antibody concentration}\label{fig:abs}
\end{figure}

\section{Proofs of Main Results}
\label{sec:proofs_theorems}

\renewcommand{\theequation}{D.\arabic{equation}}
\setcounter{equation}{0}

\begin{proof}[Theorem 1 in the main text]

\underline{Proof of part i)}. Take $\left(a,l,z\right)\in\text{supp}(A,L,Z)$ such that $a \in\mathcal{S}(l)\cup q(\mathcal{S}(l),l)$. Hence, if $h_{0}\left( a,l,w\right) $ solves the integral equation~\eqref{lat_out_eqn0}, we have
\begin{align*}
    \int h_{0}&\left( a,l,w\right)p_{W|A,L,Z}\left(
w|a,l,z\right) dw =\int  h_{0}\left( a,l,w\right) \left\{\int  p_{(W,U)|A,L,Z}\left(
w,u|a,l,z\right) du\right\}dw\\
&=\int \left\{\int h_{0}\left( a,l,w\right)  p_{W|A,L,U}\left(
w|a,l,u\right)dw \right\}p_{U|A,L,Z}\left(
u|a,l,z\right) du\quad\text{by Assumption~\ref{ass:nct0}(ii)}\\
&=\int \mathbb{E}\left(Y|A=a, L=l, U=u\right)p_{U|A,L,Z}\left(u|a,l,z\right) du\quad\quad\quad\quad\quad\quad\text{ $h_0$ solves~\eqref{lat_out_eqn0}}\\
&=\int  \mathbb{E}\left(Y|A=a, L=l, U=u, Z=z\right)p_{U|A,L,Z}\left(u|a,l,z\right) du\quad\quad\quad\text{ by Assumption~\ref{ass:nct0}(i)}\\
&= \mathbb{E}\left(Y|A=a, L=l, Z=z\right).
\end{align*}
Thus, $h_{0}\left( a,l,w\right) $ solves the integral equation ~\eqref{obs_out_eqn0}.

\underline{Proof of part ii)}. Take $\left(a,l,w\right)\in\text{supp}(A,L,W)$ satisfying $x \in\mathcal{S}(l)\cup q(\mathcal{S}(l),l)$, and such that, if $a\in q(\mc{S}(l),l)$, the map $x'\mapsto q^{-1}(a',l)$ is differentiable at $a'=a$. Hence, if $g_{0}\left( a,l,z\right) $ solves the integral
equation~\eqref{lat_tx_eqn0}, we have
    \begin{align*}
         \int g_0&\left( a,l,z\right)p_{Z|A,L,W}\left(z|a,l,w\right) dz= \int g_0\left( a, l, z\right) \left\{\int p_{(Z,U)|A,L,W}\left(z,u|a,l,w\right) du\right\}dz\\
        &=\int \left\{\int g_0\left( a,l,z\right)p_{Z|A,L,U}\left(z|a,l,u\right)dz\right\} p_{U|A,L,W}\left(u|a,l,w\right) du\quad\text{by Assumption~\ref{ass:nct0}(ii)}\\
        &\stackrel{(i)}{=}\int I\big\{ a\in q(\mathcal{S}(l),l) \big\}\frac{dq^{-1}(a,l)}{da}\cdot
\frac{p_{A|L,U}\left\{q^{-1}\left( a,l\right)
|l,u\right\} }{p_{A|L,U}\left( a|l,u\right) } p_{U|A,L,W}\left(u|a,l,w\right) du\\
&\stackrel{(\text{ii})}{=}I\big\{ a\in q(\mathcal{S}(l),l)\big\} \frac{dq^{-1}(a,l)}{da}
\int  \frac{p_{U|A,L,W}\left(u|a,l,w\right) }{p_{A|L,U,W}\left( a|l,u,w\right) } p_{A|L,U,W}\left\{q^{-1}\left( a,l\right)
|l,u,w\right\} du\\
&=I\big\{ a\in q(\mathcal{S}(l),l)\big\} \frac{dq^{-1}(a,l)}{da}
\frac{1}{p_{A|L,W}\left( a|l,w\right)}\int p_{(A,U)|L,W}\left\{q^{-1}\left( a,l\right),u
|l,w\right\} du\\
&=I\big\{ a\in q(\mathcal{S}(l),l) \big\} \frac{dq^{-1}(a,l)}{da}
\frac{p_{A|L,W}\left\{q^{-1}\left( a,l\right)
|l,w\right\} }{p_{A|L,W}\left( a|l,w\right) }\\
&=\alpha_0(a,l,w),
    \end{align*} 
where the equality in (i) follows because  $g_0$ solves~\eqref{lat_tx_eqn0} and the one in (ii) follows from Assumption~\ref{ass:nct0}(ii).
Thus, $g_{0}\left( a,l,z\right) $ solves equation~\eqref{obs_tx_eqn0}.
\end{proof}

\begin{proof}[Theorem~\ref{theo:identification0}]

\underline{Case 1}. Assumption \ref{ass:assumption7_0} holds and equation~\eqref{obs_tx_eqn0} has at least one solution. Let $h_0$ and $g^\dag$ be solutions to equations~\eqref{lat_out_eqn0} and~\eqref{obs_tx_eqn0}, respectively. Then,
\begin{align*}
    \psi_0&=\mathbb{E}\big[Y\left\{q(A,L)\right\}|(A,L)\in\mathcal{S}\big]\\
    &= \frac{\mathbb{E}\big[Y\left\{q(A,L)\right\}\cdot I\left\{(A,L)\in\mathcal{S}\right\}\big]}{P\left\{(A,L)\in\mathcal{S}
\right\} }\\
&= \frac{\mathbb{E}\left\{\mathbb{E}\big[Y\left\{q(A,L)\right\}|A,L,U\big]\cdot I\left\{(A,L)\in\mathcal{S}\right\}\right\}}{P\left\{ (A,L)\in\mathcal{S}
\right\} }\\
&= \frac{\mathbb{E}\left\{\mathbb{E}\big[h_0\left\{q(A,L),L,W\right\}|A,L,U\big]\cdot I\left\{(A,L)\in\mathcal{S}\right\}\right\}}{P\left\{ (A,L)\in\mathcal{S}
\right\} }\quad\quad \text{by Lemma~\ref{lemma:pos_cons}}\\
&= \frac{\mathbb{E}\big[h_0\left\{q(A,L),L,W\right\}\cdot I\left\{(A,L)\in\mathcal{S}\right\}\big]}{P\left\{(A,L)\in\mathcal{S}
\right\} }.
\end{align*}
By Theorem~1, $h_0$ also solves equation~\eqref{obs_out_eqn0}, then, by Lemma~\ref{lemma:equiv_ident},
\begin{align*}
   \frac{ \mathbb{E}\left\{Yg^\dag\left( A,L,Z\right)\right\}}{P\left\{(A,L)\in\mathcal{S}\right\}}
=\frac{\mathbb{E}\left[h_0\left\{ q\left( A,L\right) ,L,W\right\}\cdot I\left\{ (A,L)\in\mathcal{S} \right\}\right]}{P\left\{(A,L)\in\mathcal{S}\right\}}=\psi_0,
\end{align*}
which is the treatment bridge representation.

Now, let $h^\dag$ be any solution to equation~\eqref{obs_out_eqn0}. By Lemma~\ref{lemma:equiv_ident},
\begin{align*}
    \frac{\mathbb{E}\left[h^\dag\left\{ q\left( A,L\right) ,L,W\right\}\cdot I\left\{ (A,L)\in 
\mathcal{S} \right\}\right]}{P\left\{(A,L)\in\mathcal{S}\right\}}=\frac{ \mathbb{E}\left[Yg^\dag\left( A,L,Z\right)\right]}{P\left\{(A,L)\in\mathcal{S}\right\}}=\psi_0,
\end{align*}
which is the outcome bridge representation.

Lastly, for any $h\in\mathcal{L}^2(A,L,W)$ and any $g\in \mathcal{L}^2(A,L,Z)$, by Lemma \ref{lemma:equiv_ident},
\[
 \mathbb{E}\left\{\phi(O;h,g^\dag)\right\} = \mathbb{E}\left\{\phi(O;h^\dag,g^\dag)\right\}=\mathbb{E}\left[h^\dag\left\{ q\left( A,L\right) ,L,W\right\}\cdot I\left\{(A,L)\in\mathcal{S}\right\}\right],
\]
and
\[
\mathbb{E}\left\{\phi(O;h^\dag,g)\right\} = \mathbb{E}\left\{\phi(O;h^\dag,g^\dag)\right\}=\mathbb{E}\left\{ Yg^\dag\left( A,L,Z\right)\right\}.
\]
Hence
\[
\frac{\mathbb{E}\left\{\phi(O;h,g^\dag)\right\}}{P\left\{(A,L)\in\mathcal{S}\right\}}=\psi_0=\frac{\mathbb{E}\left\{\phi(O;h^\dag,g)\right\}}{P\left\{(A,L)\in\mathcal{S}\right\}},
\]
which is the double robust representation.

\underline{Case 2}. Assumption~\ref{ass:assumption8_0} holds and equation~\eqref{obs_out_eqn0} has at least one solution. Let $g_0$ and $h^\dag$ be solutions to equations~\eqref{lat_tx_eqn0} and~\eqref{obs_out_eqn0}, respectively. Then,
\begin{align*}
&\psi_0=\mathbb{E}\big[Y\left\{q(A,L)\right\}|(A,L)\in\mathcal{S}\big]\\
 &= \frac{\mathbb{E}\big\{\mathbb{E}\left[Y\left\{q(A,L)\right\}|A,L,U\right]\cdot I\left\{(A,L)\in\mathcal{S}\right\}\big\}}{P\left\{(A,L)\in\mathcal{S}
\right\} }\\
&=\iiint\frac{I\left\{(a,l)\in\mathcal{S}\right\}}{P\left\{(A,L)\in\mathcal{S}\right\} }\mathbb{E}\left\{Y|A=q(a,l),L=l,U=u\right\}p_{A,L,U}(a,l,u)dadldu\\
&=\iint \left[\int \frac{I\left\{(a,l)\in\mathcal{S}\right\}}{P\left\{(A,L)\in\mathcal{S}\right\} }\mathbb{E}\left\{Y|A=q(a,l),L=l,U=u\right\}p_{A|L,U}(a|l,u)da\right]p_{L,U}(l,u)dldu\\
&\stackrel{\text{(i)}}{=}\iiint\frac{I\left\{\tilde a \in q(\mathcal{S}(l),l)\right\}}{P\left\{(A,L)\in\mathcal{S}\right\} }\mathbb{E}\big(Y|A=\tilde a,L=l,U=u\big)\frac{dq^{-1}(\tilde a,l)}{d\tilde a}p_{A|L,U}\{q^{-1}(\tilde a, l)|l,u\}p_{L,U}(l,u)d\tilde adldu\\
&=\frac{1}{P\left\{(A,L)\in\mathcal{S}\right\}}\iiint\mathbb{E}\big[Y|A=\tilde a,L=l,U=u\big]\cdot\widetilde\alpha_0(\tilde a, l, u)\cdot p_{A,L,U}(\tilde a,l,u)d\tilde adldu\\
&\stackrel{\text{(ii)}}{=}\frac{1}{P\left\{(A,L)\in\mathcal{S}\right\}}\mathbb{E}\left[\mathbb{E}\big(Y|A,L,U\big)\cdot\mathbb{E}\big\{g_0(A,L,Z)|A,L,U\big\}\right]\quad\quad\quad\quad\quad\text{$g_0$ solves~\eqref{lat_tx_eqn0}}\\
&=\frac{\mathbb{E}\left\{Yg_0(A,L,Z)\right\}}{P\left\{(A,L)\in\mathcal{S}\right\}},
\end{align*}
where the equality in (i) follows from Assumption~\ref{ass:q0}, which justifies applying the change of variable $\tilde a = q(a,l)$ in $a$ for each $l$, and from the fact that $(a,l)\in\mathcal{S}$ if and only if $q(a,l)\in q(\mathcal{S}(l),l)$. For the equality in (ii), note that $\tilde a \in q(\mathcal{S}(l),l)$ implies $\tilde a = q(a,l)$ for some $a\in\mathcal{S}(l)$, and by the definitions of $\mathcal{S}$ and $\mathcal{S}(l)$, $\tilde a\in\supp{(A|L=l)}$. Then, for any $u\in\supp{(U|L)}$, Assumption~\ref{ass:pos_mtp0} implies $\tilde a\in\supp{(A|L=l, U=u)}$, and consequently $(\tilde a,l,u)\in\supp{(A,L,U)}$. Therefore, $(\tilde a,l,u)$ lies in the domain where $g_0$ solves equation~\eqref{lat_tx_eqn0}. By Theorem~1, $g_0$ also solves equation~\eqref{obs_tx_eqn0}. Thus, by Lemma~\ref{lemma:equiv_ident},
\begin{align*}
    \frac{\mathbb{E}\left[h^\dag\left\{ q\left( A,L\right) ,L,W\right\}\cdot I\left\{ (A,L)\in 
\mathcal{S} \right\}\right]}{P\left\{(A,L)\in\mathcal{S}\right\}}=\frac{ \mathbb{E}\left\{Yg_0\left( A,L,Z\right)\right\}}{P\left\{(A,L)\in\mathcal{S}\right\}}=\psi_0,
\end{align*}
which is the outcome bridge representation. From here, the other two representations are derived analogously to Case 1.
\end{proof}

\begin{proof}[Theorem \ref{theo:asymp_distr0}]  Let $\hat p:=\frac{1}{n}\sum_{i=1}^nI\left\{(A_i,L_i)\in\mc{S}\right\}$ and 
\[
U_n:=\frac{1}{\sqrt{n}}\sum_{k=1}^K\sum_{i:O_i\in \mc I_k}\left[\phi\left\{O_i;\hat h^{(-k)},\hat g^{(-k)}\right\}-\mathbb{E}\left\{\phi(O;h_0,g_0)\right\}\right].
\]
Noting that $\psi_0={p_0}^{-1}\mathbb{E}\left\{\phi(O;h_0,g_0)\right\}$, where $p_0:=P\left\{(A,L)\in\mathcal{S}\right\}$, we have
 \begin{align}
     \sqrt{n}\left(\hat\psi_{CF}^{DR}-\psi_0\right)&:=\sqrt{n}\left[\frac{1}{\hat p}\cdot\frac{1}{n}\sum_{k=1}^K\sum_{i:O_i\in \mc I_k}\phi\left\{O;\hat h^{(-k)},\hat g^{(-k)}\right\}-\psi_0\right] \nonumber\\
     &=\frac{1}{p_0}U_n+\left(\frac{1}{\hat p}-\frac{1}{p_0}\right)U_n+\sqrt{n}\left(\frac{1}{\hat p}-\frac{1}{p_0}\right)\mathbb{E}\left\{\phi(O;h_0,g_0)\right\}.
     \label{psi_hat_DR}
 \end{align}
By the Delta Method, letting $J_{\mc S}(O)=I\{(A,L)\in\mc S\}$, we have
\begin{equation}
   \sqrt{n}\: (\hat{p}^{-1}-p_0^{-1})=-p_0^{-2}\:\mathbb{G}_nJ_{\mc S},
   \label{eqn:p_hat}
\end{equation}
where for any $t(O)$, $\mathbb{G}_n (t) = \sqrt{n} \big[n^{-1} \sum_i^n t(O_i) - \mathbb{E}\{t(O)\}\big]$. Likewise, letting $n_k$ denote the cardinality of $\mc I_k$, $k=1,\dots,K$, we let $\mathbb{E}_{n_k}(t):= n_k^{-1}\sum_{i:O_i\in \mc I_k}t(O_i)$ denote the empirical mean of $t(O)$ over the sample $\mc I_k$, and $\mathbb{G}_{n_k}(t) := \sqrt{n_k}\left\{\mathbb{E}_{n_k}(t) - \int t(o) dP(o) \right\}$.
Then 
\begin{align}
    U_n&=\sum_{k=1}^K \sqrt{\frac{n_k}{n}}\:\mathbb{G}_{n_k}\left\{\phi(\cdot;\hat{h}^{(-k)},\hat{g}^{(-k)})-\phi(\cdot;h_0,g_0)\right\}+\sum_{k=1}^K \sqrt{\frac{n_k}{n}}\:\mathbb{G}_{n_k}\left\{\phi(\cdot;h_0,g_0)\right\} \nonumber
    \\
    &\quad +\sqrt{n}\sum_{k=1}^K \frac{n_k}{n}\int \left\{\phi(o;\hat{h}^{(-k)},\hat{g}^{(-k)})-\phi(o;h_0,g_0)\right\}d P(o).
    \label{U_n-Decom}
\end{align}

Now, let $\mc I^{(-k)}:=\left\{O_i:1\le i \le n, O_i\notin \mc I_k\right\}$ and let $m_k$ denote the cardinality of $\mc I^{(-k)}$. For any $\epsilon>0$, we have
\begin{align*}
D_{m_k}:=&P\Big[\Big\vert\mathbb{G}_{n_k}\left\{\phi(\cdot;\hat{h}^{(-k)},\hat{g}^{(-k)})-\phi(\cdot;h_0,g_0)\right\}\Big\vert\ge\epsilon\Big\vert \mc I^{(-k)}\Big]
\\
&\le  \mathbb{E}\Big[\Big\{ \mathbb{G}_{n_k}\left\{\phi(\cdot;\hat{h}^{(-k)},\hat{g}^{(-k)})-\phi(\cdot;h_0,g_0)\right\}\Big\}^2\Big\vert \mc I^{(-k)}\Big]/\epsilon^2
\\
&\le  \mathbb{E}\Big[ \left\{\phi(O;\hat{h}^{(-k)},\hat{g}^{(-k)})-\phi(O;h_0,g_0)\right\}^2\Big\vert \mc I^{(-k)}\Big]/\epsilon^2
\\
&=  \int \left\{\phi(o;\hat{h}^{(-k)},\hat{g}^{(-k)})-\phi(o;h_{0},g_{0})\right\}^2 dP(o)
\\
&=o_P(1),
\end{align*}
where the first inequality is by Markov's inequality, the second is because given $\mc I^{(-k)}$, \\$\mathbb{G}_{n_k}^{(k)}\big\{\phi(\cdot;\hat{h}^{(-k)},\hat{g}^{(-k)})-\phi(\cdot;h_{0},g_{0})\big\}$ is a sum of $n_k$ independent and identically distributed random variables, the second to last equality is by independence of $\mc I^{(-k)}$ and $\mc I_k$ and the last equality follows from Lemma~\ref{lemma:consist_phi}. We then conclude that $\mathbb{E}(D_{m_k})$ converges to 0 as $m_k \rightarrow \infty$ because $D_{m_k}$, $m_k=1,2,...$ is a sequence of positive random variables bounded above by 1 that converges in probability to 0. But
\begin{align*}
\mathbb{E}(D_{m_k}) &= \mathbb{E}\left\{P\Big[\Big\vert\mathbb{G}_{n_k}\left\{\phi(\cdot;\hat{h}^{(-k)},\hat{g}^{(-k)})-\phi(\cdot;h_0,g_0)\right\}\Big\vert\ge\epsilon\Big\vert \mc I^{(-k)}\Big]\right\}
\\
&=P\Big[\Big\vert\mathbb{G}_{n_k}\left\{\phi(\cdot;\hat{h}^{(-k)},\hat{g}^{(-k)})-\phi(\cdot;h_0,g_0)\right\}\Big\vert\ge\epsilon\Big].
\end{align*}
Then, assuming as we do, that $n_k/n \rightarrow 1/K$ as $n \rightarrow \infty$ we have that $m_k \rightarrow \infty$ as $n \rightarrow \infty$, and therefore $\sum_{k=1}^K \sqrt{\frac{n_k}{n}}\:\mathbb{G}_{n_k}\left\{\phi(\cdot;\hat{h}^{(-k)},\hat{g}^{(-k)})-\phi(\cdot;h_0,g_0)\right\}=o_p(1)$ as $ n \rightarrow \infty$.

Next, by Lemma~\ref{lemma:equiv_ident}, for each $k$, it holds that
\begin{align*}
    \int &\left\{\phi(o;\hat{h}^{(-k)},\hat{g}^{(-k)})-\phi(o;h_{0},g_{0})\right\}d P(o)\\
    &\quad=\int\{\hat h^{(-k)}(a,l,w)-h_0(a,l,w)\}\{g_0(a,l,z)-\hat g^{(-k)}(a,l,z)\}dP(a,l,w,z).
\end{align*}
Thus, 
$\sum_{k=1}^K \frac{n_k}{n}\int \left\{\phi(o;\hat{h}^{(-k)},\hat{g}^{(-k)})-\phi(o;h_{0},g_{0})\right\}d P(o) = p_0R_n$ where $R_n$ is defined in the statement of the theorem. From (\ref{U_n-Decom}) we thus conclude that 
\begin{equation}
    U_n=\mathbb{G}_{n}\left\{\phi(\cdot;h_0,g_0)\right\} + \sqrt{n} \: p_0 \: R_n + o_p(1)
\label{U_n_final_decom}
\end{equation}
By the Cauchy-Schwarz inequality and the norm consistency assumption on $h^{(-k)}$ and $g^{(-k)}$, we have
\[
\vert R_n\vert\le p_0^{-1} \sum_{k=1}^K \frac{n_k}{n} \Vert \hat{h}^{(-k)}-h_0\Vert_2 \: \Vert \hat{g}^{(-k)}-g_0\Vert_2 =o_p(1).
\]
Because $\mathbb{G}_{n}\left\{\phi(\cdot;h_0,g_0)\right\}=O_p(1)$ and~\eqref{eqn:p_hat} implies that $\hat{p}^{-1}-p_0^{-1}=O_p(n^{-1/2})$, it follows that
\begin{equation}
    (\hat{p}^{-1}-p_0^{-1})\:U_n=o_p(1).
    \label{p_hat times U_n}
\end{equation}

Replacing the equalities~\eqref{eqn:p_hat}, \eqref{U_n_final_decom}, and \eqref{p_hat times U_n} into~\eqref{psi_hat_DR}, we arrive at 

\begin{align*}
     \sqrt{n}\left(\psi_{CF}^{DR}-\psi_0\right)&=p_0^{-1}\mathbb{G}_{n}\left\{\phi(\cdot;h_0,g_0)\right\} + \sqrt{n} \: R_n 
     \\
     &\quad -p_0^{-2}\:\mathbb{G}_nJ_{\mc S}\,\mathbb{E}\left\{\phi(O;h_0,g_0)\right\}+o_p(1)
     \\
     &= p_0^{-1} \mathbb{G}_n \Big\{\phi(\cdot;h_0,g_0)-J_{\mc S}\:\psi_0\Big\}+ \sqrt{n} \: R_n+o_p(1).
 \end{align*}
This concludes the proof of the Theorem.
\end{proof}

\section{Supporting Lemmas}

\renewcommand{\theequation}{E.\arabic{equation}}
\setcounter{equation}{0}
\renewcommand{\thelemma}{E.\arabic{lemma}}
\setcounter{lemma}{0}

In this section, we present lemmas that are used in the proofs of Theorems~\ref{theo:identification0} and~\ref{theo:asymp_distr0}, presented in Appendix~\ref{sec:app_mtp_identification}.

Lemma \ref{lemma:change} below implies that $\alpha_0(A,L,W)$ is the Riesz representer of the linear functional given by $h\mapsto\mathbb{E}\left[ h\left\{ q\left( A,L\right) ,L,W\right\}\cdot I\left\{ (A,L)\in\mathcal{S}\right\}\right]$. This Lemma is invoked in the proof of the subsequent Lemma~\ref{lemma:equiv_ident}.

\begin{lemma} Under Assumption \ref{ass:q0}, for any $h\in \mathcal{L}^2(A,L,W)$, it holds that
\[
\mathbb{E}\left[h\left\{ q\left( A,L\right) ,L,W\right\}\cdot I\left\{ (A,L)\in\mathcal{S}\right\}\right]=E\left\{h(A,L,W)\alpha_0(A,L,W)\right\}.
\]\label{lemma:change}
\end{lemma}

\begin{proof}
   For any $h\in\mathcal{L}^2(A,L,W)$, we have
    \begin{align*}
    \iiint& h(a,l,w)\alpha_0(a,l,w) p_{A,L,W}(a,l,w) dadldw\\
        &=\iiint h(a,l,w)I\big\{ a\in q\left(\mathcal{S}(l) ,l\right) \big\} \frac{dq^{-1}\left( a,l\right)}{da}\frac{p_{A|L,W}\left\{q^{-1}\left( a,l\right)
|l,w\right\} }{p_{A|L,W}\left( a|l,w\right) }p_{A,L,W}(a,l,w)dadldw\\
 &=\iint\left[ h(a,l,w)I\big\{ a\in q\left(\mathcal{S}(l) ,l\right) \big\} \frac{dq^{-1}\left( a,l\right)}{da}p_{A|L,W}\left\{q^{-1}\left( a,l\right)
|l,w\right\}da\right]p_{L,W}(l,w)dldw\\
&\stackrel{(\text{i})}{=}\iiint h\left\{q(\tilde a,l),l,w\right\}I\left\{(\tilde a,l)\in\mathcal{S}\right\}\frac{dq^{-1}\left\{ q(\tilde a,l),l\right\}}{da}\cdot\frac{dq\left( \tilde a,l\right)}{d\tilde a}p_{A|L,W}(\tilde a|l,w){p_{L,W}(l,w)}d\tilde adldw\\
&\stackrel{(\text{ii})}{=}\iiint h\left\{q(\tilde a,l),l,w\right\}I\left\{(\tilde a,l)\in\mathcal{S}\right\}p_{A,L,W}(\tilde a,l,w)d\tilde adldw,
    \end{align*}
    where the equality in (i) follows from Assumption \ref{ass:q0}, which allows us to apply the change of variable $\tilde a = q^{-1}(a,l)$ in $a$ for each $l\in\supp{(L)}$, and from the fact that $a \in q(\mathcal{S}(l),l)$ if and only if $(q^{-1}(a,l),l)\in\mathcal{S}$. For the equality in (ii), we used that for each $l\in\supp{(L)}$,
    \[
    1=\frac{d\tilde a}{d\tilde a}=\frac{d\left[q^{-1}\left\{q(\tilde a,l),l\right\}\right]}{d\tilde a}=\frac{dq^{-1}\left\{ q(\tilde a,l),l\right\}}{da}\cdot\frac{dq\left( \tilde a,l\right)}{d\tilde a}.
    \]
    This concludes the proof.
\end{proof}

The following Lemma is invoked in the proof of Theorems~\ref{theo:identification0} and~\ref{theo:asymp_distr0}.

\begin{lemma} Let $h^\dag$ and $g^\dag$ be observed outcome and treatment bridge functions respectively. Under Assumption \ref{ass:q0}, for any  $h\in \mathcal{L}^2(A,L,W)$ and $g\in \mathcal{L}^2(A,L,Z)$, it holds that
 \begin{align*}
    \mathbb{E}\left\{\phi(O;h,g)-\phi(O;h^\dag,g^\dag)\right\}=\mathbb{E}\left[\left\{(h-h^\dag)(A,L,W)\right\}\left\{(g^\dag-g)(A,L,Z)\right\}\right].
 \end{align*}
 Furthermore,
\begin{align*}
    \mathbb{E}\left[h^\dag\left\{ q\left( A,L\right) ,L,W\right\}\cdot I\left\{ (A,L)\in 
\mathcal{S} \right\}\right]=\mathbb{E}\left\{\phi(O;h^\dag,g^\dag)\right\}=\mathbb{E}\left\{Yg^\dag\left( A,L,Z\right)\right\}.
\end{align*}

\label{lemma:equiv_ident}
\end{lemma}

\begin{proof}
Let $h^\dag$ and $g^\dag$ be observed outcome and treatment bridge functions respectively, and take any $h\in \mathcal{L}^2(A,L,W)$ and $g\in\mathcal{L}^2(A,L,Z)$. Then,
\begin{align*}
     \mathbb{E}\big\{\phi&(O;h,g)-\phi(O;h^\dag,g^\dag)\big\}\\
     =&\mathbb{E}\left[(h-h^\dag)\left\{q(A,L),L,W\right\}\cdot I\left\{(A,L)\in\mathcal{S}\right\}\right]+\mathbb{E}\left[g(A,L,Z)\left\{Y-h(A,L,W)\right\}\right]\\
     &-\mathbb{E}\Big[g^\dag(A,L,Z)\underbrace{\mathbb{E}\left\{Y-h^\dag(A,L,W)|A,L,Z\right\}}_{=0\text{ since $h^\dag$ is an outcome br fcn}}\Big]\\
     =&\mathbb{E}\left\{(h-h^\dag)(A,L,W)\alpha_0(A,L,W)\right\}+\mathbb{E}\big\{g(A,L,Z)\cdot\mathbb{E}\left(Y|A,L,Z\right)\big\}\quad\quad\text{by Lemma \ref{lemma:change}}\\
     &-\mathbb{E}\big\{g(A,L,Z)\cdot h(A,L,W)\big\}\\
      =&\mathbb{E}\Big[(h-h^\dag)(A,L,W)\underbrace{\mathbb{E}\left\{g^\dag(A,L,Z)|A,L,W\right\}}_{g^\dag \:\text{is a trx br fcn}}\Big]+\mathbb{E}\Big[g(A,L,Z)\underbrace{\mathbb{E}\left\{h^\dag(A,L,W)|A,L,Z\right\}}_{h^\dag\text{ is an outcome br fcn}}\Big]\\
      &-\mathbb{E}\left[g(A,L,Z)\cdot h(A,L,W)\right]\\
     =&\mathbb{E}\left[\left\{(h-h^\dag)(A,L,W)\right\}\left\{(g^\dag-g)(A,L,Z)\right\}\right].
\end{align*}
Taking $h\equiv h^\dag$ and $g\equiv0$ gives
\[
\mathbb{E}\left[h^\dag\left\{ q\left( A,L\right) ,L,W\right\}\cdot I\left\{ (A,L)\in 
\mathcal{S}\right\}\right]=\mathbb{E}\left\{\phi(O;h^\dag,0)\right\}= \mathbb{E}\left\{\phi(O;h^\dag,g^\dag)\right\},
\]
and taking $h\equiv0$ and $g\equiv g^\dag$ gives
\[
\mathbb{E}\left\{Yg^\dag(A,L,Z)\right\}=\mathbb{E}\left\{\phi(O;0,g^\dag)\right\}= \mathbb{E}\left\{\phi(O;h^\dag,g^\dag)\right\}.
\]
This concludes the proof.
\end{proof}

The following Lemma is invoked in the proof of Theorem~\ref{theo:identification0}.

\begin{lemma} Under Assumptions~\ref{ass:consist0}--\ref{ass:pos_mtp0}, for any $(a,l,u)\in\text{supp}(A,L,U)$ such that $a\in\mathcal{S}(l)$, it holds that
\[
\mathbb{E}\left[ Y\left\{ q\left( A,L\right) \right\} |A=a,L=l,U=u\right] =\mathbb{E}\left\{ Y|A=q\left( a,l\right), L=l, U=u\right\}.
\]
Moreover, if Assumption~\ref{ass:nct0}(ii) holds, then for any latent outcome bridge function $h_0$ it holds that
\[
\mathbb{E}\left[ Y\left\{ q\left( A,L\right) \right\} |A=a, L=l, U=u\right] =\mathbb{E}\left[h_0\left\{q(A,L), L, W\right\}|A=a, L=l, U=u\right].
\]
\label{lemma:pos_cons}
\end{lemma}

\begin{proof}

Take $(a,l,u)\in\supp{(A,L,U)}$ such that $a\in\mathcal{S}(l)$. By the definitions of $\mathcal{S}$ and $\mathcal{S}(l)$, $q(a,l)\in\supp{(A|L=l)}$. Then, by Assumption \ref{ass:pos_mtp0}, $q(a,l)\in\supp{(A|L=l,U=u)}$, which implies that $(q(a,l),l,u)\in\supp{(A,L,U)}$. Then,
\begin{align*}
    \mathbb{E}\left[ Y\left\{ q\left( A,L\right) \right\} |A=a,L=l,U=u\right]&= \mathbb{E}\left[ Y\left\{ q\left( a,l\right) \right\} |A=a,L=l,U=u\right]\\
&=  \mathbb{E}\left[ Y\left\{ q\left( a,l\right) \right\} |A=q(a,l),L=l,U=u\right] \\
&= \mathbb{E}\left\{ Y |A=q(a,l),L=l,U=u\right\},
\end{align*}
where the second equality follows from Assumption~\ref{ass:rand0} and the last one from Assumption~\ref{ass:consist0}.

Now, let $h_0$ be a solution to equation~\eqref{lat_out_eqn0}. Since $(q(a,l),l,u)\in\supp{(A,L,U)}$ and $q(a,l)\in q(\mathcal{S}(l),l)$, $(q(a,l),l,u)$ lies in the domain where $h_0$ solves equation~\eqref{lat_out_eqn0}. Then,
\begin{align*}
 \mathbb{E}\left[ Y |A=q(a,l),L=l,U=u\right]&=\int h_0\{q(a,l),l,w\}p_{W|A,L,U}\{w|q(a,l),l,u\}dw\\
    &=\int h_0\{q(a,l),l,w\}p_{W|A,L,U}(w|a,l,u)dw\quad\quad\text{by Assumption~\ref{ass:nct0}(ii)}\\
&=\mathbb{E}\big[h_0\{q(A,L),L,W\}|A=a,L=l,U=u\big].
\end{align*}
This concludes the proof.
\end{proof}

Lemma~\ref{lemma:consist_phi} below is invoked in the proof of Theorem~\ref{theo:asymp_distr0}.

\begin{lemma} Under the assumptions of Theorem~\ref{theo:asymp_distr0}, for each $k$, it holds that
\[
\big\Vert\phi(O;\hat h^{(-k)},\hat g^{(-k)})-\phi(O;h_0,g_0)\big\Vert_2=o_p(1).
\]\label{lemma:consist_phi}
\end{lemma}
\begin{proof}
    From the definition of $\phi$, we have
\begin{align*}
    \phi(O;\hat h,\hat g)-\phi(O;h_0,g_0)=&  \big(\hat h-h_0\big)\left\{q(A,L),L,W\right\}\cdot I\left\{(A,L)\in\mathcal{S}\right\}+Y\left(\hat g-g_0\right)(A,L,Z)\\
    &- \hat h(A,L,W)\hat g(A,L,Z)+h_0(A,L,W)g_0(A,L,Z),
\end{align*}
and by the triangle inequality
\begin{align*}
    \sqrt{\mathbb{E}\left[\left\{ \phi(O;\hat h,\hat g)-\phi(O;h_0,g_0)\right\}^2\right]}\le& \sqrt{\mathbb{E}\left[\left\{ \left(\hat h-h_0\right)\left\{q(A,L),L,W\right\}\cdot I\left\{(A,L)\in\mathcal{S}\right\}\right\}^2\right]}\\
    &+\sqrt{\mathbb{E}\left[\left\{Y\left(\hat g-g_0\right)(A,L,Z)\right\}^2\right]}
    \\
    &+\sqrt{\mathbb{E}\left[\left\{\hat h(A,L,W)\hat g(A,L,Z)- h_0(A,L,W) g_0(A,L,Z)\right\}^2\right]}
\end{align*}
As shown in Appendix A of \cite{olivas2025source}, in their verification of a particular assumption for their Example 2, the bound $|\alpha_0(A,L,W)|\le B$ implies that
\begin{align*}
\sqrt{\mathbb{E}\left[\left\{ \left(\hat h-h_0\right)\left\{q(A,L),L,W\right\}\cdot I\left\{(A,L)\in\mathcal{S}\right\}\right\}^2\right]}\le \sqrt{B} \big\Vert \hat h-h_0 \big\Vert_2 = O(\big\Vert \hat h-h_0 \big\Vert_2 ).
\end{align*}
Also, since $|Y|\le B$, it follows that
\[
\sqrt{\mathbb{E}\left[\left\{Y\left(\hat g-g_0\right)(A,L,Z)\right\}^2\right]} \le B \left\Vert\hat g-g_0\right\Vert_2 =O(\left\Vert\hat g-g_0\right\Vert_2).
\]
Furthermore, the term
\[
\sqrt{\mathbb{E}\left[\left\{\hat h(A,L,W)\hat g(A,L,Z)- h_0(A,L,W) g_0(A,L,Z)\right\}^2\right]}=\big\Vert \hat h\cdot\hat g-h_0\cdot g_0\big\Vert_2
\]
is upper bounded by
\begin{align*}
    \min&\Big\{\big\Vert \hat h(\hat g-g_0)\big\Vert_2+\big\Vert g_0(\hat h-h_0)\big\Vert_2, \big\Vert \hat g(\hat h-h_0)\big\Vert_2+\big\Vert h_0(\hat g-g_0)\big\Vert_2\Big\}\\
    &\le\min\Big\{\Vert\hat h\Vert_\infty\big\Vert \hat g-g_0\big\Vert_2+\Vert g_0\Vert_\infty\big\Vert \hat h-h_0\big\Vert_2,\Vert\hat g\Vert_\infty\big\Vert \hat h-h_0\big\Vert_2+\Vert h_0\Vert_\infty\big\Vert \hat g-g_0\big\Vert_2\Big\}
    \\
    &=O\left(\big\Vert \hat h-h_0\big\Vert_2+\big\Vert \hat g-g_0\big\Vert_2\right)
\end{align*}
where the last equality follows because by assumption $\Vert \hat h\Vert_\infty+\Vert g_0\Vert_\infty\le B \quad \text{or} \quad\Vert \hat g\Vert_\infty+\Vert h_0\Vert_\infty\le B$.

Thus, we conclude that 
\begin{align*}
    \sqrt{\mathbb{E}\left[\left\{ \phi(O;\hat h,\hat g)-\phi(O;h_0,g_0)\right\}^2\right]}
    =O\left(\big\Vert \hat h-h_0\big\Vert_2+\big\Vert \hat g-g_0\big\Vert_2\right)
    \end{align*}
which is $o_p(1)$ since $\Vert \hat h-h_0\Vert_2^2=o_p(1)$, $\Vert \hat g-g_0\Vert_2^2=o_p(1)$.    
\end{proof}

\section{Bridge Functions for the Data-Generating Process and Policy Introduced in Appendix~\ref{sec:dgp}}
\label{sec:webapp_G}

\renewcommand{\theequation}{F.\arabic{equation}}
\setcounter{equation}{0}

\renewcommand{\thefigure}{F.\arabic{figure}}
\setcounter{figure}{0}

\subsection{Derivation of an Approximate Outcome Bridge Function}\label{sec:deriv_out_br}

The data-generating process introduced in Appendix~\ref{sec:dgp}, generates the outcome with conditional probability
\[
\mathbb{E}\left(Y\mid A,L,U,W\right)=\left[1+\exp\left\{-(\mu+\beta_9L+\beta_{10}U+\beta_{11}A+\beta_{12}W+\gamma A^2)\right\}\right]^{-1}.
\]
This motivates the following parametric model for a latent outcome bridge function:
\[
h(a,l,w;\varphi)=\kappa(\varphi)\left[1+\exp\left\{-(\varphi_0+\varphi_1a+\varphi_2l+\varphi_3w+\varphi_4a^2)\right\}\right]^{-1},
\]
where $\varphi=(\varphi_0,\varphi_1,\varphi_2,\varphi_3,\varphi_4)^\top$.

For $h(a,l,w;\varphi)$ to act as a latent outcome bridge function, it must satisfy equation~\eqref{lat_out_eqn0}. In this section, we show that when $\beta_{10}/\beta_5$ and $\beta_{12}$ are small, a closed-form expression for $\varphi$ can be obtained as a function of the data-generating parameters $(\beta_1,\dots,\beta_{12},\mu,\gamma)$, so that
\[
\mathbb{E}\left(Y\mid A,L,U\right)\approx\mathbb{E}\left[h(A,L,U;\varphi)\mid A,L,U\right].
\]

Let $f(v,t):=v\left\{1+\exp(-t)\right\}^{-1}$ so that
\[
h(a,l,w;\varphi)=f\big(\kappa(\varphi),\varphi_0+\varphi_1a+\varphi_2l+\varphi_3w+\varphi_4a^2\big)
\]
and
\[
\mathbb{E}(Y\mid A=a,L=l,U=u,W=w)=f(1,\mu+\beta_9l+\beta_{10}u+\beta_{11}a+\beta_{12}w+\gamma a^2).
\]
 To approximate $\mathbb{E}[h(A,L,W;\varphi)|A,L,U]$ and $\mathbb{E}(Y| A,L,U)=\mathbb{E}[\mathbb{E}(Y| A,L,U,W)| A,L,U]$, we apply a second-order Taylor approximation of $f$ around $(v_0,t_0)$:
\begin{align*}
    f(v,t)&\approx f(v_0,t_0)+(v-v_0)\frac{\partial f (v_0,t_0)}{\partial v
    }+(t-t_0)\frac{\partial f(v_0,t_0)}{\partial t}+\frac{(t-t_0)^2}{2}\frac{\partial^2 f(v_0,t_0)}{\partial t^2}\\&\quad+(v-v_0)(t-t_0)\frac{\partial^2 f(v_0,t_0)}{\partial v\partial t}\\
    &= f(v_0,t_0)+(v-v_0)\frac{f(v_0,t_0)}{v_0}+(t-t_0)f(v_0,t_0)\left\{1-\frac{f(v_0,t_0)}{v_0}\right\}\\
    &\quad+\frac{(t-t_0)^2}{2}f(v_0,t_0)\left\{1-\frac{f(v_0,t_0)}{v_0}\right\}\left\{1-\frac{2f(v_0,t_0)}{v_0}\right\}\\
    &\quad+(v-v_0)(t-t_0)\frac{f(v_0,t_0)}{v_0}\left\{1-\frac{f(v_0,t_0)}{v_0}\right\}.
\end{align*}

Define $\widetilde{W}:=W-\beta_4 L - \beta_5 U $, so that $\widetilde{W}\sim\mathcal{TN}_c(0,1)$ and is independent of $(A,L,U)$. Then
\[
\varphi_0+\varphi_1A+\varphi_2L+\varphi_3W+\varphi_4A^2
= T_0 + \varphi_3 \widetilde{W},
\]
where  
$T_0 = \varphi_0 + \varphi_1A + (\varphi_2+\varphi_3\beta_4)L + \varphi_3 \beta_5 U + \varphi_4A^2$.

Applying the Taylor approximation with $(v_0,t_0)=(1,T_0)$ and taking conditional expectations gives
\begin{align*}
\mathbb{E}&\left\{h(A,L,W;\varphi)|A,L,U\right\}\\
&\quad\quad=\mathbb{E}\left\{f\big(\kappa(\varphi),T_0+\varphi_3\widetilde{W}\big)|A,L,U\right\}\\
&\quad\quad\approx f(1,T_0)+\left\{\kappa(\varphi)-1\right\}f(1,T_0)+\frac{\varphi_3^2}{2}f(1,T_0)\{1-f(1,T_0)\}\{1-2f(1,T_0)\}\mathbb{E}[\widetilde{W}^2]\\
&\quad\quad\approx \left\{\kappa(\varphi)+\frac{\varphi_3^2}{2}\mathbb{E}[\widetilde{W}^2]\right\}f(1,T_0)\\
&\quad\quad=\left\{\kappa(\varphi)+\frac{\varphi_3^2}{2}\mathbb{E}[\widetilde{W}^2]\right\}\left[1+\exp\left\{-(\varphi_0+\varphi_1A+(\varphi_2+\varphi_3\beta_4)L+\varphi_3\beta_5U+\varphi_4A^2)\right\}\right]^{-1}.
\end{align*}

A parallel derivation produces the corresponding approximation for $\mathbb{E}(Y\mid A,L,U)$:
\[
\left(1 + \frac{\beta_{12}^2}{2}\mathbb{E}[\widetilde{W}^2]\right)
\!\left[1+\exp\!\left\{-(\mu+\beta_{11}A+(\beta_9+\beta_{12}\beta_4)L+(\beta_{10}+\beta_{12}\beta_5)U+\gamma A^2)\right\}\right]^{-1}.
\]

Equating these expressions suggests the approximations $\varphi\approx\big(\mu,\, \beta_{11},\,\beta_{9}-\frac{\beta_4\beta_{10}}{\beta_5},\, \beta_{12}+\frac{\beta_{10}}{\beta_5},\,\gamma\big)$ and $\kappa(\varphi)\approx1+(\beta_{12}^2-\varphi_3^2)\mathbb{E}[\widetilde{W}^2]/2$.

When $\mathbb{E}[\widetilde{W}^2]$ is known, this yields an explicit approximation for $\kappa(\varphi)$. However, this expression may be negative—particularly when $\beta_5$ is small and $\beta_{10}/\beta_5$ has the same sign as $\beta_{12}$—because $\varphi_3\approx\beta_{12}+\beta_{10}/\beta_5$. A negative value of $\kappa(\varphi)$ can produce unstable estimators.

To avoid this issue, we instead set $\kappa(\varphi)=1+\frac{\varphi_3^2}{2}\mathbb{E}[\widetilde{W}^2]$. When $\beta_{10}/\beta_5$ and $\beta_{12}$ are small, taking
\[
\varphi=\Big(\mu+\log\Big\{1+\frac{\beta_{12}^2}{2}\mathbb{E}[\widetilde{W}^2]\Big\}-\log\Big\{1+\Big(\beta_{12}+\frac{\beta_{10}}{\beta_5}\Big)^2\mathbb{E}[\widetilde{W}^2]\Big\},\, \beta_{11},\,\beta_{9}-\frac{\beta_4\beta_{10}}{\beta_5},\, \beta_{12}+\frac{\beta_{10}}{\beta_5},\,\gamma\Big)^\top
\]
gives a function $h(a,l,w;\varphi)$ that approximately solves equation (\ref{lat_out_eqn0}).

While the derivation is strictly valid when 
$\beta_{12}$ and $\beta_{10}/\beta_5$ are small, our numerical experiments indicate that the proposed parametric form for 
$h$ also performs well in more challenging settings (e.g., $\beta_8=0.3$ and $\beta_{12}=-0.3$; see Figures~\ref{fig:num_exp_param_bp_case2} and~\ref{fig:num_exp_param_cov_case2}).

\subsection{Derivation of a Treatment Bridge Function}\label{sec:deriv_tr_br}

Under the data-generating process and policy~\eqref{q_general} introduced in Appendix~\ref{sec:dgp}, when $\beta_8=0$, the function $\widetilde\alpha_0(a,l,u)$ defined in Assumption~\ref{ass:assumption8_0} can be written as
\begin{align*}
    \widetilde\alpha_0(a,l,u)&=I\left\{a\in q\big(\mc{S}(l)\big)\right\}\frac{d(q^{\delta,\varepsilon,r})^{-1}(a,l)}{da}\frac{p_{A|L,U}\left((q^{\delta,\varepsilon,r})^{-1}(a,l)|l,u\right)}{p_{A|L,U}(a|l,u)}\\
    &=\rho_0(a)\exp\left[-\frac{1}{2}\left\{(q^{\delta,\varepsilon,r})^{-1}(a)-a\right\}\left\{(q^{\delta,\varepsilon,r})^{-1}(a)+a-2\left(\beta_6l+\beta_7u\right)\right\}\right]\\
     &=\rho_0(a)\exp\left[\frac{\delta V(a)}{2}\left\{2a-\delta V(a)-2\left(\beta_6l+\beta_7u\right)\right\}\right]\\
      &=\rho_0(a)\exp\left[\delta V(a)\left\{a-\beta_6l-\beta_7u-\frac{\delta}{2} V(a)\right\}\right]
\end{align*}
with $\rho_0(a):=I\left\{c+\delta\le a\le d-r\varepsilon\right\}+\frac{\delta}{\varepsilon}I\left\{d-\varepsilon< a\le d-r\varepsilon\right\}$ and, recall, $V(a)=I\{c+\delta\le a\le  d-\varepsilon\}+\frac{d-a}{\varepsilon}\cdot I\{d-\varepsilon< a\le d-r\varepsilon\}$. Since the latent treatment bridge equation~\eqref{obs_tx_eqn0} has $\widetilde\alpha_0(a,l,u)$ on its left-hand side, the structure of $\widetilde\alpha_0$ motivates the following parametric model for the latent treatment bridge function:
\[
g(a,l,z;\eta)=\rho(a;\eta)\exp\left[\left\{\eta_0a+\eta_1l+\eta_2z+\eta_3V(a)\right\}V(a)\right],
\]
where $\eta=(\eta_0,\eta_1,\eta_2,\eta_3)^\top$.

 Using the moment generating function of $Z|A,L,U$,
\[
\mathbb{E}\left\{\exp(tZ)|A,L,U\right\}=\frac{\Phi(3-t)-\Phi(-3-t)}{\Phi(3)-\Phi(-3)}\exp\left(\left[\beta_2L+\beta_3U\right]t+\frac{t^2}{2}\right),
\]
for $t$ potentially depending on $(a,l,u)$, it follows that
\begin{align*}
    \mathbb{E}\left\{g(A,L,Z;\eta)|A,L,U\right\}&=\rho(A;\eta)\exp\left[\left\{\eta_0A+\eta_1L+\eta_3V(A)\right\}V(A)\right]\cdot\mathbb{E}\left\{\eta_2V(A)Z|A,L,U\right\}\\
    &=\rho_1(A;\eta)\exp\left[\left\{\eta_0A+\left(\eta_1+\eta_2\beta_2\right)L+\eta_2\beta_3U+\left(\eta_3+\frac{\eta_2^2}{2}\right)V(A)\right\}V(A)\right],
\end{align*}
where $\rho_1(a;\eta):=\frac{\Phi(3-\eta_2V(a))-\Phi(-3-\eta_2V(a))}{\Phi(3)-\Phi(-3)}\rho(a;\eta)$.

Hence, taking
\[
\rho(a;\eta)=\frac{\Phi(3)-\Phi(-3)}{\Phi\big\{3-\eta_2V(a)\big\}-\Phi\big\{-3-\eta_2V(a)\big\}}\cdot\rho_0(a)
\]
and
\[
\eta =\left[\delta,-\delta\left(\beta_6-\frac{\beta_2\beta_7}{\beta_3}\right), -\delta\frac{\beta_7}{\beta_3}, -\frac{\delta^2}{2}\left(1+\frac{\beta_7^2}{\beta_3^2}\right) \right]^\top,
\]
ensures that the proposed function $\hat g(a,l,z;\eta)$ exactly solves the integral equation \eqref{lat_tx_eqn0}.

When $\beta_8\neq0$, using the identity $p_{A|L,U}(a|l,u)=\int p_{A|L,U,Z}(a|l,u,z)p_{Z|L,U}(z|l,u)dz$, it can be shown that
\begin{align*}
     \widetilde\alpha_0(a,l,u)&=\tilde\rho_0(a)\exp\left[\frac{\delta }{1+\beta_8^2}V(a)\left\{a-(\beta_6+\beta_2\beta_8)l-(\beta_7+\beta_3\beta_8)u-\frac{\delta}{2}V(a)\right\}\right],
\end{align*}
where
\begin{align*}
    \tilde \rho_0(x)&=\rho_0(a)\cdot \frac{\theta\big(a-\delta V(a),l,u\big)}{\theta(a,l,u)},
\end{align*}
and
\[
\theta(a,l,u):=\int_{-3+\beta_l+\beta_3u}^{3+\beta_2l+\beta_3u}\frac{\exp\left(-\frac{1+\beta_8^2}{2}\left[z-\frac{1}{1+\beta_8^2}\big\{\beta_8a+(\beta_2-\beta_6\beta_8)l+(\beta_3-\beta_7\beta_8)u\big\}^2\right]\right)}{\Phi(2-\beta_6l-\beta_7u-\beta_8z)-\Phi(-2-\beta_6l-\beta_7u-\beta_8z)}dz.
\]
Similarly, for the parametric function $g$, we have
\begin{align*}
\mathbb{E}\big[g(&A,L,Z;\eta)|A,L,U\big]=\\
&=\rho(a;\eta)\exp\left[\left\{\eta_0A+\eta_1L+\eta_3V(A)\right\}\:V(A)\right]\cdot \mathbb{E}\left[\eta_2V(A)Z|A,L,U\right]\\
    &=\rho_1(a;\eta)\exp\Bigg(\Bigg[\left(\eta_0+\frac{\eta_2\beta_8}{1+\beta_8^2}\right)A+\left\{\eta_1+\frac{\eta_2(\beta_2-\beta_6\beta_8)}{1+\beta_8^2}\right\}L\\
&\quad\quad\quad\quad\quad\quad\quad\quad\quad\quad+\frac{\eta_2(\beta_3-\beta_7\beta_8)}{1+\beta_8^2}U+\left\{\eta_3+\frac{\eta_2^2}{2(1+\beta_8^2)}\right\}V(A)\Bigg]V(A)\Bigg),
\end{align*}
where
\begin{align*}
    \tilde \rho_1(a;\eta):=\rho(a;\eta)\cdot  \frac{\theta\left(a+\frac{\eta_2V(a)}{\beta_8},l,u\right)}{\theta(a,l,u)}.
\end{align*}
For fixed $(a,l,u)$, define $\xi(\beta_8):=\frac{\theta\big(a-\delta V(a),l,u\big)}{\theta\left(a+\frac{\eta_2V(a)}{\beta_8},l,u\right)}$. When $\beta_8$ is small, $\xi'(0)\approx 0$, and a first-order Taylor expansion gives
\[
\xi(\beta_8)\approx \xi(0)=\frac{\Phi(3)-\Phi(-3)}{\Phi\big\{3-\eta_2V(a)\big\}-\Phi\big\{-3-\eta_2V(a)\big\}}.
\]

Hence, taking
\[
\eta = \left[\frac{\delta\beta_3}{\beta_3-\beta_7\beta_8}, -\frac{\delta(\beta_3\beta_6-\beta_2\beta_7)}{\beta_3-\beta_7\beta_8},-\frac{\delta(\beta_7+\beta_3\beta_8)}{\beta_3-\beta_7\beta_8},-\frac{\delta^2}{(\beta_3-\beta_7\beta_8)^2}\left\{\frac{\beta_3^2+\beta_7^2}{2}\right\}\right],
\]
and
\[
\rho(a;\eta)=\frac{\Phi(3)-\Phi(-3)}{\Phi\big\{3-\eta_2V(a)\big\}-\Phi\big\{-3-\eta_2V(a)\big\}}\cdot \rho_0(a),
\]
yields a good approximation to the true bridge function for small $\beta_8$.

\begin{figure}
\centerline{\includegraphics[width=6.5in]{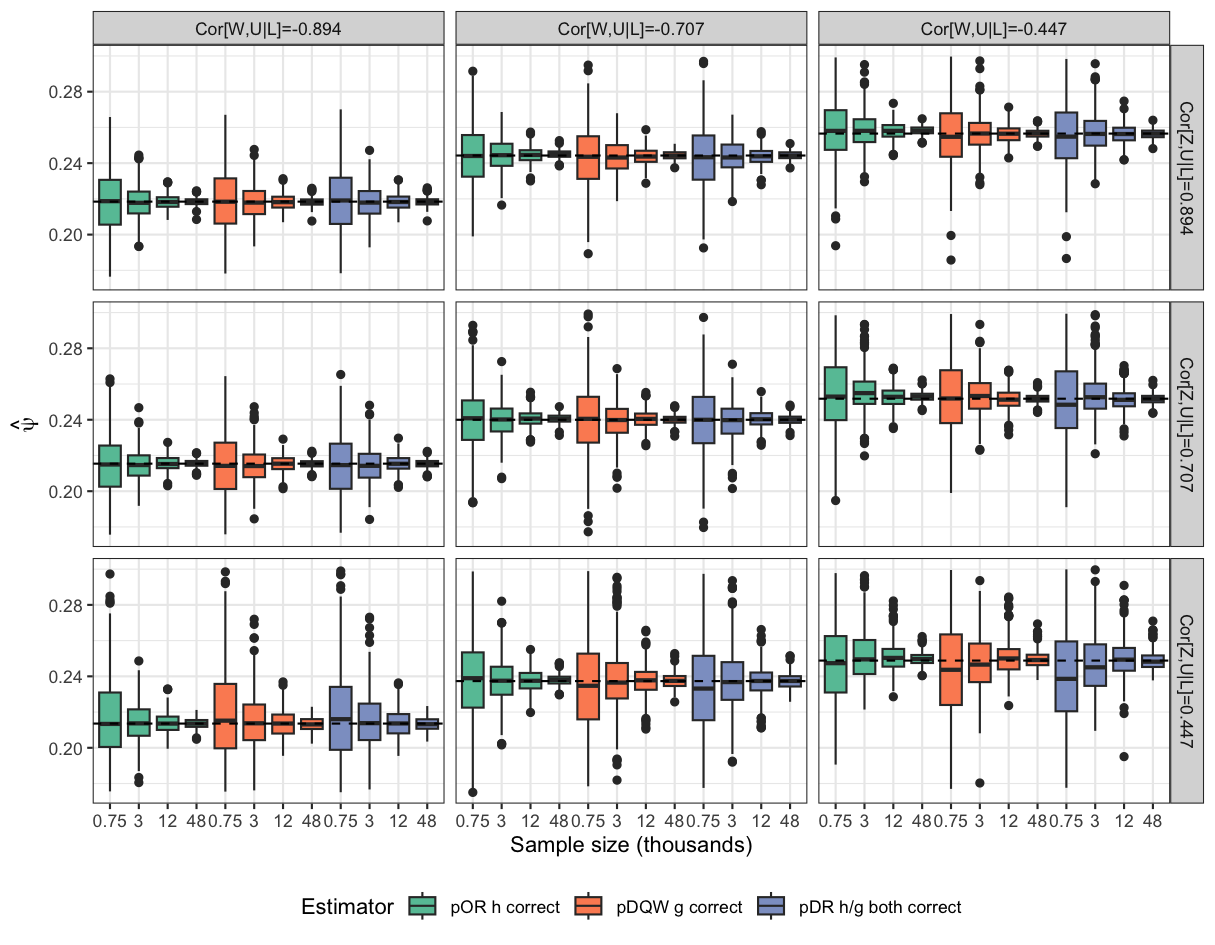}}
    \caption{Boxplots of the \textit{correctly specified} parametric estimators from numerical experiments under a more challenging data-generated mechanism, where $Z$ directly affects $A$ and $W$ directly affects $Y$. Results are shown across varying conditional correlations of the proxies $Z$ and $W$ with the unmeasured confounder $U$, and sample sizes $n=750, 3000, 12000$, and 48000. Dashed lines indicate the true parameter values.}\label{fig:num_exp_param_bp_case2}
\end{figure}

\begin{figure}
\centerline{\includegraphics[width=6.5in]{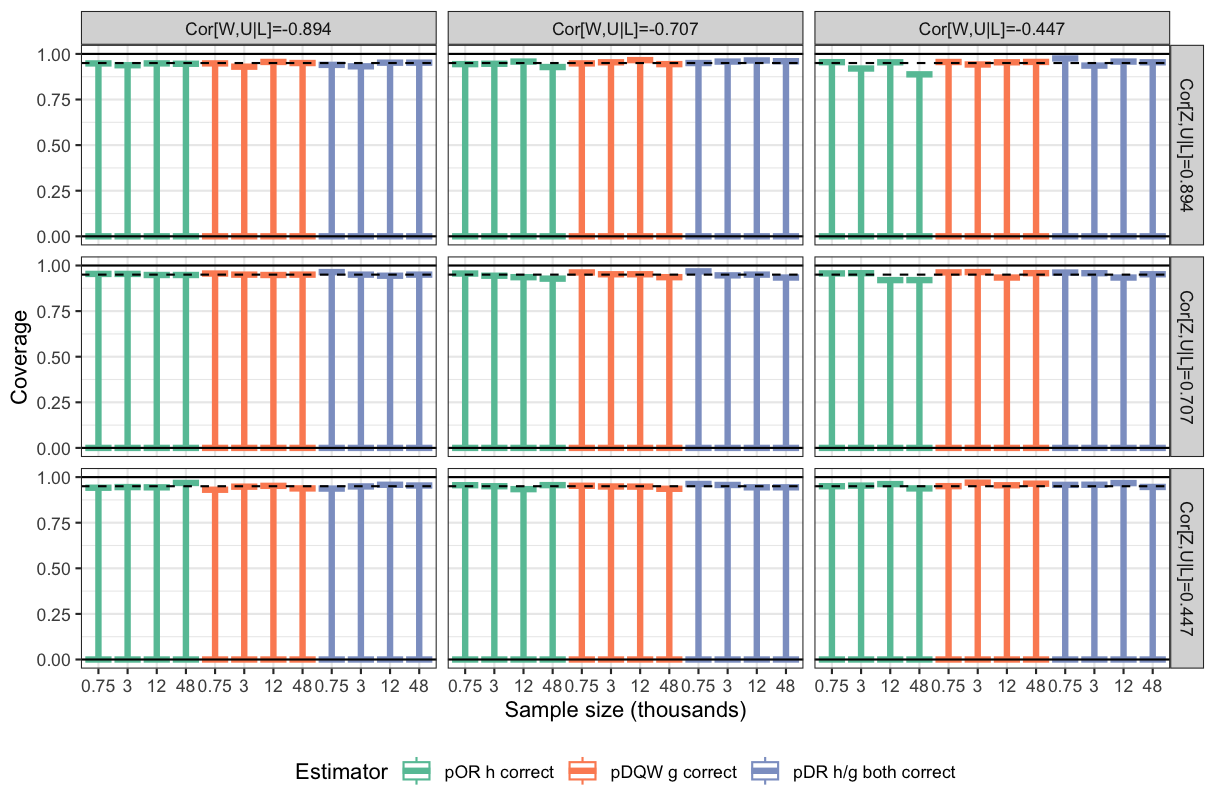}}
    \caption{Empirical coverage of 95\% confidence intervals of \textit{correctly specified}  parametric estimators from numerical experiments under a more challenging data-generated mechanism, where $Z$ directly affects $A$ and $W$ directly affects $Y$. Results are shown across varying conditional correlations of the proxies $Z$ and $W$ with the unmeasured confounder $U$, and sample sizes $n=750, 3000, 12000$, and 48000. Dashed lines indicate the true parameter values. The dashed line indicates the nominal coverage level of 0.9.}\label{fig:num_exp_param_cov_case2}
\end{figure}

While this derivation justifies the parametric model for small $\beta_8$, the same functional form can be used in practice for moderate values of $\beta_8$. Our numerical experiments, using this parametric model under a data-generating mechanism with $\beta_8 = 0.3$ and $\beta_{12}=-0.3$, demonstrate good finite-sample performance for estimating $\psi_0$ (see Figures~\ref{fig:num_exp_param_bp_case2} and~\ref{fig:num_exp_param_cov_case2}).

\bibliographystyle{apalike}
\bibliography{references}

\begin{thebibliography}{}

\bibitem[Boyd and Vandenberghe, 2004]{boyd2004convex}
Boyd, S. and Vandenberghe, L. (2004).
\newblock {\em Convex optimization}.
\newblock Cambridge university press.

\bibitem[Carrasco et~al., 2007]{carrasco2007linear}
Carrasco, M., Florens, J.-P., and Renault, E. (2007).
\newblock Linear inverse problems in structural econometrics estimation based on spectral decomposition and regularization.
\newblock {\em Handbook of econometrics}, 6:5633--5751.

\bibitem[Chernozhukov et~al., 2018]{chernozhukov2018double}
Chernozhukov, V., Chetverikov, D., Demirer, M., Duflo, E., Hansen, C., Newey, W., and Robins, J. (2018).
\newblock Double/debiased machine learning for treatment and structural parameters.
\newblock {\em The Econometrics Journal}, 21(1):C1–C68.

\bibitem[Cui et~al., 2024]{cui2024semiparametric}
Cui, Y., Pu, H., Shi, X., Miao, W., and Tchetgen~Tchetgen, E. (2024).
\newblock Semiparametric proximal causal inference.
\newblock {\em Journal of the American Statistical Association}, 119(546):1348--1359.

\bibitem[D{\'\i}az et~al., 2023]{diaz2023nonparametric}
D{\'\i}az, I., Williams, N., Hoffman, K.~L., and Schenck, E.~J. (2023).
\newblock Nonparametric causal effects based on longitudinal modified treatment policies.
\newblock {\em Journal of the American Statistical Association}, 118(542):846--857.

\bibitem[D{\'\i}az~Mu{\~n}oz and Van Der~Laan, 2012]{diaz2012population}
D{\'\i}az~Mu{\~n}oz, I. and Van Der~Laan, M. (2012).
\newblock Population intervention causal effects based on stochastic interventions.
\newblock {\em Biometrics}, 68(2):541--549.

\bibitem[Dikkala et~al., 2020]{dikkala2020minimax}
Dikkala, N., Lewis, G., Mackey, L., and Syrgkanis, V. (2020).
\newblock Minimax estimation of conditional moment models.
\newblock {\em Advances in Neural Information Processing Systems}, 33:12248--12262.

\bibitem[Feng et~al., 2021]{feng2021correlates}
Feng, S., Phillips, D.~J., White, T., Sayal, H., Aley, P.~K., Bibi, S., Dold, C., Fuskova, M., Gilbert, S.~C., Hirsch, I., et~al. (2021).
\newblock Correlates of protection against symptomatic and asymptomatic sars-cov-2 infection.
\newblock {\em Nature medicine}, 27(11):2032--2040.

\bibitem[Fleming and Powers, 2012]{fleming2012biomarkers}
Fleming, T.~R. and Powers, J.~H. (2012).
\newblock Biomarkers and surrogate endpoints in clinical trials.
\newblock {\em Statistics in medicine}, 31(25):2973--2984.

\bibitem[Fong et~al., 2023]{fong2023immune}
Fong, Y., Huang, Y., Benkeser, D., Carpp, L.~N., {\'A}{\~n}ez, G., Woo, W., McGarry, A., Dunkle, L.~M., Cho, I., Houchens, C.~R., et~al. (2023).
\newblock Immune correlates analysis of the prevent-19 covid-19 vaccine efficacy clinical trial.
\newblock {\em Nature Communications}, 14(1):331.

\bibitem[Fong et~al., 2022]{fong2022immune}
Fong, Y., McDermott, A.~B., Benkeser, D., Roels, S., Stieh, D.~J., Vandebosch, A., Le~Gars, M., Van~Roey, G.~A., Houchens, C.~R., Martins, K., et~al. (2022).
\newblock Immune correlates analysis of the ensemble single ad26. cov2. s dose vaccine efficacy clinical trial.
\newblock {\em Nature Microbiology}, 7(12):1996--2010.

\bibitem[Garreau et~al., 2017]{garreau2017large}
Garreau, D., Jitkrittum, W., and Kanagawa, M. (2017).
\newblock Large sample analysis of the median heuristic.
\newblock {\em arXiv preprint arXiv:1707.07269}.

\bibitem[Ghassami et~al., 2022]{ghassami2022minimax}
Ghassami, A., Ying, A., Shpitser, I., and Tchetgen, E.~T. (2022).
\newblock Minimax kernel machine learning for a class of doubly robust functionals with application to proximal causal inference.
\newblock In {\em International conference on artificial intelligence and statistics}, pages 7210--7239. PMLR.

\bibitem[Gilbert et~al., 2022a]{gilbert2022covid}
Gilbert, P.~B., Donis, R.~O., Koup, R.~A., Fong, Y., Plotkin, S.~A., and Follmann, D. (2022a).
\newblock A covid-19 milestone attained—a correlate of protection for vaccines.
\newblock {\em New England Journal of Medicine}, 387(24):2203--2206.

\bibitem[Gilbert et~al., 2022b]{gilbert2022immune}
Gilbert, P.~B., Montefiori, D.~C., McDermott, A.~B., Fong, Y., Benkeser, D., Deng, W., Zhou, H., Houchens, C.~R., Martins, K., Jayashankar, L., et~al. (2022b).
\newblock Immune correlates analysis of the mrna-1273 covid-19 vaccine efficacy clinical trial.
\newblock {\em Science}, 375(6576):43--50.

\bibitem[Haneuse and Rotnitzky, 2013]{haneuse2013estimation}
Haneuse, S. and Rotnitzky, A. (2013).
\newblock Estimation of the effect of interventions that modify the received treatment.
\newblock {\em Statistics in medicine}, 32(30):5260--5277.

\bibitem[Hejazi et~al., 2021]{hejazi2021efficient}
Hejazi, N.~S., van~der Laan, M.~J., Janes, H.~E., Gilbert, P.~B., and Benkeser, D.~C. (2021).
\newblock Efficient nonparametric inference on the effects of stochastic interventions under two-phase sampling, with applications to vaccine efficacy trials.
\newblock {\em Biometrics}, 77(4):1241--1253.

\bibitem[Huang et~al., 2023]{huang2023stochastic}
Huang, Y., Hejazi, N.~S., Blette, B., Carpp, L.~N., Benkeser, D., Montefiori, D.~C., McDermott, A.~B., Fong, Y., Janes, H.~E., Deng, W., et~al. (2023).
\newblock Stochastic interventional vaccine efficacy and principal surrogate analyses of antibody markers as correlates of protection against symptomatic covid-19 in the cove mrna-1273 trial.
\newblock {\em Viruses}, 15(10):2029.

\bibitem[Kallus et~al., 2022]{kallus2022causal}
Kallus, N., Mao, X., and Uehara, M. (2022).
\newblock Causal inference under unmeasured confounding with negative controls: A minimax learning approach.
\newblock {\em arXiv preprint arXiv:2103.14029}.

\bibitem[Kress, 2014]{kress2014}
Kress, R. (2014).
\newblock {\em Linear Integral Equations}, volume~82 of {\em Applied Mathematical Sciences}.
\newblock Springer, 3 edition.

\bibitem[Miao et~al., 2018]{miao2018identifying}
Miao, W., Geng, Z., and Tchetgen~Tchetgen, E.~J. (2018).
\newblock Identifying causal effects with proxy variables of an unmeasured confounder.
\newblock {\em Biometrika}, 105(4):987--993.

\bibitem[Olivas-Martinez and Rotnitzky, 2025]{olivas2025source}
Olivas-Martinez, A. and Rotnitzky, A. (2025).
\newblock Source-condition analysis of kernel adversarial estimators.
\newblock {\em arXiv preprint arXiv:2508.17181}.

\bibitem[Plotkin, 2010]{plotkin2010correlates}
Plotkin, S.~A. (2010).
\newblock Correlates of protection induced by vaccination.
\newblock {\em Clinical and vaccine immunology}, 17(7):1055--1065.

\bibitem[Robins, 1986]{robins1986new}
Robins, J. (1986).
\newblock A new approach to causal inference in mortality studies with a sustained exposure period—application to control of the healthy worker survivor effect.
\newblock {\em Mathematical modelling}, 7(9-12):1393--1512.

\bibitem[Robins et~al., 2008]{robins2008higher}
Robins, J., Li, L., Tchetgen, E., van~der Vaart, A., et~al. (2008).
\newblock Higher order influence functions and minimax estimation of nonlinear functionals.
\newblock In {\em Probability and statistics: essays in honor of David A. Freedman}, volume~2, pages 335--422. Institute of Mathematical Statistics.

\bibitem[Sadoff et~al., 2022]{sadoff2022final}
Sadoff, J., Gray, G., Vandebosch, A., C{\'a}rdenas, V., Shukarev, G., Grinsztejn, B., Goepfert, P.~A., Truyers, C., Van~Dromme, I., Spiessens, B., et~al. (2022).
\newblock Final analysis of efficacy and safety of single-dose ad26. cov2. s.
\newblock {\em New England Journal of Medicine}, 386(9):847--860.

\bibitem[Sch{\"o}lkopf et~al., 2001]{scholkopf2001generalized}
Sch{\"o}lkopf, B., Herbrich, R., and Smola, A.~J. (2001).
\newblock A generalized representer theorem.
\newblock In {\em International conference on computational learning theory}, pages 416--426. Springer.

\bibitem[Shi et~al., 2020]{shi2020multiply}
Shi, X., Miao, W., Nelson, J.~C., and Tchetgen~Tchetgen, E.~J. (2020).
\newblock Multiply robust causal inference with double-negative control adjustment for categorical unmeasured confounding.
\newblock {\em Journal of the Royal Statistical Society Series B: Statistical Methodology}, 82(2):521--540.

\bibitem[Sholukh et~al., 2020]{sholukh2020evaluation}
Sholukh, A.~M., Fiore-Gartland, A., Ford, E.~S., Hou, Y.~J., Tse, V., Kaiser, H., Park, A., Lempp, F.~A., Germain, R.~S., Bossard, E., et~al. (2020).
\newblock Evaluation of sars-cov-2 neutralization assays for antibody monitoring in natural infection and vaccine trials.
\newblock {\em medRxiv}, pages 2020--12.

\bibitem[Tchetgen~Tchetgen et~al., 2024]{tchetgen2024introduction}
Tchetgen~Tchetgen, E.~J., Ying, A., Cui, Y., Shi, X., and Miao, W. (2024).
\newblock An introduction to proximal causal inference.
\newblock {\em Statistical Science}, 39(3):375--390.

\bibitem[Van~der Vaart, 2000]{van2000asymptotic}
Van~der Vaart, A.~W. (2000).
\newblock {\em Asymptotic statistics}, volume~3.
\newblock Cambridge university press.

\end{thebibliography}

\end{document}